\documentclass[nofootinbib, secnumarabic, amsmath, nobibnotes,10pt,aps,pra]{revtex4-1}

\usepackage{graphicx}

\newtheorem{problem}{Problem}
\newtheorem{solution}{Solution}

\newcommand{\ket}[1]{\ensuremath{|#1\rangle}}
\newcommand{\bra}[1]{\ensuremath{\langle#1|}}
\newcommand{\braket}[2]{\langle#1|#2\rangle}

\newcommand{\fref}[1]{Fig. \ref{#1}}
\newcommand{\sref}[1]{Sec. \ref{#1}}
\newcommand{\eref}[1]{Eq. (\ref{#1})}
\newcommand{\tref}[1]{Table (\ref{#1})}
\newcommand{\Fref}[1]{Figure \ref{#1}}

\newcommand{\Eref}[1]{Equation (\ref{#1})}

\newcommand{\twoconds}[2]{\begin{smallmatrix} #1 \\ #2 \end{smallmatrix}}

\newcommand{\abs}[1]{\left| #1 \right|} 
\newcommand{\avg}[1]{\left\langle #1 \right\rangle} 

\newcommand{\matrixel}[3]{\left\langle #1 \vphantom{#2#3} \right|
#2 \left| #3 \vphantom{#1#2} \right\rangle} 

\begin{document}

\begin{figure}
\centering
\includegraphics*[width=0.90\textwidth]{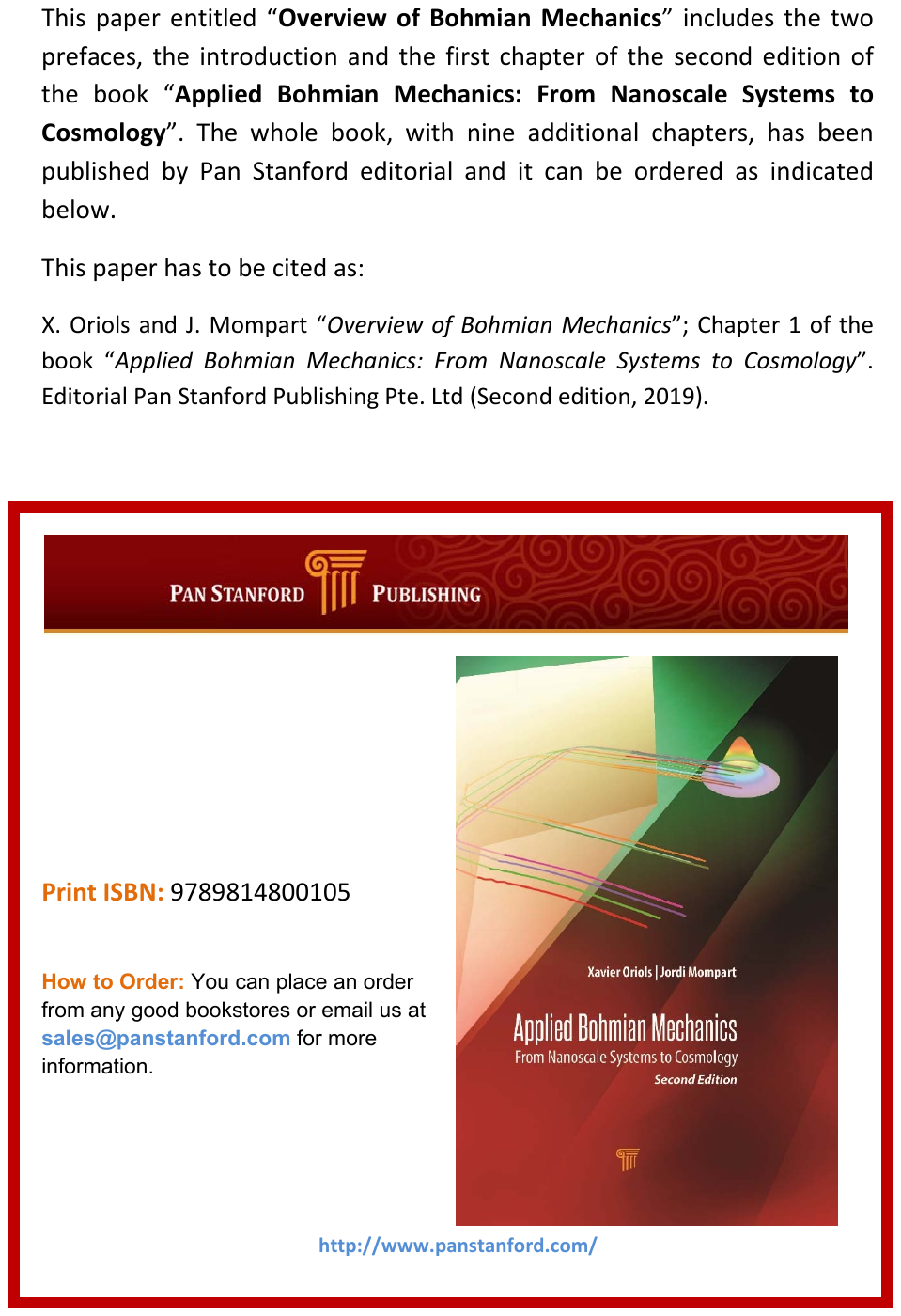}
\nonumber
\end{figure}

\title{Overview of Bohmian Mechanics}%

\author{Xavier Oriols\textsuperscript{a} and Jordi Mompart\textsuperscript{b}}
\email[E-mail: ]{xavier.oriols@uab.cat; jordi.mompart@uab.cat}
\affiliation{\textsuperscript{a}Departament d'Enginyeria Electr\`{o}nica, Universitat Aut\`{o}noma de Barcelona, 08193, Bellaterra, SPAIN \\\textsuperscript{b}Departament de F\'isica, Universitat Aut\`{o}noma de Barcelona, 08193 Bellaterra, SPAIN}

\begin{abstract}
This chapter provides a fully comprehensive overview of the Bohmian formulation of quantum phenomena. It starts with a historical review of the difficulties found by Louis de Broglie, David Bohm  and John Bell to convince the scientific community about the validity and utility of Bohmian mechanics. Then, a formal explanation of Bohmian mechanics for non-relativistic single-particle quantum systems is presented. The generalization to many-particle systems, where correlations play an important role, is also explained.
After that, the measurement process in Bohmian mechanics is discussed. It is emphasized that Bohmian mechanics exactly reproduces the mean value and temporal and spatial correlations obtained from the standard, i.e., `orthodox', formulation. The ontological characteristics of the Bohmian theory provide a description of measurements in a natural way, without the need of introducing stochastic operators for the wavefunction collapse. Several solved problems are presented at the end of the chapter giving additional mathematical support to some particular issues.
A detailed description of computational algorithms to obtain Bohmian trajectories from the numerical solution of the Schr\"odinger or the Hamilton--Jacobi equations are presented in an appendix.
The motivation of this chapter is twofold. First, as a didactic introduction of the Bohmian formalism which is used in the subsequent chapters. Second, as a self-contained summary for any newcomer interested in using Bohmian mechanics in their daily research activity.
\end{abstract}

\maketitle

\tableofcontents

\section{Preface first edition\label{preface}}
\subsection*{New cutting edge ideas come from outside of the main stream}

Most of our collective activities are regulated by other people who decide whether they are well done or not. One has to learn some arbitrary symbols to write understandable messages, or to read those from others. Human rules over collective activities govern the evolution of our culture. On the contrary, natural systems, from atoms to galaxies, evolve independently of the human rules. Men cannot modify physical laws. We can only try to understand them. Nature itself judges, through experiments, whether a plausible explanation for some natural phenomena is correct or not. Nevertheless, in forefront research where the unknowns start to become understandable, the new knowledge is still unstable, somehow immature. It is supported by few experimental evidences or the evidences are still subjected to different interpretations. Certainly, novel research grows up closely tied to the economical, sociological or historical circumstances of the involved researchers.  A period of time is needed in order to distil new knowledge, separating pure scientific arguments from cultural influences.

The past and the present status of Bohmian mechanics cannot be understood without these cultural considerations. The Bohmian formalism was proposed by Louis de Broglie even before the standard, that is Copenhagen, explanation of quantum phenomena was established.
Bohmian mechanics provides an explanation of quantum phenomena in terms of particles guided by waves.
One object cannot be a wave and a particle simultaneously, but two can. Specially, if one of the objects is a wave and the other is a particle. 
Unfortunately, Louis de Broglie, himself, abandoned these ideas. Later, in the fifties, David Bohm clarified the meaning and applications of this original explanation of quantum phenomena. Bohmian mechanics agrees with all non-relativistic quantum experiments done up to now. However, it remains almost ignored by most of the scientific community. In our opinion, there are no scientific arguments to support its marginal status, but only cultural reasons. One of the motivations for writing this book is helping in the maturing process that Bohmian mechanics still needs.

Certainly, the distilling process of Bohmian mechanics is being quite slow. Anyone interested enough to walk this causal road of quantum mechanics can be easily confused by many misleading signposts that have been raised in the scientific literature, not only by its detractors, but unfortunately very often also by some of its advocates. Following opinions from other reputed physicists (we are easily persuaded by those scientists with authority) is far from being a proper scientific strategy.

In any case, since the mathematical structure of Bohmian mechanics is quite simple, it can be easily learned by anyone with only a basic knowledge of classical and quantum mechanics who makes the necessary effort to build his own scientific opinion based on logical deductions, free from cultural influences. The introductory chapter of this book, including a thorough list of exercises and easily programmable codes, provides a reasonable and objective source of information in order to achieve this later goal, even for undergraduate students.

Curiously, the fact that Bohmian mechanics is ignored and remains mainly unexplored is an attractive feature for some adventurous scientists. They know that very often \emph{new cutting edge ideas come from outside of the main stream} and find in Bohmian mechanics a useful tool in their research activity. On the one hand, it provides an explanation of quantum mechanics, in terms of trajectories, that results to be very useful in explaining the dynamics of quantum systems, being thus also a source of inspiration to look for novel quantum phenomena. On the other hand, since it provides an alternative mathematical formulation, Bohmian mechanics offers new computational tools to explore physical scenarios that presently are computationally inaccessible, such as many particle solutions of the Schr\"odinger equation. In addition, Bohmian mechanics sheds light on the limits and extensions of our present understanding of quantum mechanics towards other paradigms such as relativity or cosmology, where the internal structure of Bohmian mechanics in terms of well-defined trajectories is very attractive. With all these previous motivations in mind, this book provides nine chapters (apart from the introduction in the first chapter) with practical examples showing how Bohmian mechanics helps us in our daily research activity.

Obviously, there are other books focused on Bohmian mechanics. However, many of them are devoted to the foundations of quantum mechanics emphasizing the difficulties or limitations of the Copenhagen interpretation for providing an ontological description of our world. On the contrary, this book is not focus on the foundations of quantum mechanics, but on the discussion about the practical application of the ideas of de Broglie and Bohm to understand and compute the quantum world. Several examples of such practical applications written by leading experts in different fields, with an extensive updated bibliography, are provided here. The book, in general, is addressed to students in physics, chemistry, electrical engineering, applied mathematics, nanotechnology, as well as to both theoretical and experimental researchers who seek new computational and interpretative tools for their everyday research activity. We hope that the newcomers to this causal explanation of quantum mechanics will use Bohmian mechanics in their research activity so that Bohmian mechanics will become more and more popular for the broad scientific community. If so, we expect that, in the near future, Bohmian mechanics will be taught regularly at the Universities, not as the unique and revolutionary way of understanding quantum phenomena, but as an additional and useful interpretation of all quantum phenomena in terms of quantum trajectories. In fact, Bohmian mechanics has the ability of removing most of the mysteries of the Copenhagen interpretations and, somehow, simplifying (or demystifying) quantum mechanics. We will be very glad if this book can contribute to shorten the time needed to achieve all these goals.

Finally, we want to acknowledge many different people that has allowed us to embark into and successfully finish this book project. First of all, Alfonso Alarc\'on and Albert Benseny who became involved in the book project from the very beginning, as two additional editors. We also want to thank the rest of the authors of the book for accepting our invitation to participate in this project and writing their chapters according to the general spirit of the book. Due to page limitations, only nine examples of practical applications of Bohmian mechanics in forefront research activity are presented in this book. Therefore, we want to apologize to many other researchers who could have certainly been also included in the book. We also want to express our gratitude to \emph{Pan Stanford Publishing} for accepting our book project and for their kind attention during the publishing process.

\section{Preface second edition:\label{preface2}}
\subsection*{Quantum engineering: a giant with feet of clay}

More than five years after the first edition of our book on applied Bohmian mechanics, our original motivations for writing it are still very present. Certainly, today, a lot of publicity about the abilities of the Bohmian theory is still needed among the scientific community. In fact, over these last few years, many research programs have been devoted to the so-called quantum engineering with the goal of developing new materials, new sensors, or new computing strategies based on pure quantum phenomena. Thus, this book can be understood as a promotional presentation on how the Bohmian theory, among others, can help in the design and development of such applications. However, Bohmian mechanics is not a mere computational tool in terms of quantum trajectories, but a complete and ontological theory that provides a consistent explanation on how Nature works. In this regard, this book can also be seen as a useful exercise to sincerely question our present understanding of the physical laws that govern the quantum world. After a bit of reflection on this point, many of us will probably conclude that our knowledge about the fundamentals of the quantum world are much more immature and imprecise than what we previously thought.  In this sense, we want this book to be a warning on the the risks of constructing the new and exciting discipline of quantum engineering as a \emph{giant with feet of clay}. 

The beginning of the 20th century saw the first quantum revolution where novel and original theories were developed to understand unexpected non-classical phenomena. What determines the structure of the periodic table? Why are some materials metals and some dielectrics, while others behave like semiconductors? Nowadays, having established answers to these basic questions, a second quantum revolution is starting to take place, focusing on actively capitalizing on our quantum knowledge to alter the face of the physical world, developing a myriad of new quantum technologies. The difference between these two quantum revolutions is just the difference between science and engineering. The first revolution tried to properly understand our physical surroundings, the natural objects around us, while the second one intends to manipulate these surroundings to our own benefit. This is the typical evolution of most scientific disciplines. When scientific knowledge is mature enough, and the necessary technological means are available, engineers can use this knowledge for practical applications.

It is a common belief in our society that quantum theory, after more than a century, is ready to take a leap towards the engineering field. We have certainly outstanding technological means to manipulate quantum systems, even individual atoms, at the nanometer and femtosecond scales. Therefore, many national or international research organizations are focusing their programs towards the effective development of quantum technologies, trying to ensure that money spent on science has a direct impact on our society and its challenges for a better life. This is indeed a legitimate and compelling goal. However, is the quantum theory mature enough to blindly jump from science towards engineering? The pressure from society (in terms of research programs, grants, citing indices, etc.) is so effective that it forces most of the scientists to forget about the maturity of the quantum theory and just focus on (what really matters) the fast development of practical applications in the new and exciting field of quantum engineering.

We argue that the development of quantum engineering cannot be done at the price of forgetting the need for a deeper understanding of the physical laws governing the quantum world. One of the \emph{forgotten} discussions by the new generation of quantum engineers is the \emph{measurement problem}, which remains inside the backbone of the quantum theory. The measurement problem is manifested in the orthodox theory by its failure in explaining which physical interactions among particles constitute a measurement and which do not. In fact, there are many more examples of the immaturity of our quantum knowledge. Our inability to properly describe many-body systems due to their exponential complexity (the so-called many-body problem) makes that most of our understanding is based on a \emph{puerile} single-particle description. We do not have a clear physical picture on the quantum-to-classical transition. What makes a quantum system to behave classically in some circumstances? The fact that there are several quantum theories which are empirically equivalent but radically different at the ontological level is a clear evidence of our bad understanding. The Copenhagen theory is the most extensively investigated and presently the one with more support among the scientific community. Others include spontaneous collapse theories or the many-worlds theory. The one studied in this book, Bohmian mechanics, provides a description of quantum phenomena by particles choreographed by the wave function. In general, neither of these theories is more mature than the orthodox one, but they remove the need of an observer, which relaxes some of the difficulties to understand the measurement at a quantum level. Some of these alternative theories are not free of problems, including the quantum-to-classical transition and the many-body problem, while others still need to be dealt with. We do not mean to imply that these alternative theories are better (how does one quantify \emph{better} here?), but that there is still a lot of work needed to certify that our comprehension of the quantum world is unproblematic. 

Let us try to exemplify the risks of developing quantum engineering alone without worrying on its fundamentals. In the orthodox theory, every time we make a measurement a random process occurs. But, as we do not really know what makes a physical interaction to be a measurement, we really do not know the origin of such randomness. In the Bohmian theory, for example, this randomness comes from an uncertainty in the initial position of the particles. With further efforts to clarify the quantum theories, we can perhaps achieve a better understanding of the origin of quantum randomness and then, the exciting new building of application developed along the new discipline of quantum cryptography, based on the unavoidable presence of such intrinsic quantum randomness, will simple melt as a \emph{giant with feet of clay}. The reader can argue that there are a lot of scientific works supporting the actual status of quantum cryptography. Perhaps this particular warning is completely unfounded and quantum cryptography will certainly remain as robust as we know today. But, perhaps not. It is enlightening to remember here the theorem that John von Neumann stated in 1932 about the impossibility of explaining quantum mechanics with hidden variables (such as quantum trajectories). This theorem remained an unquestionable truth, and part of the essence of the quantum world, until David Bohm (with an explanation of quantum phenomena in terms of waves and particles) showed that the theorem was wrong (as its own preliminary assumptions precludes the existence of Bohmian trajectories). The curious spectacle is not that John von Neumann (an outstanding scientist in many disciplines) made a mistake in a theorem, but that the community (with the exception of Grete Hermann in 1935 that was totally ignored) blindly accepted the theorem for almost half a century.

There are many more examples which certify that our understanding of the quantum world is still immature. The wave function, the basic element in most theories, can be prepared for instance by forcing the quantum system into its ground state, but it cannot be directly measured in a single experiment. The wave function can be measured through a weak protocol, but also Bohmian velocities can be measured though such protocol. We do not even know what is really the wave function at the most fundamental level: a law? a field? a probability transporter? 

In summary, the quantum world is so complicated that one century has not been enough for the scientific community to clearly elaborate an unproblematic description of the laws of quantum mechanics. We are not arguing here that research on quantum engineering needs to stop. Just the contrary. The development of the quantum engineering and the research on the foundations of the quantum theory have to evolve intimately connected to benefice from each other in achieving better practical applications and a deeper understanding of the quantum world. Otherwise, we will build a \emph{giant with feet of clay}. We hope that the present book can be viewed as a modest contribution in both directions.

\section{Introduction \label{introduction}}

The beginning of the twentieth century brought surprising non-classical phenomena. Max Planck's explanation of the black body radiation \cite{om.Planck-BlackBody}, the work of Albert Einstein on the photoelectric effect \cite{om.Einstein-Photoelectric} and the Niels Bohr's model to account for the electron orbits around the nuclei \cite{om.bohr,om.bohr2,om.bohr3} established what is now known as the \textit{old quantum theory}.
To describe and explain these effects, phenomenological models and theories were first developed, without rigorous justification. In order to provide a complete explanation for the underlying physics of such new non-classical phenomena, physicists were forced to abandon classical mechanics to develop novel, abstract and imaginative formalisms.

In 1924, Louis de Broglie suggested in his doctoral thesis that matter, apart from its intrinsic particle-like behavior, could exhibit also a wave-like one \cite{om.dB_AnnPhys}. Three years later he proposed an interpretation of quantum phenomena based on non-classical trajectories guided by a \emph{wave field}~\cite{om.debroglie1927b}.
This was the origin of the pilot-wave formulation of quantum mechanics that we will refer as Bohmian mechanics to account for the following work of David Bohm \cite{om.bohm1952a,om.bohm1952b}. In the Bohm formulation, an individual quantum system is formed by a point particle and a guiding wave.
Contemporanely, Max Born and Werner Heisenberg, in the course of their collaboration in Copenhagen with Niels Bohr, provided an original formulation of quantum mechanics without the need of trajectories \cite{om.Born1926,om.Heisenber1925}. This was the origin of the so-called Copenhagen interpretation of quantum phenomena and, since it is the most accepted formulation, it is basically the only one explained at most universities. Thus, it is also known as the \emph{orthodox} formulation of quantum mechanics. In the Copenhagen interpretation, an individual quantum system exhibits its wave or its particle nature depending on the experimental arrangement.

The present status of Bohmian mechanics among the scientific community is quite marginal (the quantum chemistry community is an encouraging exception). Most researchers do not know about it or believe that is not fully-correct. There are others that know that quantum phenomena can be interpreted in terms of trajectories, but they think that this formalism cannot be useful in their daily research activity. Finally, there are few researchers, the authors of this book among them, who think that Bohmian mechanics is a useful tool to make progress in front-line research fields involving quantum phenomena.

The main (non-scientific) reason why still many researchers believe that there is something wrong with Bohmian mechanics can be illustrated with Hans Christian Andersen's tale \emph{The Emperor's new clothes}. Two swindlers promise the Emperor the finest clothes that, as they tell him,  are invisible to anyone who is unfit for their position. The Emperor cannot see the (non-existing) clothes, but pretends that he can for fear of appearing stupid.  The rest of the people do the same. Advocates of the Copenhagen interpretation have attempted to produce \emph{impossibility proofs} in order to demonstrate that Bohmian mechanics is incompatible with quantum phenomena \cite{om.impossibility_proofs}. Most researchers, who are not aware of the incorrectness of such \emph{proofs}, might conclude that there is some controversy with the Bohmian formulation of quantum mechanics and they prefer not to support it, for fear of appearing discordant. At the end of the tale, during the course of a procession, a small child cries out ``\textit{the Emperor is naked!}''. In the tale of quantum mechanics, David Bohm \cite{om.bohm1952a,om.bohm1952b} and John Bell \cite{om.Bell1987} were the first to exclaim to the scientific community ``\textit{Bohmian mechanics is a correct interpretation of quantum phenomena that exactly coincides with the predictions of the orthodox interpretation!}''.

\subsection{What is a quantum theory}
\label{sec_onto}

Albert Einstein, in the paper entitled ``Physics and reality'' \cite{einstein}, pointed out the possibility of living in a bizarre world without comprehensible explanations for natural phenomena. He wrote: \emph{``The fact that [the world] is comprehensible is a miracle"}. Similarly, Eugene Wigner wrote: \emph{``The unreasonable effectiveness of mathematics in the natural science ....is a wonderful gift which we neither understand nor deserve"} \cite{wigner}. Both reflections were inspired by the previous work of the German philosopher Immanuel Kant who wrote the very same idea almost two centuries before: \emph{``The eternal mystery of the world is its comprehensibility."}. Fortunately, it seems that we  live in a comprehensible world.

Kant divided scientific knowledge into three  categories: appearance, reality and theory. Appearance is the content of our sensory experience of natural phenomena, which is the empirical outcome of an experiment. Reality is what lies behind all natural phenomena. A theory is a human model that tries to mirror both appearance and reality. A useful theory might predict the outcome of an experiment in a laboratory or the observation of a phenomenon in Nature. Empiricists believe on experimental outcomes (what Kant called appearance) and refuse to speculate about a deeper reality. On the other hand, realists believe that good physical theories explain, or at least provide clues about, the reality of our comprehensible world. Most researchers are a combination of both stereotypes, with variable proportions.

As all human creations, there are successful and unsuccessful theories. When in 1864 James Clerk Maxwell conjectured that light was an electromagnetic vibration, it was believed that all waves had to vibrate in some medium. The medium in which light presumably travels was named  \emph{luminiferous ether}. During almost a century eminent scientists believed blindly on that concept. Nowadays, the \emph{luminiferous ether} plays no role at all in modern physical theories\cite{om.herbert}. The atomicity of matter is an example of a very successful theory. It was introduced by the British chemist John Dalton in 1808 to explain why some chemical substances need to combine in some fixed ratios. During one century it was thought that atoms were a crazy idea. Marcelin Berthelot said \emph{``who [has] ever seen a gas molecule  or an atom?''}, expressing the disdain that many chemists felt for the unseen atoms, which were inaccessible to experiments \cite{om.herbert}. Even their defenders saw little hope of ever directly verifying the atomic hypothesis. Nowadays, the fact that everything is made of atoms is one of the most precious knowledge that we get on how Nature works\cite{feynmann}, and their images  are even routinely seen in the screens of scanning tunneling microscopes \cite{binning}. 

A quantum theory is a human explanation of quantum phenomena. All quantum theories have associated their own intangible reality. The so-called ontology of the theory.  For example, the ontology of the Bohmian theory is very simple: everything is build by point particles guided (``choreographed'') by waves.  The different quantum theories available today (Copenhagen, Bohmian, many worlds, spontaneous collapse, etc.) are indeed inspired by radically different realities, but all of them provide the same empirical predictions on quantum phenomena. In Kant's words, all of them provide the same explanation of the appearance of our world. As we repeatedly stressed, up to know, in spite of many attempts, there is no experimental evidence that can discern between Bohmian and Copenhagen realities (ontologies).

In fact, for practical applications, even wrong theories can be very useful. Most natural phenomena that affect our ordinary life can be exclusively explained in terms of classical mechanics. However, today, we know that the reality behind the classical theory is wrong because it does not provide accurate predictions for some natural phenomena, like relativistic (with particles with high velocities) or quantum (atomistic dimensions) experiments.  Surprisingly, the fact that the classical theory is a wrong theory does not demerit its extraordinary utility and our confidence on its predictions within its range of validity\footnote{We take \emph{classical} planes expecting that they will follow a deterministic trajectory, e.g., from Barcelona to Paris. However, we know that quantum uncertainty precludes us to affirm that there is only one possible trajectory for the fly departing from Barcelona. Even after doing our best to fix the initial conditions of the physical degrees of the plane, there is still an unavoidable quantum randomness implying that several trajectories are possible. Of course, the differences between trajectories are so small at a macroscopic level that the pilot can easily certify that we will arrive to Paris.}. The same is true for most physical theories at a practical level. Even if we could demonstrate in the future that either the Copenhagen or the Bohmian theories is wrong (or both), the practical utility of these theories in their range of validity would not dismiss.

\subsection{How Bohmian mechanics helps}\label{sec_why}

Although there is no experimental evidence against Bohmian mechanics, many researchers believe that Bohmian mechanics is not a useful tool to do research. In the words of Steven Weinberg, in a private exchange of letters with Sheldon Goldstein \cite{om.Weinberg}: \textit{``In any case, the basic reason for not paying attention to the Bohm approach is not some sort of ideological rigidity, but much simpler --- it is just that we are all too busy with our own work to spend time on something that doesn't seem likely to help us make progress with our real problems.''}.

The history of science seems to give credit to Weinberg`s sentence. In spite of the controversies that have always been associated with the Copenhagen interpretation since its birth a century ago, its mathematical and computational machinery has enabled physicists, chemists and (quantum) engineers to calculate and predict the outcome of a vast number of experiments, while the contribution of Bohmian mechanics during the same period is much less significative. The differences are due to the fact that Bohmian mechanics remains mainly unexplored.

Contrarily to Weinberg's opinion, we believe that Bohmian mechanics \emph{can help us make progress with our real problems}. There are, at least, three clear reasons why one could be interested in studying quantum problems with Bohmian mechanics:
\begin{enumerate}
\item \textbf{Bohmian \emph{explaining}:} Even when the Copenhagen mathematical machinery is used to compute observable results, the Bohmian interpretation ofently offers different interpretational tools. We can find descriptions of electron dynamics such as \emph{``an electron crosses a resonant tunneling barrier and interacts with another electron inside the well''}. However, according to the orthodox theory, we can only talk about the properties of an electron (for example, its position) when we measure it. If we do not measure it, the electron has no property. Thus, an electron crossing a tunneling region is not rigorously supported within orthodox quantum mechanics, but it is within the Bohmian picture.
Thus, in contrast to the Copenhagen theory, Bohmian mechanics allows for an easy visualization of quantum phenomena in terms of trajectories that has important demystifying or clarifying consequences. In fact, Bohmian mechanics allows for an unambiguous\footnote{About the ambiguity of the orthodox explanation of quantum mechanics and the unambiguity of Bohmian mechanics, J. Bell wrote\cite{om.Bell1987} (page 111): ``I will try to interest you in the de Broglie - Bohm version of non-relativistic quantum mechanics. It is, in my opinion, very instructive. It is experimentally equivalent to the usual version insofar as the latter is unambiguous.''} description of measured and unmeasured properties of particles (an electron crossing a tunneling barrier is a description of unmeasured properties). Bohmian mechanics provides a single-event description of the experiment, while Copenhagen quantum mechanics accounts for its statistical or ensemble explanation. We will present several examples in chapters 2 and 3 emphasizing all these points. \\

\item \textbf{Bohmian \emph{computing}:} Although the predictions of the Bohmian interpretation reproduce the ones of the orthodox formulation of quantum mechanics, its mathematical formalism is different. In some systems, the Bohmian equations might provide better computational tools than the ones obtained from the orthodox machinery, resulting in a reduction of the computational time, an increase in the number of degrees of freedom directly simulated, etc.  We will see examples of these computational issues in quantum chemistry in chapters 4 and 5, as well as in quantum electron transport in Chap. 6.\\

\item \textbf{Bohmian \emph{thinking}:} From a more fundamental point of view, alternative formulations of quantum mechanics can provide alternative routes to look for the limits and possible extensions of the quantum theory. In particular, Chap. 7 presents the route to connect Bohmian mechanics with geometrical optics and beyond opening the way to apply the powerful computational tools of quantum mechanics to classical optics, and even to electromagnetism. The natural extension of Bohmian mechanics to the relativistic regime and to quantum field theory are presented in Chap. 8, while Chap. 9 and Chap. 10 discusses its application to cosmology.
\end{enumerate}

The fact that all measurable results of the orthodox quantum mechanics can be exactly reproduced with Bohmian mechanics (and vice versa) is the relevant point that completely justifies why Bohmian mechanics can be used for \textbf{\emph{explaining}} or \textbf{\emph{computing}} different quantum phenomena in physics, chemistry, electrical engineering, applied mathematics, nanotechnology, etc. In the scientific literature, the \textbf{Bohmian \emph{computing}} technique to find the trajectories (without directly computing the wavefunction) is also known as a \emph{syntectic} technique, while the \textbf{Bohmian \emph{explaining}} technique (where the wavefunction is directly computed first) is referred as the \emph{analytic} technique \cite{om.wyatt2005}. Furthermore, the fact that Bohmian mechanics is a theory without observers is an attractive feature for those researchers interested in \textbf{\emph{thinking}} about the limits or extensions of the quantum theory.

In order to convince the reader about the practical utility of Bohmian mechanics for \textbf{\emph{explaining}}, \textbf{\emph{computing}} or \textbf{\emph{thinking}}, we will not present elaborated mathematical developments or philosophical discussions, but provide practical examples. Apart from the first chapter, devoted to an overview of Bohmian mechanics, the book is divided into nine additional chapters with several examples on the practical application of Bohmian mechanics to different research fields, ranging from atomic systems to cosmology. These examples will clearly show that the previous quotation by Weinberg does not have to be always true.

\subsection{On the name Bohmian mechanics}

Any possible newcomer to Bohmian mechanics can certainly be quite confused and disoriented by the large list of names and slightly different explanations of the original ideas of de Broglie and Bohm that are present in the scientific literature. Different researchers use different names. Certainly, this is an indication that the theory is still not correctly settled down among the scientific community.

In his original works \cite{om.dB_AnnPhys,om.debroglie1927b}, de Broglie used the term \emph{pilot-wave theory} \cite{om.Valentini2006}, to emphasize the fact that wave-fields guided the motion of point particles.
After de Broglie abandoned his theory, Bohm rediscovered it in the seminal papers entitled \emph{``A suggested interpretation of the quantum theory in terms of hidden variables''} \cite{om.bohm1952a,om.bohm1952b}.
The term \emph{hidden variables}\footnote{Note that the term \emph{hidden variables} can also refer to other (local and non-local) formulations of quantum mechanics.}, refering to the  positions of the particles, was perhaps pertinent in 1952, in the context of the \emph{impossibility proofs} \cite{om.impossibility_proofs}.
Nowadays, these words might seem inappropriate because they suggest something metaphysical on the trajectories\footnote{Sometimes it is argued that the name \emph{hidden variables} is because Bohmian trajectories cannot be measured directly. However, what is not directly measured in experiments is the (complex) wavefunction amplitude, while the final positions of particles can be directly measured, for example, by the imprint they leave on a screen. John S. Bell wrote \cite{om.Bell1987} (page 201): \emph{``Absurdly, such theories are known as 'hidden variable' theories. Absurdly, for there it is not in the wavefunction that one finds an image of the visible world, and the results of experiments, but in the complementary 'hidden'(!) variables.''}}.

To give credit to both de Broglie and Bohm, some researchers refer to their works as the \emph{de Broglie--Bohm theory}\footnote{In fact, even de Broglie and Bohm were not the original names of the scientists' families.  Louis de Broglie's family, which included dukes, princes, ambassadors and marshals of France, changed their original Italian name Broglia to de Broglie when they established in France in the seventeenth century \cite{om.valentini2009Solvay}. David Bohm's father,  Shmuel D\"um,  was born in the Hungarian town of Munk\'acs and, was sent to America when he was young. Upon landing at Ellis Island, he was told by an immigration official that his name, D\"um would mean ``stupid'' in English. The official himself decided to change the name to Bohm \cite{om.infinite_potential}.}\cite{om.Holand1993}.
Some reputed researchers argue that de Broglie and Bohm did not provide the same exact presentation of the theory \cite{om.Valentini2006,om.valentini2008}. While de Broglie presented a first order development of the quantum trajectories (integrated from the velocity), Bohm himself did a second order (integrated from the acceleration) emphasizing the role of the quantum potential. The differences between both approaches appear when one considers initial ensembles of trajectories which are not in quantum equilibrium\footnote{Quantum equilibrium assumes that the initial positions and velocities of Bohmian trajectories are defined compatible with the initial wavefunction. Then the trajectories computed from Bohm's or de Broglie's formulations will become identical. However, one can select completely arbitrary initial positions from the (first order) de Broglie explanation and arbitrary initial velocities and positions from the (second order) Bohm work, see \sref{om.sec_single.6}.}. Except for this issue, which will not be addressed in this book, both approaches are identical.

Many researchers prefer to use the name \emph{Bohmian mechanics} \cite{om.Bohmian1996}. It is perhaps the most popular name. We know directly from his alive collaborators, Basile Hiley and David Peat \cite{om.davidpeat}, that this name irritated David Bohm and he said about its own work ``\textit{it's Bohmian non-mechanics}''. He argued that the {\emph{'quantum potential'}} is a non-local potential that depends on the relative shape of the wavefunction and thus it is completely different from other mechanical (such as the gravitational or the electrostatic) potentials which decrease with distance. See this particular discussion in the last chapters of Bohm and Hiley's book entitled \emph{The Undivided Universe: An Ontological Interpretation of Quantum Theory} \cite{om.Bohm1993}. He preferred the names \emph{causal} or \emph{ontological interpretation} of quantum mechanics \cite{om.Holand1993,om.Bohm1993}. The latter names emphasize the foundational aspects of its formulation of quantum mechanics.

Finally, another very common term is \emph{quantum hydrodynamics} \cite{om.wyatt2005} that underlines the fact that Bohmian trajectories provide a mathematical relationship between the Schr\"odinger equation and fluid dynamics. In fact, this name is more appropriate when one refers to the Madelung theory \cite{om.Madelung}, which is considered as a precursor of Bohm's work, see \sref{om.sec_intro.9}.

From all these different names, we choose \textbf{Bohmian mechanics} because it is short and clearly specifies what we are referring to. It has the inconvenience of not giving credit to the initial work of Louis de Broglie.
Although it might be argued that Bohm merely reinterpreted the prior work of de Broglie, we think that he was the first person to genuinely understand its significance and implications. As we mentioned, Bohm himself disliked this name. However, as any work of art, the explanation of the quantum phenomena done in the 1952 Bohm's paper does not completely belong to the author\footnote{For example, Erwin Schr\"odinger, talking about quantum theory, wrote: \emph{``I don't like it, and I'm sorry I ever had anything to do with it''}, but his opinion did not influence the great applicability of his famous equation in the orthodox theory.}, but has become part of our scientific heritage. It has happened many times during the history of science that the mathematical equation developed by a scientist contains much more physical substance than what he/she imagined at the beginning. 
In any case, we understand Bohmian mechanics as a generic name that includes all those works inspired from the original ideas of Bohm and de Broglie. In the figure below, we plot the numbers of citations per year for the 1952 Bohm's seminal papers~\cite{om.bohm1952a,om.bohm1952b}, certifying the exponentially growing influence of these papers, which is not the case for the original work of de Broglie~\cite{om.debroglie1927b}.

\begin{figure}
\centering
\includegraphics*[width=13cm]{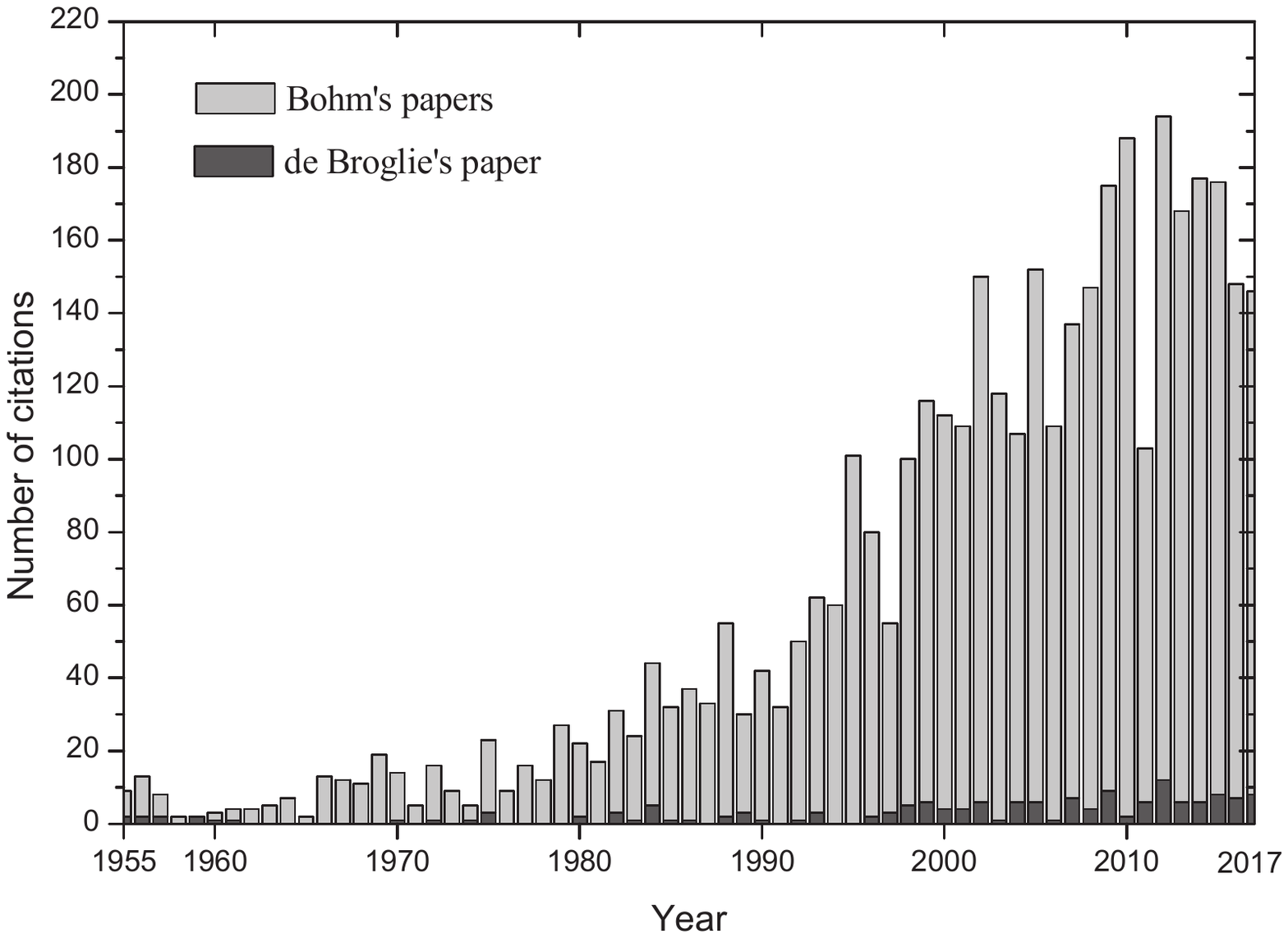}
\caption{Number of citations per year for (a) the two 1952 David Bohm's papers entitled ``A suggested interpretation of the quantum theory in terms of hidden variables'' \cite{om.bohm1952a,om.bohm1952b} and (b) Louis de Broglie's paper ``La m\'ecanique ondulatorie et la structure atomique de la mati\`ere et du rayonnement'' ~\cite{om.debroglie1927b}. Data retreived from ISI Web of Knowledge \cite{om.isiweb} in December 2017.}
\nonumber
\label{om_Datos_Bohm}
\end{figure}

\subsection{On the book contents}

The book contains ten chapters. The first chapter provides an accessible introduction to Bohmian mechanics. The rest of chapters present practical examples on the applicability of Bohmian mechanics. Let us start mentioning the cover of this second edition of the book. It represents the wave and particle nature of electrons according to the Bohmian theory. In particular, we see the Bohmian trajectories of an electron which suffers Klein tunneling when impinging on a triangular potential barrier of a graphene structure. The wave packet of the electron corresponding to the bispinor solution of the Dirac equation (electron with positive and negative energies) is also plotted. 

\textbf{Chapter one} is the longest one and it is entitled \textbf{Overview of Bohmian mechanics}. It is written by Xavier Oriols and Jordi Mompart, the editors of the book, both from the Universitat Aut\`{o}noma de Barcelona, Spain. This chapter is intended to be an introduction to any newcomer interested in Bohmian mechanics.  Only basic concepts of classical and quantum mechanics are assumed. The chapter is divided into four different sections. First, the historical development of Bohmian mechanics is explained. Then, Bohmian mechanics for single particle and for many particle systems (with spin and entanglement discussions) are presented. Finally, the topic of Bohmian measurement is addressed in section four. The chapter contains also a list of solved problems and easily implementable codes for computation of Bohmian trajectories.

\textbf{Chapter two} is entitled \textbf{Hydrogen photoionization with strong lasers}. It is written by Albert Benseny from the Okinawa Institute of Science and Technology in Japan; Antonio Pic\'on and Luis Plaja from the Universidad de Salamanca, Spain; Jordi Mompart from the Universitat Aut\`{o}noma de Barcelona, Spain and Luis Roso from the CLPU, the Laser Center for Ultrashort and Ultraintense Pulses, in Salamanca. They discuss the dynamics of a single hydrogen atom interacting with a strong laser. In particular, the Bohmian trajectories of these electrons represent an interesting illustrating view, with new calculation methods (i.e., \textbf{Bohmian \emph{computing}}), of both  the above threshold ionization and the harmonic generation spectra problems. They do also present a full three dimensional model to discuss the dynamics of Bohmian trajectories when the light beam and the hydrogen atom exchange angular momentum. The chapter does also provide a practical example on how Bohmian mechanics is computed, with an analytical (i.e., \textbf{Bohmian \emph{explaining}}) procedure, when full (scalar and vector potentials) electromagnetic fields are considered.

The title of \textbf{chapter three} is \textbf{Atomtronics: Coherent control of atomic flow via adiabatic passage} and it is written by Albert Benseny from the Okinawa Institute of Science and Technology in Japan; Joan Bagud\`{a}, Xavier Oriols, and Jordi Mompart from the Universitat Aut\`{o}noma de Barcelona, Spain; and Gerhard Birkl from the Institut f\"ur Angewandte Physik, from the Technische Universit\"at Darmstadt in Germany. Here, it is discussed an efficient and robust technique to coherently transport a single neutral atom, a single hole, or even a Bose--Einstein condensate between the two extreme traps of the triple-well potential. The dynamical evolution of this system with the direct integration of the Schr\"odinger equation presents a very counterintuitive effect: by slowing down the total time duration of the transport process it is possible to achieve atomic transport between the two extrem traps with a very small (almost negligible) probability to populate the middle trap. The analytical (i.e., \textbf{Bohmian \emph{explaining}}) solution of this problem with Bohmian trajectories enlightens the role of the \emph{particle conservation law} in quantum systems showing that the negligible particle presence is due to a sudden particle acceleration yielding, in fact, ultra-high atomic velocities. The Bohmian contribution opens the discussion about the possible detection of these high velocities or the need for a relativistic formulation to accurately describe such a simple quantum system.

\textbf{Chapter four}, entitled \textbf{Bohmian pathways into Chemistry: A brief overview}, is prepared by \'Angel S. Sanz, from the Universidad Complutense de Madrid, Spain, and deals with the issue of how the \textbf{Bohmian computing} abilities have been explored and exploited in Chemistry over decades. Interestingly, contrary to Physics, Bohmian mechanics has always found a better accommodation and acceptance within different areas of Chemistry, where the pedagogical advantages mentioned by John Bell have been widely recognized.
Because providing an exhaustive account on the applications (both as a problem solver and as a computational tool) where Bohmian mechanics has been of relevance within Chemistry would exceed the scope of the chapter, it has been prepared in a way that may serve the reader as a guide to acquire a general perspective (or impression) on how this trajectory-based quantum approach has permeated the different traditional levels or pathways to approach the problems of interest in Chemistry.

\textbf{Chapter five}, whose title is \textbf{Adaptive quantum Monte Carlo approach states for high-dimensional systems}, is written by Eric R. Bittner, Donald J. Kouri, Sean Derrickson, from the University of Houston; and Jeremy B. Maddox, from the Western Kentucky University, in USA. They provide one particular example on the success of Bohmian mechanics in the chemistry community. In this chapter, the authors explain their \textbf{Bohmian \emph{computing}} development for knowing the \emph{ab initio} quantum mechanical structure, energetics and thermodynamics of multi-atoms systems. They use a variational approach that finds the quantum ground sate (or even excited states at finite temperature) using a statistical modeling approach for determining the best estimate of a quantum potential for a multi-dimensional system.

\textbf{Chapter six} is entitled \textbf{Nanoelectronics: Quantum electron transport}. It is written by Enrique Colom\'es, Devashish Pandey,  Alfonso Alarc\'on  and Xavier Oriols from the Universitat Aut\`{o}noma de Barcelona, Spain; Zhen Zhan from the Wuhan University, in China; Guillem Albareda from Max Planck Institute for the Structure and Dynamics of Matter in Germany and  Fabio Lorenzo Traversa from University of California, in USA. The authors explain the ability of their own many-particle \textbf{Bohmian \emph{computing}} algorithm to understand and model nanoscale electron devices. In particular, it is shown that the adaptation of Bohmian mechanics to electron transport in open systems (with interchange of particles and energies) leads to a quantum Monte Carlo algorithm, where randomness appears because of the uncertainties in the number of electrons, their energies and the initial positions of (Bohmian) trajectories. A general, versatile and time-dependent 3D electron transport simulator for nanoelectronic devices, named BITLLES (Bohmian Interacting Transport in Electronic Structures), is presented showing its ability for a full prediction (DC, AC, fluctuations) of the electrical characteristics of any nanoelectronic device. The BITLLES simulator is also applied to graphene structures (by solving the Dirac equation) as reflected in the cover of this book.

\textbf{Chapter seven}, entitled \textbf{Beyond the eikonal approximation in classical optics and quantum physics}, is written by Adriano Orefice, Raffaele Giovanelli and Domenico Ditto from the Universit\`{a} degli Studi di Milano, Italy. It is devoted to discuss how \textbf{Bohmian \emph{thinking}} can also help in optics, exploring the fact that the time-independent Schr\"{o}dinger equation is strictly analogous to the Helmholtz equation appearing in classical wave theory. Starting from this equation they obtain indeed, without any omission or approximation, a Hamiltonian set of ray-tracing equations providing (in stationary media) the exact description in term of rays of a family of wave phenomena (such as diffraction and interference) much wider than that allowed by standard geometrical optics, which is contained as a simple limiting case. They show in particular that classical ray trajectories are ruled by a wave potential presenting the same mathematical structure and physical role of Bohm's quantum potential, and that the same equations of motion obtained for classical rays hold, in suitable dimensionless form, for quantum particle dynamics, leading to analogous trajectories and reducing to classical dynamics in the absence of such a potential.

\textbf{Chapter eight}, entitled \textbf{Relativistic quantum mechanics and quantum field theory}, is written by Hrvoje Nikoli\'c from the Rudjer Bo\v{s}kovi\'c Institute, Croatia. This chapter presents a clear example on how a \textbf{Bohmian \emph{thinking}} on superluminal velocities and nonlocal interactions helps in extending the quantum theory towards relativity and quantum field theory. A relativistic-covariant formulation of relativistic quantum mechanics of fixed number of particles (with or without spin) is presented, based on many-time wavefunctions and on an interpretation of probabilities in the spacetime. These results are used to formulate the Bohmian interpretation of relativistic quantum mechanics in a manifestly relativistic covariant form and are also generalized to quantum field theory. The corresponding Bohmian interpretation of quantum field theory describes an infinite number of particle trajectories. Even though the particle trajectories are continuous, the appearance of creation and destruction of a finite number of particles results from quantum theory of measurements describing entanglement with particle detectors.

\textbf{Chapter nine}, whose title is \textbf{Sub-quantum accelerating universe}, is written by Pedro F. Gonz\'alez-D\'{i}az from the Instituto de F\'isica Fundamental, Consejo Superior de Investigaciones Cient\'ificas, Spain and and Alberto Rozas-Fern\'andez from the Instituto de Astrof\'{\i}sica e Ci\^{e}ncias do Espa\c{c}o, in Portugal. Contrarily to the general belief, quantum mechanics does not only govern microscopic systems, but it has influence also on the cosmological domain. However, the extension of the Copenhagen version of quantum mechanics to cosmology is not free from conceptual difficulties: the probabilistic interpretation of the wavefunction of the whole universe is somehow misleading because we cannot make statistical ``measurements'' of different realizations of our universe. This chapter deals with two new cosmological models describing the accelerating universe in the spatially flat case.  Also in this chapter there is a discussion on the quantum cosmic models that result from the existence of a nonzero entropy of entanglement. In such a realm, they obtain new cosmic solutions for any arbitrary number of spatial dimensions, studying the stability of these solutions, as well as the emergence of gravitational waves in the realm of the most general models. \\

Finally, \textbf{chapter ten} entitled \textbf{Bohmian quantum gravity and cosmology}, is written by Nelson Pinto-Neto from the Centro Brasileiro de Pesquisas F\'{\i}sicas, in Brazil, and by Ward Struyve from the Ludwig-Maximilians-Universit\"at M\"unchen, in Germany. This chapter is another enlightening example on the utility of \textbf{Bohmian \emph{thinking}} concerning the nature of space-time and mass in physical theories. The authors discuss how many conceptual problems that appear in a description of gravity in quantum mechanical terms, such as the measurement problem and the problem of time, can be overcome by adopting a Bohmian perspective.  In addition to solving conceptual problems, the authors show that \textbf{Bohmian \emph{computing}} in quantum cosmology gives new types of semi-classical approximations to quantum gravity, and approximations for quantum perturbations moving in a quantum background. \\

\section{Historical Development of Bohmian Mechanics}\label{om.sec_intro}

In general, the history of quantum mechanics is explained in
textbooks as a chronicle where each step follows naturally from the
preceding one. However, it was exactly the opposite. The
development of quantum mechanics was a zigzagging route full of
misunderstandings and personal disputes. It was a painful history, where scientists were
forced to abandon well-established classical concepts and to explore
new and imaginative routes. Most of the new routes went nowhere.
Others were simply abandoned. Some of the explored routes were
successful in providing new mathematical formalisms capable of
predicting experiments at the atomic scale. Even such successful
routes were painful enough, so relevant scientists, such as Albert
Einstein or Erwin Schr\"odinger, decided not to support them. In
this section we will briefly explain the history of one of these
routes: Bohmian mechanics. It was first proposed by Louis de Broglie
\cite{om.debroglie1923}, who abandoned it soon afterward, and
rediscovered by David Bohm \cite{om.bohm1952a,om.bohm1952b} many years
later, and it has been ignored by most of the scientific community
since then. We will discuss the historical development of
Bohmian mechanics to understand its present status. Also, we will
introduce the basic mathematical aspects of the theory, while the
formal and rigorous structure will be presented in the subsequent 
sections.

\subsection{Particles and waves}\label{om.sec_intro.1}

The quantum theory revolves around the notions of particles and waves.
In classical physics, the concept of a \textit{particle} is very
useful for the description of many natural phenomena. A particle is
directly related to a \textit{trajectory}\footnote{In order to avoid
confusion, let us emphasize that in the orthodox formulation of
quantum mechanics, the concept of a particle is not directly related
to the concept of a trajectory. For example, the electron is a
particle but there is not trajectory in the orthodox ontology , as we will discuss
later.} $\vec{r}_{i}[t]$ that defines its position as a continuous
function of time, usually found as a solution of a set of
differential equations. For example, the planets can be considered
particles orbiting around the sun, whose orbits are determined by
the classical Newton gravitational laws.

In classical mechanics, it is natural to think that the total number
of particles (e.g., planets in the solar system) is conserved, and
the particle trajectories must be continuous in time: if a particle
goes from one place to another, then, it  has to go through all the trajectory
positions between these two places. This condition can be summarized
with a \textit{local} conservation law:
\begin{equation}
\label{om.difcurrent_density}
\frac{\partial\rho(\vec{r},t)}{\partial t} + \vec{\nabla} \vec{j}(\vec{r},t) = 0
\end{equation}
where $\rho(\vec{r},t)$ is the density of particles and
$\vec{j}(\vec{r},t)$ is the particle current density. For an
ensemble of point particles at positions $\vec{r}_i[t]$ with
velocites $\vec{v}_i[t]$, it follows that $\rho(\vec{r},t) =
\sum\delta(\vec{r}-\vec{r}_{i}[t])$ and $\vec{j}(\vec{r},t) =
\sum\vec{v}_{i}[t] \delta(\vec{r}-\vec{r}_{i}[t])$, with
$\delta(\vec{r})$ being the Dirac delta function, satisfying \eref{om.difcurrent_density}. We have used the property that ${\partial \delta(\vec{r}-\vec{r}_{i}[t])}/{\partial t}= -\vec{\nabla} \delta(\vec{r}-\vec{r}_{i}[t]) {d \vec{r}_{i}[t]}/{dt}$.

However, the total number of planets in the solar system could be conserved in another (quite different) way.
A phenomenon where a planet disappearing (instantaneously) from its orbit and appearing (instantaneously) at another point far away from its original location would certainly conserve the number of planets but it would violate \eref{om.difcurrent_density}. We must then think of \eref{om.difcurrent_density} as a law for the local conservation of particles.

Fields, and particularly waves, also appear in many explanations of
physical phenomena. The concept of a field was initially introduced
to deal with the interaction of distant particles.  For example,
there is an interaction between the electrons in an emitting radio
antenna at the top of a mountain and those in the receiving antenna
at home. Such interaction can be explained through the use of an
electromagnetic field. Electrons in the transmitter generate an
electromagnetic field, a radio wave, that propagates through the
atmosphere and \textit{arrives} at our antenna, affecting its
electrons. Finally, a loudspeaker transforms the electron motion
into music at home.

The simplest example of a wave is the so-called plane wave:
\begin{equation} \label{om.planewave}
\psi(\vec{r},t) = e^{i(\omega t- \vec{k} \cdot \vec{r})}
\end{equation}
where the angular frequency $\omega$ and the wave vector $\vec{k}$
refer respectively to its temporal and spatial behavior. In
particular, the angular frequency $\omega$ specifies when the
temporal behavior of such wave is repeated. The value of
$\psi(\vec{r_1},t_1) $ at position $\vec{r_1}$ and time $t_1$ is
identical to $\psi(\vec{r_1},t_2) $ when $t_2 = t_1 + 2 \pi
n/\omega$ for $n$ integer. The angular frequency $\omega$ can be
related to the linear frequency $\nu$ as $ \omega = 2 \pi \nu$.
Analogously, the wave vector $\vec{k}$ determines the spatial
repetition of the wave, that is, the wavelength $\lambda$. The value
of $\psi(\vec{r_1},t_1)$ at position $\vec{r_1}$ and time $t_1$ is
identical to $\psi(\vec{r_2},t_1)$ when $\vec{k}\cdot\vec{r_2} =
\vec{k}\cdot\vec{r_1} + 2 \pi n$ with $n$ integer. Unlike a
trajectory, a wave is defined at all possible positions and times.
Waves can be a scalar or a vectorial function and take real or
complex values. For example, \eref{om.planewave} is a scalar complex
wave of unit amplitude. The waves' dynamical evolution is determined by a set of
differential equations. In our \textit{broadcasting} example,
Maxwell equations define the electromagnetic field of the emitted
radio wave that is given by two vectorial functions, one for the
electric field and one for the magnetic field.

Whenever the differential equations that govern the fields are
linear, one can apply the superposition principle to explain what
happens when two or more fields (waves) traverse simultaneously the
same region. The modulus of the total field at each position is
related to the amplitudes of the individual waves. In some cases,
the modulus of the sum of the amplitudes is much smaller than the
sum of the modulus of the amplitudes; this is called destructive
interference. In other cases, it is roughly equal to the sum of the
modulus of the amplitudes; this is called constructive  interference.

\subsection{Origins of the quantum theory}\label{om.sec_intro.2}

At the end of the nineteenth century, Sir Joseph John Thomson discovered the electron, and in 1911, Ernest Rutherford, a New Zealander student working in Thomson's laboratory, provided experimental evidence that inside atoms, electrons orbited around a nucleus in a similar manner as planets do around the sun.
Rutherford's model of the atom was clearly in contradiction with well-established theories, since classical electromagnetism predicted that orbiting electrons should radiate, gradually lose energy, and spiral inward.
Something was missing in the previous explanations, since it seemed that the electron behavior inside an atom could not be explained in terms of classical trajectories. Therefore, alternative ideas needed to be explored to understand atom stability.

In addition, at that time, classical electromagnetism was unable to
explain the radiated spectrum of a black body, which is an idealized
object that emits a temperature-dependent spectrum of light (like a
big fire with different \textit{colors}, depending on
the flame temperature). The predicted continuous intensity spectrum
of this radiation became unlimitedly large in the limit of large frequencies,
resulting in an unrealistic emission of infinite power, which was
called the ultraviolet catastrophe. However, the measured radiation
of a black body did not behave in this way, indicating that a wave
description of the electromagnetic field was also incomplete.

In summary, at the beginning of the twentieth century, it was clear that natural phenomena such as atom stability or black-body radiation, were not well explained in terms of a particle or a wave description alone. It seemed necessary to merge both concepts.

In 1900, Max Planck suggested \cite{om.Planck-BlackBody} that black
bodies emit and absorb electromagnetic radiation in discrete
energies $h\nu$, where $\nu$ is the frequency of the emitted
radiation and $h$ is the (now-called) Planck constant. Five years
later, Einstein used this discovery in his explanation of the
photoelectric effect \cite{om.Einstein-Photoelectric}, suggesting
that light itself was composed of \textit{light quanta} or
\textit{photons}\footnote{In fact, the word \textit{photon} was not
coined until 1926, by Gilbert Lewis \cite{om.gilbert1926}.} of
energy $h\nu$. Even though this theory solved the black-body
radiation problem, the fact that the absorption and emission of
light by atoms are discontinuous was still in conflict with the
classical description of the light-matter interaction.

In 1913, Niels Bohr \cite{om.bohr,om.bohr2,om.bohr3} wrote a
revolutionary paper on the hydrogen atom, where he solved the
(erroneously predicted in classical terms) instability by
postulating that electrons can only orbit around atoms in some
particular \textit{nonradiating} orbits. Thus, atom radiation occurs only when electrons \textit{jump} from one orbit to another of lower
energy. His \textit{imaginative} postulates were in full agreement
with the experiments on spectral lines. Later, in 1924, de Broglie
proposed in his PhD dissertation that all particles (such as
electrons) exhibit wave-like phenomena like interference or
diffraction \cite{om.debroglie1923}. In particular, one way to
arrive at Bohr's hypothesis is to think that the electron orbiting
around the proton is a stationary wave. Since we know that the probability of finding the electron far from the proton is zero, we can impose such spatial boundary conditions on the shape of such a stationary wave. We will obtain that only very particular shapes of the waves (associated to very particular energies) are allowed. Physics at the
atomic scale started to be understandable by mixing the concepts of
particles and waves. All these advances were later known as the
\textit{old quantum theory}. The word \textit{quantum} referred to
the minimum unit of any physical entity (e.g., the energy) involved
in the interactions at such atomistic scales.

\subsection{``Wave or particle?'' vs. ``wave and particle''}\label{om.sec_intro.4}

In the mid-1920s, theoreticians found themselves in a difficult
situation when attempting to advance Bohr's ideas. A group of atomic
theoreticians centered on Bohr, Max Born, Wolfgang Pauli, and Werner
Heisenberg suspected that the problem went back to trying to
understand electron trajectories within atoms. In under two years, a
series of unexpected discoveries brought about a scientific
revolution \cite{om.waerden}.

Heisenberg wrote his first paper on quantum mechanics in 1925
\cite{om.Heisenber1925} and two years later stated his uncertainty
principle \cite{om.Heisenber1927}. It was him, with the help of Born
and Pascual Jordan, who developed the first version of quantum
mechanics based on a matrix formulation
\cite{om.Born1926,om.Heisenber1925,om.Heisenber1925b,om.Heisenber1925c}.

In 1926, Schr\"odinger published \textit{An Undulatory Theory of the Mechanics of Atoms and Molecules} \cite{om.scho1926}, where, inspired by de Broglie's work \cite{om.dB_AnnPhys,om.debroglie1923,om.debroglie1927b}, he described material points (such as electrons or protons) in terms of a wave solution of the following (wave) equation:
\begin{equation}
\label{om.scho}
i \hbar \frac{\partial \psi(\vec{r},t)}{\partial t} = -\frac{\hbar^2}{2m} \nabla^2 \psi(\vec{r},t) + V(\vec{r},t) \psi(\vec{r},t)
\end{equation}
where $V(\vec{r},t)$ is the potential energy \textit{felt} by the electron, and the wave (field) $\psi(\vec{r},t)$ was called the wave function. Schr\"odinger, at first, interpreted his wave function as a description of the electron charge density $q \abs{\psi(\vec{r},t)}^{2}$ with $q$ the electron charge. Later, Born refined the interpretation of Schr\"odinger and defined $|\psi(\vec{r},t)|^{2}$ as the probability density of \textit{finding} the electron in a particular position $\vec{r}$ at time $t$ \cite{om.waerden}.

Schr\"odinger's wave version of quantum mechanics and Heisenberg's
matrix mechanics were apparently incompatible, but they were
eventually shown to be equivalent  by
Wolfgang Ernst Pauli and Carl Eckart, independently \cite{om.waerden,om.waerden2}.

In order to  explain the
physics behind quantum systems, the concepts of waves and particles
should be merged in some way. Two different routes appeared:
\begin{enumerate}
\item \textbf{Wave or particle?} The concept of a trajectory was, consciously or unconsciously, abandoned by most of the young scientists (Heisenberg, Pauli, Dirac, Jordan$,\ldots).$ They started a new route, the \textit{wave or particle?} route, where depending on the experimental situation, one has to choose between a wave or a particle behavior. Electrons are associated basically to probability (amplitude) waves. The particle nature of the electron appears when we measure the position of the electron. In Bohr's words, an object cannot be both a wave and a particle at the same time; it must be either one or the other, depending upon the situation. This approach, mainly supported by Bohr, is one of the pillars of the Copenhagen, or orthodox, interpretation of quantum mechanics.

\item \textbf{Wave and particle}: Louis de Broglie, on the other hand, presented an explanation of quantum phenomena where the wave and particle concepts merge at the atomic scale, by assuming that a pilot-wave solution of \eref{om.scho} guides the electron trajectory. This is what we call the Bohmian route. One object cannot be a wave and a particle at the same time, but two can.
\end{enumerate}

The differences between the two routes can be easily seen in the
interpretation of the double-slit experiment. A beam of electrons
with low intensity (so that electrons are injected one by one)
impinges upon an opaque surface with two slits removed on it. A
detector screen, on the other side of the surface, detects the
position of electrons. Even though the detector screen responds to
particles, the pattern of detected particles shows the interference
fringes characteristic of waves. The system exhibits, thus, the
behavior of both waves (interference patterns) and particles (dots
on the screen).

According to the \textit{wave or particle?} route, first the electron presents a wavelike nature alone when the wave function (whose squared modulus gives the probability density of \textit{finding} a particle when a position measurement is done) \textit{travels} through both slits. Suddenly, the wave function \textit{collapses} into a delta function at a (random) particular position on the screen. The particle-like nature of the electron appears, while its wavelike nature disappears. Since the screen positions where \textit{collapses} occur follow the probability distribution dictated by the squared modulus of the wave function, a wave interference pattern appears on the detector screen.

According to the \textit{wave and particle} route, the wave function (whose squared modulus means the particle probability density of \textit{being} at a certain position, regardless of the measurement process) travels through both slits. At the same time, a well-defined trajectory is associated with the electron. Such a trajectory passes through only one of the slits. The final position of the particle on the detector screen and the slit through which the particle passes is determined by the initial position of the particle. Such an initial position is not controllable by the experimentalist, so there is an appearance of randomness in the pattern of detection. The wave function guides the particles in such a way that they avoid those regions in which the interference is destructive and are attracted to the regions in which the interference is constructive, giving rise to the interference pattern on the detector screen. Let us quote the enlightening summary of Bell \cite{om.Bell1987}:

\begin{quote}
Is it not clear from the smallness of the scintillation on the screen that we have to do with a particle? And is it not clear, from the diffraction and interference patterns, that the motion of the particle is directed by a wave? De Broglie showed in detail how the motion of a particle, passing through just one of two holes in screen, could be influenced by waves propagating through both holes. And so influenced that the particle does not go where the waves cancel out, but is attracted to where they cooperate. This idea seems to me so natural and simple, to resolve the wave-particle dilemma in such a clear and ordinary way, that it is a great mystery to me that it was so generally ignored.
\end{quote}

Now, with almost a century of perspective and the knowledge that both routes give exactly the same experimental predictions, it seems that such great scientists took the \textit{strangest} route. Let us imagine that a student asks his or her professor, ``What is an electron?'' The answer of a (Copenhagen) professor could be, ``The electron is not a wave nor a particle. But, do not worry! You do not have to know what an electron is to (compute observable results) pass the exam.''\footnote{For example, in the book \textit{Quantum Theory}, \cite{om.bohmbook} written by Bohm before he formulated Bohmian mechanics in 1952, he wrote, when talking about the wave-particle duality: ``We find a strong analogy here to the fable of the seven blind men who ran into an elephant: One man felt the trunk and said that `an elephant is a rope'; another felt the leg and said that `an elephant is obviously a tree,' and so on.''} If the student insists, the professor might reply, ``Shut up and calculate.''\footnote{This quote is sometimes attributed to Dirac, Richard Feynman, or David Mermin \cite{om.mermin,om.mermin2}. It recognizes that the important content of the orthodox formulation of quantum theory is the ability to apply mathematical models to real experiments.}

Another example of the vagueness of the orthodox formulation can be illustrated by the question that Einstein posed to Abraham Pais: ``Do you really think the moon is not there if you are not looking at it?'' The answer of a Copenhagen professor, such as Bohr, would be, ``I do not need to answer such a question, because you cannot ask me such question experimentally''. This answer is technically correct because, from an orthodox point of view, the property of the position of an object is undefined unless we measure it. But, knowing now that an explanation of quantum phenomena can be formulated with well defined positions of particles independently of being measured or not, the previous answer seems a bit impertinent. 

On the other hand, an alternative (Bohmian) professor would answer,
``Electrons are particles whose trajectories are guided by a pilot
field which is  the wave function solution of the Schr\"odinger
equation. There is some uncertainty in the initial conditions of the
trajectories, so that experiments have also some uncertainties.''
With such a simple explanation, the student would understand
perfectly the role of the wave and the particle in the description
of quantum phenomena. Furthermore, in the Bohmian interpretation,
the position of an electron (or the moon) while we are not looking
at it, is always defined, even though it is a hidden variable  for experimentalists.

One of the reasons that led the proponents of quantum mechanics to
choose the \textit{wave or particle?} route is that the predictions
about the positions of electrons are uncertain because the wave
function is \textit{spread out} over a volume. This effect is known
as the \textit{uncertainty principle}: it is not possible to
measure, simultaneously, the exact position and velocity (momentum)
of a particle. Therefore, scientists preferred to look for an
explanation of quantum effects without the concept of a trajectory
that seemed unmeasurable. They constructed a theory to explain the quantum world where the concept of trajectory was not present in the ontology. However, their argumentation to neglect the use of trajectories is, somehow, unfair and
unjustified, since it relies on the ``principle'' that the ontology of a physical
theory should not contain entities that cannot be
observed.\footnote{From a philosophical point of view, this is known
as ``positivism'' or ``empiricism'' discussed in the introduction and it can be understood as a nonphysical
limitation on the possible kinds of theories that we could choose to
explain quantum phenomena. For example, the wave function cannot be
measured directly in a single experiment but only from
an ensemble of experiments. However, there is no doubt that the
(complex) wave function comes to be a very useful
concept to understand quantum phenomena. Identically, in the de
Broglie and Bohm interpretation, the trajectories cannot be directly measured, but they can also be a very interesting tool for
understanding quantum phenomena.}

In addition, everyone with experience on Fourier transforms of
conjugate variables recognizes the quantum uncertainty principle as
a trivial effect present in any wave theory where the momentum of a
particle depends on the slope of its associated wave function. Then,
a very localized particle would have a very sharp wave function.
In this case such a wave function would have a great slope that implies
a large range of possible momenta. On the contrary, if the wave
function is built from a quite small range of momenta, then it will
have a large spatial dispersion.

\subsection{Louis de Broglie and the fifth Solvay Conference}\label{om.sec_intro.5}

Perhaps the most relevant event for the development of the quantum
theory was the fifth Solvay Conference, which took place from
October 24--29, 1927, in Brussels \cite{om.valentini2009Solvay}. As
on previous occasions, the participants stayed at \textit{H\^{o}tel
Britannique}, invited by Ernest Solvay, a Belgian chemist and
industrialist with philanthropic purposes due to the exploitation of
his numerous patents. There, de Broglie presented his recently
developed pilot wave theory and how it could account for quantum
interference phenomena with electrons \cite{om.valentini2009Solvay}.
He did not receive an enthusiastic reaction from the illustrious
audience gathered for the occasion. In the following months, it
seems that he had some difficulties in interpreting quantum
measurement with his theory and decided to avoid his new pilot wave theory. In fact, one
(nonscientific) reason that perhaps forced de Broglie to give up on his
theory was that he worked isolated, having little contact with the
main research centers in Berlin, Copenhagen, Cambridge, or Munich.
By contrast, most of the Copenhagen contributors worked with fluid
and constant collaborations among them.

Finally, let us mention that the elements of the pilot wave theory
(electrons guided by waves) were already in place in de Broglie's
thesis in 1924 \cite{om.debroglie1923}, before either matrix or wave
mechanics existed. In fact, Schr\"odinger used the de Broglie phases
to develop his famous equation (see \eref{om.scho}). In addition, it
is important to remark that de Broglie himself developed a
single-particle and a many-particle description of his pilot waves,
visualizing also the \textit{nonlocality} of the latter
\cite{om.valentini2009Solvay}. Perhaps, his remarkable contribution
and influence have not been fairly recognized by scientists and
historians because he abandoned his own ideas rapidly without properly
defending them \cite{om.valentini2009Solvay,om.Broglie1956}.

\subsection{Albert Einstein and locality}\label{om.sec_intro.6}

Not even Einstein gave explicit support to  the pilot wave theory \cite{om.waerden}. It remains almost unknown that in 1927, the same year that de Broglie published his pilot wave theory \cite{om.debroglie1927b}, Einstein worked out an alternative version of the pilot wave with trajectories determined by many-particle wave functions. However, before the paper appeared in print, Einstein phoned the editor to withdraw it. The paper remains unpublished, but its contents are known from a manuscript \cite{om.einsteinhidden1927,om.Hollaneinstein}.

It seems that Einstein, who was unsatisfied with the Copenhagen
approach, did not like the pilot wave approach either because both
interpretations have this notion of action at a distance: particles
that are far away from each other can profoundly and
\textit{instantaneously} affect each other. As the father of the
theory of relativity, he believed that action at a distance cannot
travel faster than the speed of light. Let Bohm explain the
difficulties of Einstein with both the Bohmian and the orthodox
interpretations \cite{om.Bohminterview}:\vspace*{-6pt}\\

\begin{quote}
In the fifties, I sent [my Quantum Theory book] around to various
quantum physicists - including Niels Bohr, Albert Einstein, and
Wolfgang Pauli. Bohr didn't answer, but Pauli liked it. Albert
Einstein sent me a message that he'd like to talk with me. When we
met he said the book had done about as well as you could do with
quantum mechanics. But he was still not convinced it was a
satisfactory theory. Einstein's objection was not merely that it was
statistical. He felt it was a kind of abstraction; quantum mechanics
got correct results but left out much that would have made it
intelligible. I came up with the causal interpretation (that the
electron is a particle, but it also has a field around it. The
particle is never separated from that field, and the field affects
the movement of the particle in certain ways). Einstein didn't like
it, though, because the interpretation had this notion of action at
a distance: Things that are far away from each other profoundly
affect each other. He believed only in local action.\vspace*{-6pt}\\
\end{quote}

Einstein, together with Boris Podolsky and Nathan Rosen, presented
objections to the orthodox quantum theory in the famous EPR article in 1935,
entitled ``Can Quantum-Mechanical Description of Physical Reality Be
Considered Complete?'' \cite{om.Einstein_rosen1935}. There, they
argued that on the basis of the absence of action at a distance,
quantum theory must be incomplete. In other words, quantum theory is
either nonlocal or incomplete. Einstein believed that locality was a
fundamental principle of physics, so he adhered to the view that
quantum theory was incomplete. Einstein died in 1955, convinced that
a \textit{correct} reformulation of quantum theory would preserve
local causality. We will see later that he was wrong in this particular point.

\subsection{David Bohm and why the ``impossibility proofs'' were wrong?}\label{om.sec_intro.7}

Perhaps the first utility of Bohm's work was the demonstration that the mentioned von Neumann theorem about the ``impossible proofs'' had limited validity. In 1932, von Neumann put quantum theory on a firm theoretical basis \cite{om.impossibility_proofs}. Some of the earlier works lacked mathematical rigor, and he put the entire theory into the setting of operator algebra. In particular,  von Neumann studied the following question: ``If the present mathematical formulation of the quantum theory and its usual probability interpretation are assumed to lead to absolutely correct results for every experiment that can ever be done, can quantum-mechanical probabilities be understood in terms of any conceivable distribution over hidden parameters?'' von Neumann answered this question negatively.
His conclusions, however, relied on the fact that he implicitly restricted his proof to an excessively narrow class of hidden variables, excluding Bohm's hidden variables model.
In other words, Bohmian mechanics is a counterexample that disproves von Neumann's conclusions, in the sense that it is possible to obtain the very same predictions of orthodox quantum mechanics with a hidden variables theory \cite{om.Holand1993,om.Bell1987}.

Bohm's formulation of quantum mechanics\footnote{Apart from these
works, the history of science has recognized many other relevant
contributions by Bohm \cite{om.infinite_potential}. As a
postgraduate at Berkeley, he developed a theory of plasmas,
discovering the electron phenomenon now known as Bohm diffusion. In
1959, with his student Yakir Aharonov, he discovered the
Aharonov--Bohm effect, showing how a magnetic field could affect a
region of space in which the field had been shielded, although its
vector potential did not vanish there. This showed for the first
time that the magnetic vector potential, hitherto a mathematical
convenience, could have real physical (quantum) effects.} appeared
after the orthodox formalism was fully established. Bohm was,
perhaps, the first person to genuinely understand the significance
and fundamental implications of the description of quantum phenomena
with trajectories guided by waves. Ironically, in 1951, Bohm wrote a
book, \textit{Quantum Theory} \cite{om.bohmbook}, where he provided
``proof that quantum theory is inconsistent with hidden variables''
(see page 622 in [21]). In fact, he wrote in a footnote in that section,
``We do not wish to imply here that anyone has ever produced a
concrete and successful example of such a [hidden variables] theory,
but only state that such theory is, as far as we know,
conceivable.'' Furthermore, the book does also contain an unusually
long chapter devoted to the \textit{quantum theory of the process of
measurement}, where Bohm discusses how the measurement itself can be
described from the time evolution of a many-particle wave function, rather than
invoking the \textit{wave function collapse}. It seems that Bohm
became dissatisfied with the orthodox approach that he had written
in his book and began to develop his own causal formulation of
quantum theory, which he published in 1952
\cite{om.bohm1952a,om.bohm1952b}.

The original papers of Bohm in 1952 \cite{om.bohm1952a,om.bohm1952b} provide
a formal justification of the guidance equation developed 25 years
before by de Broglie. Instead of reproducing his exact mathematical
development in terms of the quantum Hamilton--Jacobi equation, here, we
discuss a very simple explanation of the guidance equation. By a
simple mathematical manipulation of the (wave) equation,
\eref{om.scho}, we can find the local (particle) continuity
equation, \eref{om.difcurrent_density}, discussed at the beginning
of this section.\footnote{See the formal demonstration in
\sref{om.sec_conservation} or in Ref. \cite{om.cohen}.} From the
standard definition of the current density $\vec{j}(\vec{r},t)$, as
a product of the particle density $\rho(\vec{r},t) =
|\psi(\vec{r},t)|^2$ and the velocity $\vec{v}(\vec{r},t)$, we can
exactly obtain the guidance equation that was predicted by de
Broglie and Bohm for the particle velocity:
\begin{equation}
\label{om.guidance}
\frac{d\vec{r}(t)}{dt} = \vec{v}(\vec{r},t) = \frac{\vec{j}(\vec{r},t)} {\rho(\vec{r},t)}
\end{equation}

There was another important point explained by Bohm in 1952. If one considers an ensemble of trajectories whose initial positions at time $t = 0$ are distributed according to $\rho(\vec{r},0) = |\psi(\vec{r},0)|^2$, such an ensemble of trajectories will reproduce $\rho(\vec{r},t) = |\psi(\vec{r},t)|^2$ at any other time if the trajectories follow the guidance equation, \eref{om.guidance}. Therefore, we are able to exactly reproduce the time evolution of the wave function solution of \eref{om.scho} from an ensemble of trajectories guided by waves. Let Bohm himself explain this revolutionary point in the first page of his original reference \cite{om.bohm1952a}:\\
\begin{quote}
The usual [Copenhagen] interpretation of the quantum theory is based
on an assumption [\ldots] that the physical state of an individual
system is completely specified by a wave function that determines
only the probabilities of actual results that can be obtained in a
statistical ensemble of similar experiments. [\ldots] In contrast,
this alternative interpretation permits us to conceive of each
individual system as being in a precisely definable state, whose
changes with time are determined by definite laws, analogous to (but
not identical with) the classical equations of motion. Quantum
mechanical probabilities are regarded (like their counterpart in
classical mechanics) as only a practical necessity and not as a
manifestation of an inherent lack of complete determination in the
properties of matter at the quantum level.\\
\end{quote}
Bohm's original papers do also provide a different path to find the trajectories by
introducing the polar form of the wave function $\psi(\vec{r},t) =
R(\vec{r},t) e^{i S(\vec{r},t)/\hbar}$ into the (nonrelativistic)
Schr\"odinger equation, \eref{om.scho}. Let us emphasize that
$R(\vec{r},t)$ and $S(\vec{r},t)$ are real (not complex) functions. Then, after a
quite simple manipulation, one obtains from the real part a quantum
Hamilton--Jacobi equation:
\begin{equation}
\label{om.hamitlon2}
 \frac{\partial S(\vec{r},t)}{\partial t} + \frac{1}{2m} \left(\vec{\nabla} S(\vec{r},t)\right)^{2} + V(\vec{r},t) + Q(\vec{r},t) = 0
\end{equation}
where we have defined the quantum potential $Q(\vec{r},t)$ as:
\begin{equation}
\label{om.hamitlon3}
 Q(\vec{r},t) = -\frac{\hbar^{2}}{2 m} \frac{\vec{\nabla}^{2}R(\vec{r},t)}{R(\vec{r},t)}
\end{equation}
which is the only difference with respect to the classical
Hamilton--Jacobi equation if the velocity is defined as $\vec{v} =
\vec{\nabla}S / m$.\footnote{See \sref{sec.quantum_HJ.om} for the
detailed formal demonstration.} Such an alternative explanation
provides an additional justification for the guidance equation,
\eref{om.guidance} (see problem \ref{om.p5}). It allows a
second-order interpretation of Bohmian trajectories in terms of
acceleration, forces, and energies. In particular, in this
second-order point of view, the new quantum potential $Q(\vec{r},t)$
is responsible for the deviations of the Bohmian trajectories from
the classical behavior that can be expected from the classical
potential $V(\vec{r},t)$. As we will discuss later, the quantum
potential of a system of $N$ particles will be the responsible for
the (nonclassical) nonlocal causality of Bohmian mechanics.

Bohm completed the work of de Broglie in two fundamental aspects. First, as explained before, he demonstrated that Bohmian mechanics leads to exactly the same predictions as the ones obtained by orthodox quantum mechanics. Second, he provided a theory of measurement. He developed an explanation of the measurement problem without invoking the \textit{wave function collapse}. The theory of Bohmian measurement will be discussed in \sref{om.sec_measurement}. Some authors argue that if we could change history, allowing Bohm to help de Broglie defend his pilot wave theory in the Solvay Conference, Bohmian mechanics would now be certainly taught at universities \cite{om.nikolic2008a}. The enlightening work of Bohm, however, appeared 25 years too late, once the Copenhagen interpretation of quantum phenomena was already too well established.

After the Solvay Conference, Bohr, Heisenberg, and their colleagues
spread the new interpretation around the world and convinced the
vast majority of the physics community that the Copenhagen theory worked with extraordinary precision. A lot of
young physicists were attracted to European institutes to study with
the ``fathers'' of this new theory, and during the second quarter of
the twentieth century, as good disciples, they spread the Copenhagen
interpretation over the entire globe. In his 1976 Nobel lecture,
Murray Gell-Mann referred to this question: ``Niels Bohr brainwashed
an entire generation of physicists into believing that the problem
[of the interpretation of quantum mechanics] had been solved fifty
years ago'' \cite{om.hard}.

Finally, we summarize the importance of Bohm's work with another magistral quote from Bell that appears in a 1982 paper entitled ``On the Impossible Pilot Wave'' and collected in his famous book \textit{Speakable and Unspeakable in Quantum Mechanics} \cite{om.Bell1987}:
\begin{quote}
But in 1952 I saw the impossible done. It was in papers by David Bohm. Bohm showed explicitly how parameters could indeed be introduced, into nonrelativistic wave mechanics, with the help of which the indeterministic description could be transformed into a deterministic one. More importantly, in my opinion, the subjectivity of the ``orthodox'' version, the necessary reference to the ``observer,'' could be eliminated. Moreover, the essential idea was on that had been advanced already by de Broglie in 1927, in his ``pilot wave'' picture. But why then had Born not told me of this ``pilot wave''? If only to point out what was wrong with it? Why did von Neumann not consider it? More extraordinarily, why did people go on producing ``impossibility'' proofs after 1952, and as recently as 1978? When even Pauli, Rosenfeld, and Heisenberg, could produce no more devastating criticism of Bohm's version than to brand it as ``metaphysical'' and ``ideological''? Why is the pilot wave picture ignored in textbooks? Should it not be taught, not as the only way, but as an antidote to the prevailing complacency? To show that vagueness, subjectivity, and indeterminism, are not forced on us by experimental facts, but by deliberate theoretical choice?
\end{quote}

\subsection{John Bell and nonlocality} \label{om.sec_intro.8}

Along this introduction, we have already mentioned Bell's positive opinion on Bohmian mechanics. Furthermore, his opinion about the Copenhagen interpretation was that the orthodox theory is ``unprofessionally vague and ambiguous'' \cite{om.Bell1987,om.bell1990,om.bell1982,om.Bell1964} in so far as its fundamental dynamics is expressed in terms of ``words which, however, legitime and necessary in application, have no place in a formulation with any pretension to physical precision'' \cite{om.bell1990}.

Bell spent most of his professional career at the European
Organization for Nuclear Research (CERN), working almost exclusively
on theoretical particle physics and on accelerator design, but found
time to pursue a major avocation investigating the foundations of
quantum theory \footnote{J. Bell defined himself as ``I am a quantum engineer, but on sundays I have principles.'' Underground colloquium, March 1983.}. As seen in many of his quotes used in this
introduction, his didactic ability to defend Bohmian mechanics
against many unjustified attacks has been of extraordinary
importance for maintaining the work of de Broglie and Bohm alive
among the scientific community. Fortunately, Bell himself had his
own reward from this unbreakable support of Bohmian mechanics. His
outstanding work on locality and causality was directly inspired by
his deep knowledge of Bohmian mechanics. Bell's theorem has been
called ``the most profound discovery of science''
\cite{om.stapp1977}.

Bell's most relevant contribution to physics is probably the
demonstration that quantum mechanics is nonlocal, contrarily to what
Einstein expected. In 1964, inspired by the EPR paper
\cite{om.Einstein_rosen1935} and Bohm's work on \textit{nonlocal}
hidden variables, Bell elaborated a theorem establishing clear
mathematical inequalities, now known as Bell inequalities, for
experimental results that would be fulfilled by \textit{local}
theories but would be violated by \textit{nonlocal} ones
\cite{om.Bell1964}. Over the past 30 years, a great number of Bell
test experiments have been conducted. These experiments have
confirmed that Bell's inequalities are violated (see, for example,
Ref. \cite{om.aspect1982}). Therefore, we have to conclude that
quantum experimental results cannot be explained with local hidden
variable theories. According to Bell, we must accept the real
existence, in nature, of faster-than-light causation.

The experimental violation of Bell's inequalities gave direct support, not only to the Copenhagen interpretation, but also to Bohm's formulation of quantum theory, since both are nonlocal theories.

In the Bohmian case, as we have discussed, the quantum potential of $N$ \textit{entangled} particles is defined in a $3N$ configuration space so that an action on the first particle can have an \textit{instantaneous} (i.e., faster-than-light) causal effect on the last particle. In Bohmian mechanics, one can understand that the quantum potential is responsible for the \textit{instantaneous} nonlocal changes on the trajectories of quantum particles.
Let Bell explain this point in his own words \cite{om.Bell1987}:\vspace*{-6pt}\\
\begin{quote}
That the guiding wave, in the general case, propagates not in
ordinary three-dimensional (3D) space but in a multidimensional
configuration space is the origin of the notorious ``nonlocality''
of quantum mechanics. It is a merit of the de Broglie--Bohm version
[of quantum mechanics] to bring this out so explicitly that it
cannot be ignored.\vspace*{-6pt}\\
\end{quote}

Unfortunately, as already happened with the von Neumann theorem \cite{om.impossibility_proofs}, there is an historical misunderstanding about the consequences of Bell's theorems on Bohmian mechanics. Let us mention just one example that appeared in a prestigious journal in 2000 \cite{om.100years}:\vspace*{-6pt}\\
\begin{quote}
In the mid-1960s John S. Bell showed that if hidden variables existed, experimentally observed probabilities would have to fall below certain limits, dubbed Bell's inequalities. Experiments were carried out by a number of groups, which found that the inequalities were violated. Their collective data came down decisively against the possibility of hidden variables.\vspace*{-6pt}\\
\end{quote}
The author of this sentence has omitted the adjective ``local'' when
he mentions hidden variables. Therefore, a confident reader, who has
no time to read Bell's and Bohm's works, will understand that
Bohmian mechanics is refuted by Bell's theorem. However, it is
exactly the contrary. Bell's inequalities give direct support to
Bohmian mechanics. Unfortunately, this misunderstanding appeared,
and continues to appear, in many scientific articles, propagating
into textbooks, websites, etc., provoking further comments and
replies in the scientific literature. This discussion can give the
impression that there still exists some controversy about the
validity of Bohmian mechanics for all non-relativistic quantum phenomena or that there is ``something unclear''
about  it,
\footnote{See, for example, the experience
of J. T. Cushing \cite{om.erors2} or some recent works,
``demonstrating'' that Bohmian mechanics was wrong
\cite{om.bifoton2}, and the comment to the work
\cite{om.bifoton1}.} which is clearly not the case.\enlargethispage{-1pc}

\subsection{Quantum hydrodynamics}\label{om.sec_intro.9}

It is sometimes claimed that ideas similar to those developed by de Broglie were put forward by Madelung in 1926 \cite{om.Madelung}. What Madelung proposed, however, was to regard an electron with mass $m$ and wave function $\psi(\vec{r},t)$ not as a particle with a determined trajectory but as a continuous fluid with mass density $m \abs{\psi(\vec{r},t)}^2$ \cite{om.wyatt2005,om.Takabayasi}. In Madelung's hydrodynamic interpretation of \eref{om.scho}, the fluid velocity coincides mathematically with de Broglie's guiding \eref{om.guidance}, but the ontological interpretation is quite different.

Let us return again to the double-slit experiment to understand Madelung's point of view. Orthodox quantum mechanics does not predict what happens to a single electron crossing a double slit, but it predicts what is the statistical probability of detecting electrons when we consider an infinite ensemble of such experiments. In contrast, the proposal from de Broglie and Bohm intends to predict what happens, in principle, to a single electron. At the end of the day, however, because it is not possible to determine the initial position of such a single electron with better uncertainty than that obtained from the initial wave packet spatial dispersion, Bohmian mechanics also provides statistical results.
In this regard, Madelung was not interested in dealing with a single-electron trajectory but with the ensemble. When we deal with a very large (infinite) number of particles (trajectories), the particle and current densities that appear in \eref{om.difcurrent_density} are no longer discrete functions (sum of deltas) anymore but they have to be interpreted as continuous functions. Aerodynamics and hydrodynamics are examples of such continuous material systems. The concept of an individual trajectory becomes irrelevant in these disciplines but \eref{om.difcurrent_density} is still present. In summary, from an ontological point of view, Madelung's proposal is completely orthodox. However, from a mathematical point of view, its formalism (and computational abilities) is very similar to de Broglie's proposal. Therefore, a reader who does not feel comfortable with the ideas of Bohm and de Broglie can assume Madelung's point of view and use many of the concepts explained in this book as a mathematical computational tool.

\vspace*{-6pt}
\subsection{Is Bohmian mechanics a useful theory?}\label{om.sec_intro.10}

Our historical introduction to Bohmian mechanics ends here. The
reader has certainly noticed that we have some preference for the
Bohmian interpretation over the Copenhagen, orthodox, explanation of
quantum phenomena. Obviously, we have a profound and
sincere respect for the Copenhagen interpretation of quantum
phenomena and its extraordinary computing capabilities. As discussed in in the section "What Is a Quantum Theory" in the introduction, a quantum theory connects the ontological and empirical planes. The scientific method is based on experiments. From an empirical point of view, 
what we have to expect from a physical theory that describes
natural phenomena is quite pragmatic. As Bohr said (see page 228 in
\cite{om.bohr1949}):\\
\begin{quote}
In my opinion, there could be no other way to deem a logically consistent mathematical formalism than by demonstrating the departure of its consequence from experience or by proving that its predictions did not exhaust the possibilities of observations.\\
\end{quote}
Curiously, similar empirical arguments were used by Bohm to defend his causal theory against its many detractors (see page 18 in \cite{om.Bohm1953b}):\\
\begin{quote}
In conclusion, the author would like to state that we would admit only two valid reasons for discarding a theory that explains a wide range of phenomena. One is that the theory is not internally consistent, and the second is that it disagrees with experiments.\\
\end{quote}\vspace*{-12pt}
In summary, from all the discussions done up to here, the only
statement that we will use in the rest of the book is that one
cannot discard an interpretation of quantum phenomena in terms of
wave functions and trajectories because it is \textit{internally consistent}
and it \textit{agrees with experiments}. Once such a statement is
accepted, other practical questions about Bohmian mechanics appear
naturally, for example, is Bohmian mechanics a useful
\textit{computational tool} for predicting the results of quantum experiments?
Although it provides identical predictions as the orthodox
formalism, are there advantages in
\textit{understanding/visualizing/explaining} quantum experiments in
terms of quantum trajectories? Or, even, is it a recommended
formalism for \textit{thinking} about limits and extensions of
(nonrelativistic) quantum theory? This book, through several
examples in the following chapters, will try to convince the reader
that the answer to these three questions is affirmative.

\vspace*{-1pc}
\section{Bohmian Mechanics for a Single Particle}\label{om.sec_single}

After the previous historical introduction, we start with the formal
presentation of Bohmian mechanics for a single particle. Usually,
new scientific knowledge is built from small variations of old
ideas. This explains why the initial development of quantum theory
(see \sref{om.sec_intro.4}) was so traumatic. Quantum explanations
did not evolve from a small variation of classical mechanics but
from radically new ideas. The languages of both theories were
completely different. Classical theory provides an explanation of a
physical experiment on particles in terms of a well-defined trajectory,
while the orthodox quantum theory needs a wave function. Here, we
will see that both a wave description of an ensemble of classical
particles and a description of quantum systems with trajectories are
possible. Starting by setting a common language for both classical
and quantum theories will certainly improve our understanding on the
similarities and differences between them.\footnote{It is very rare
to find such descriptions in the standard literature but, in our
opinion, it is very important to be able to compare classical and
quantum mechanics on an equal footing.}

\subsection{Preliminary discussions}\label{om.sec_single.1}

For a reader without any previous knowledge on quantum mechanics,
the following sections would be not only a presentation on Bohmian
mechanics, but also on quantum mechanics in general. The \textit{Copenhagen} and the \textit{Bohmian} wave functions have exactly the
same time evolution (when the same configuration space is used in both cases). In addition, the \textit{Bohmian} wave function is
complemented by a \textit{Bohmian} trajectory which appears in a natural
way when developing a trajectory-based explanation of quantum
mechanics as we will discuss in section 1.2.3.

The initial formulation of quantum mechanics with Bohmian
trajectories developed by de Broglie and Bohm was perfomed for dynamical
systems with associated velocities much slower than the speed of
light. Although it is, in principle, possible to extend Bohmian mechanics to
relativistic systems (see, for example, chapter 8 and Refs.
\cite{om.extra6,om.extra7,om.ward}), we will only deal in this
chapter with nonrelativistic systems.

In order to simplify as much as possible our mathematical notation,
we will first study a single spinless particle \textit{living} in a
one-dimensional (1D) space. The spatial  degree of freedom of the particle will be represented by $x$. The generalization
of all the arguments mentioned in this section into a single
particle in a 3D space is quite simple. However, the practical
solution of a single-particle Schr\"odinger equation in a 3D space with $x$, $y$ and $z$ degrees of freedom has considerable computational difficulties, as we will discuss in
\sref{om.sec_many}
and appendix A.1. In \sref{om.sec_many.4} we will consider the
role of the spin.

\subsection{Creating a wave equation for classical mechanics}\label{om.sec.single_2}

In this section we will derive a wave equation similar, although not
identical, to the Schr\"odinger equation for an ensemble of
classical trajectories. Certainly, this approach allows us to
compare the quantum and classical theories by using a very similar
language.

\subsubsection{Newton's second law}

Our starting point to derive such a classical wave equation will be
Newton's second law \cite{om.Feynmann1963}. Let us consider a particle with
mass $m$ in a classical system that moves under the action of a
potential $V(x)$, where $x$ is the position coordinate. Here, we
assume the potential to be time independent to simplify the
mathematical treatment. We define the particle trajectory as $x[t]$
and its velocity as $v[t] = dx[t]/dt\equiv\dot{x}[t]$. Since we are
considering a classical system, the trajectory of the particle will
be obtained from its acceleration, $a[t] =
d^2x[t]/dt^2\equiv\ddot{x}[t]$, from Newton's second law:
\begin{equation} \label{om.Newton1D}
m \ddot{x}[t] = \left [-\frac{\partial V(x)}{\partial x} \right ] _{x = x[t]}
\end{equation}
Since \eref{om.Newton1D} is a second-order differential equation, we need to specify both the initial position $x[t_0] = x_0$ and the initial velocity $v[t_0] = v_0$ of the particle.

\subsubsection{Hamilton's principle}

Apart from \eref{om.Newton1D}, there are other alternative ways to describe a classical system. For example, according to
Hamilton's principle \cite{om.Feynmann1963}, the trajectory $x_{\rm
p}[t]$\footnote{The suffix ``p'' means \textit{physical} in order to
distinguish from nonphysical trajectories, ``np,'' but it will be
omitted, when unnecessary.} solution of \eref{om.Newton1D} between
two different times, $t_0$ and $t_f$, provides a stationary value
for the action function, $S(x[t];x_0,t_0;x_f,t_f)$, where $x_0 =
x[t_0]$ and $x_f = x[t_f]$. Hereafter, whenever possible, we will
omit the dependence of the action function on $x_0$, $t_0$, $x_f$,
and $t_f$:
\begin{equation} \label{om.Hamilton_principle}
\left[\frac{\delta S(x[t])}{\delta x[t]}\right]_{x[t] = x_{\rm p}[t]} = 0
\end{equation}
where the action function is defined as:
\begin{equation} \label{om.action_function1D}
 S(x[t]) = \int_{t_0}^{t_f} L(x[t],\dot{x}[t]) dt
\end{equation}
The function $L(x,v)$ is the Lagrangian function:
\begin{equation} \label{om.lagrangian}
L(x,v) = K(v)-V(x)
\end{equation}
with $K(v) = m v^2/2$ the kinetic energy of the particle. The Lagrangian equation can also be defined as $L(x,v) = p v-H(x,v)$, where $H(x,p) = K(v) + V(x)$ is the Hamiltonian function\footnote{These definitions of Lagrangian and Hamiltonian functions are valid for the simple system described here. In any case, a different definition does not change the main results developed here.}
and $p = mv$ is the (linear) momentum.

\begin{figure}
\centering
\includegraphics{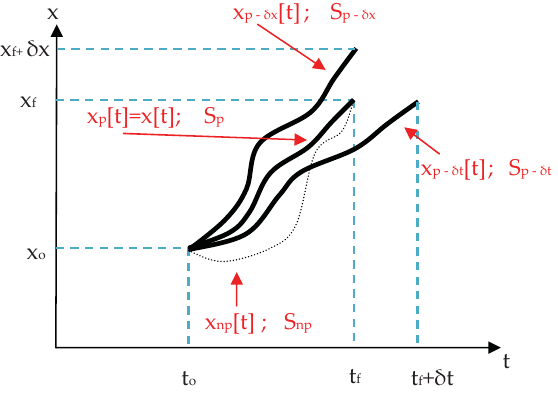}
\caption{Schematic representation of physical (solid lines) and nonphysical (dotted
lines) trajectories in the $(x,t)$ plane. The trajectory $x_{\rm
p}[t]$ has the initial time $t_0$, the final time $t_f$, the initial
position $x_{\rm p}[t_0] = x_0$, and the final position $x_{\rm
p}[t_f] = x_f$. The trajectory $x_{{\rm p} - \delta x}[t]$ is a
physical trajectory with identical initial and final conditions as
$x_{\rm p}[t]$, except for a different final position $x_f + \delta
x$. The trajectory $x_{{\rm p} - \delta t}[t]$ is a physical
trajectory with identical initial and final conditions as $x_{\rm
p}[t]$ but reaching the final point at a larger time $t + \delta
t$.} \label{om_fig_action}
\end{figure}

Let us discuss the meaning of a stationary (or extremal) value of
the integral in \eref{om.action_function1D}. We denote the physical
trajectory solution of \eref{om.Newton1D} as $x_{\rm p}[t]$ and
choose a slightly different trajectory $x_{\rm np}[t] = x_{\rm p}[t]
+ \delta x[t]$ with the same initial and final conditions, that is,
the same $x_0$, $t_0$, $x_f$, and $t_f $. Therefore,
we have $\delta x[t_0] = 0$ and $\delta x[t_f] = 0$. See
\fref{om_fig_action} for a schematic representation of both
trajectories. \Eref{om.Hamilton_principle} means that the value of
the action function $S_{\rm np} = S(x[t] + \delta
x[t];x_0,t_0;x_f,t_f)$ will always be greater\footnote{Strictly
speaking, the physical trajectory can also correspond to a maximum
of the action function. Such a maximum value is also an (stationary)
extremal value of \eref{om.Hamilton_principle}.} than $S_{\rm p}$;
that is, $S_{\rm np}>S_{\rm p}$. Thus, the trajectory that provides
a stationary value for the action function is the physical
trajectory, $x_{\rm p}[t]$, while $x_{\rm np}[t] = x_{\rm p}[t] +
\delta x[t]$ is a nonphysical solution incompatible with
\eref{om.Newton1D}.

The numerical evaluation of \eref{om.action_function1D} to find
$S_{\rm p}$ requires the prior knowledge of the trajectory $x_{\rm
p}[t]$. If we do not know the trajectory, we would have to evaluate
$S(x[t];x_0,t_0;x_f,t_f)$ for all possible trajectories (starting
from $x_0$ and $t_0$ and ending at $x_f$ and $t_f$) and choose the
one that provides a stationary value. Certainly, it seems that
\eref{om.Hamilton_principle} has little practical utility. However,
it provides an interesting starting point to develop our classical
wave equation.

\subsubsection{Lagrange's equation}

From Hamilton's principle it is possible to derive the Lagrange's
equation \cite{om.Feynmann1963} that gives us the differential
equation that the physical trajectory satisfies.

Let us define a trajectory $x_{\rm np}[t] = x_{\rm p}[t] + \delta x[t]$ by adding a small arbitrary displacement $\delta x[t]$ to the physical trajectory.
In particular we fix $\delta x[t_0] = 0$ and $\delta x[t_f] = 0$, so both trajectories have the same initial and final conditions. A Taylor expansion of the Lagrange function of \eref{om.lagrangian} around the physical trajectory $x_{\rm p}[t]$ reads:
\begin{equation} \label{om.cal_var1}
L(x_{\rm np}[t], \dot{x}_{\rm np}[t]) = L(x_{\rm p}[t],\dot{x}_{\rm p}[t]) + \left[ \frac {\partial L} {\partial x} \delta x[t] + \frac {\partial L} {\partial \dot{x}} \dot{\delta x} [t] \right]_{\twoconds{x = x_{\rm p}[t]}{\dot{x} = \dot{x}_{\rm p}[t]}}
\end{equation}
The action function of \eref{om.action_function1D} with the Lagrangian of \eref{om.cal_var1} gives two contributions. The first term of the right-hand side (r.h.s.) of \eref{om.cal_var1} is $S_{\rm p}$. After performing an integration by parts on the second part and using $\delta x[t_0] = 0$ and $\delta x[t_f] = 0$, we obtain:
\begin{eqnarray} \label{om.cal_var3}
 S(x_{\rm np}[t];x_0,t_0;x_f,t_f) &=& S_{\rm p} + \int_{t_0}^{t_f}
 \delta x[t]\times\left(\left[ \frac {\partial L} {\partial x}\right]_{\twoconds{x = x_{\rm p}[t]}{\dot{x} = \dot{x}_{\rm p}[t]}} - \frac {d} {dt} \left[\frac {\partial L} {\partial \dot{x}} \right]_{\twoconds{x = x_{\rm p}[t]}{\dot{x} = \dot{x}_{\rm p}[t]}} \right) dt\nonumber\\
\end{eqnarray}
Hamilton's principle tells us that the action function must take a
stationary value for the physical trajectory. This is equivalent to impose that small variations
around the physical trajectory do not modify the value of the
action, that is, for any (small) variation $\delta x[t]$, $S_{\rm
np} = S_{\rm p}$. Thus:
\begin{equation} \label{om.lagrange_equation}
\left[ \frac {\partial L(x,\dot{x})} {\partial x}
\right]_{\twoconds{x = x_{\rm p}[t]}{\dot{x} = \dot{x}_{\rm p}[t]}}
- \frac {d} {dt} \left[ \frac {\partial L(x,\dot{x})} {\partial
\dot{x}} \right]_{\twoconds{x = x_{\rm p}[t]}{\dot{x} = \dot{x}_{\rm
p}[t]}} = 0
\end{equation}

\Eref{om.lagrange_equation} is the Lagrange equation and gives, for a classical system, a differential equation that classical trajectories must fulfill.
In order to enlighten the meaning of the Lagrange equation, \eref{om.lagrange_equation}, we can see that by substituting ${\partial L}/{\partial x} = -{\partial V(x)}/{\partial x}$ and $ {\partial L}/{\partial \dot{x}} = m v$ in \eref{om.lagrangian}, we recover:
\begin{equation}
\label{om.lagrange_equation_Newton}
\left[ \frac {\partial V(x,t)} {\partial x} \right]_{\twoconds{x = x_{\rm p}[t]}{\dot{x} = \dot{x}_{\rm p}[t]}} - \frac {d} {dt} \left[ m \frac {dx[t]} {dt} \right]_{\twoconds{x = x_{\rm p}[t]}{\dot{x} = \dot{x}_{\rm p}[t]}} = 0
\end{equation}
which is the original Newton's second law of \eref{om.Newton1D}.
In fact, what we have done is to  check that Newton's second law is included in the fundamental Hamilton's principle.

\vspace*{-6pt}
\subsubsection{Equation for an (infinite) ensemble of trajectories}

The formalism based on the action function allows us to find not only a single physical trajectory but also the equation for an (infinite) ensemble of physical trajectories with slightly different initial or final conditions. For example, let us define:
\begin{equation}
\label{om.action_function1D_dt}
 S_{{\rm p} - \delta t} = S(x_{{\rm p} - \delta t}[t]) = \int_{t_0}^{t_f + \delta t} L(x_{{\rm p} - \delta t}[t],\dot{x}_{{\rm p} - \delta t}[t]) dt
\end{equation}
$x_{{\rm p} - \delta t}[t]$ being a trajectory with identical initial and final positions as $x_{\rm p}[t]$, but taking a longer time $t_f + \delta t $ (see trajectory $x_{{\rm p} - \delta t}[t]$ in \fref{om_fig_action}).
The new physical trajectory $x_{{\rm p} - \delta t}[t]$ can be written as $x_{{\rm p} - \delta t}[t] = x_{\rm p}[t] + \delta x[t]$ for $t \in [t_0,t_f]$. For larger times, $x_{{\rm p} - \delta t}[t_f + \delta t] = x_{\rm p}[t_f + \delta t] + \delta x[t_f + \delta t]$ with $x_{\rm p}[t_f + \delta t] = x_{\rm p}[t_f] + \dot{x}_{\rm p}[t_f] \delta t$, where we have done a first-order Taylor expansion of $x_{\rm p}[t]$ around $t_f$ for a small $\delta t$. Since $x_{{\rm p} - \delta t}[t_f + \delta t] = x_{\rm p}[t_f]$, we obtain:
\begin{equation}
\label{om.cal_var4}
\delta x[t_f + \delta t] = -\dot{x}_{\rm p}[t_f] \delta t
\end{equation}
By following the same steps done after the Taylor expansion in \eref{om.cal_var1}, we obtain:
\begin{eqnarray}
\label{om.cal_var5}
 S_{{\rm p} - \delta t} &=& \int_{t_0}^{t_f + \delta t} L dt + \left[ \frac {\partial L} {\partial \dot{x}} \delta x[t] \right]_{t_0}^{t_f + \delta t}+ \int_{t_0}^{t_f + \delta t} \delta x[t] \left(\left[ \frac {\partial L} {\partial x} \right]_{x = x[t]}^{\dot{x} = \dot{x}[t]} + \frac {d} {dt} \left[ \frac {\partial L} {\partial \dot{x}} \right]_{x = x[t]}^{\dot{x} = \dot{x}[t]} \right) dt\qquad
\end{eqnarray}
For the first term of the r.h.s. of \eref{om.cal_var5}, we obtain:
\begin{equation}
\label{om.cal_var6}
\int_{t_0}^{t_f + \delta t} L dt = \int_{t_0}^{t_f} L dt + \int_{t_f}^{t_f + \delta t} L dt = S_{\rm p} + L \delta t
\end{equation}
For the second term of the r.h.s. of \eref{om.cal_var5}:
\begin{equation}
\label{om.cal_var7}
\left[ \frac {\partial L} {\partial \dot{x}} \delta x[t] \right]_{t_0}^{t_f + \delta t} = -\left[ \frac {\partial L} {\partial \dot{x}} \right]_{x = x[t_f + \delta t]}^{\dot{x} = \dot{x}[t_f + \delta t]} \dot{x}(t_f) \delta t = -p[t_f + \delta t] \dot{x}[t_f] \delta t
\end{equation}
where we use \eref{om.cal_var4}, $\delta x[t_0] = 0$ and ${\partial L}/{\partial \dot{x}} = m v = p$, $p = m v$ being the particle momentum.

Finally, since $x[t]$ is also a physical trajectory that fulfills
Lagrange equation, 
 from  $t_0$ to $t_f +
\delta t$, we obtain for the third term:
\begin{equation}
\label{om.cal_var8}
\int_{t_0}^{t_f + \delta t} \delta x[t] \left[ \left[ \frac {\partial L} {\partial x} \right]_{x = x[t]}^{\dot{x} = \dot{x}[t]} + \frac {d} {dt} \left[ \frac {\partial L} {\partial \dot{x}} \right]_{x = x[t]}^{\dot{x} = \dot{x}[t]} \right]dt = 0
\end{equation}
Putting together Eqs. (\ref{om.cal_var6}) (\ref{om.cal_var7}), and (\ref{om.cal_var8}), we obtain:
\begin{equation}
\label{om.cal_var9}
 S_{{\rm p} - \delta t} = S_{\rm p} + L \delta t - p \dot{x} \delta t = S_{\rm p} - H \delta t
\end{equation}
So, with ${\partial S}/{\partial t_f} = \lim_{\delta t \rightarrow 0}(S_{{\rm p} - \delta t} - S_{\rm p})/\delta t$ we can conclude that:
\begin{equation}
\label{om.HJ1}
\frac{\partial S(x_{\rm p}[t])} {\partial t_f} = -H(x_{\rm p}[t],\dot{x}_{\rm p}[t])
\end{equation}
In summary, if $S_{\rm p}$ is the stationary value of the action function for $x_{\rm p}[t]$, then the variation of the new stationary value of another physical trajectory $x_{{\rm p} - \delta t}[t]$, which has identical initial and final conditions but a slightly modified final time, is equal to the Hamiltonian (with a negative sign) evaluated at the final time of the trajectory $x_{\rm p}[t]$. See problem \ref{om.p1} to discuss a particular example.

Next, we will see that the evaluation of the value $S_{{\rm p} - \delta x}$ when we modify the final position $x_f + \delta x$ without modifying the initial and final times leads also to an interesting result (see \fref{om_fig_action}). Notice that the final position of the new physical trajectory $x_{{\rm p} - \delta x}[t] = x[t] + \delta x[t]$ means:
\begin{eqnarray}
\label{om.cal_var10}
\delta x[t_0] = 0  ; \; \; \delta x[t_f] = \delta x_f
\end{eqnarray}
By following the same steps done after the Taylor expansion in \eref{om.cal_var1}, we obtain now:
\begin{eqnarray}
\label{om.cal_var11}
\hspace*{-24pt}S_{{\rm p} - \delta x} &=& \int_{t_0}^{t_f} L dt + \left[ \frac
 {\partial L} {\partial \dot{x}} \delta x[t] \right]_{t_0}^{t_f} + \int_{t_0}^{t_f} \delta x[t] \left(\left[ \frac {\partial L} {\partial x} \right]_{x = x[t]}^{\dot{x} = \dot{x}[t]} + \frac {d} {dt} \left[ \frac {\partial L} {\partial \dot{x}} \right]_{x = x[t]}^{\dot{x} = \dot{x}[t]} \right) dt
\end{eqnarray}\vspace*{-12pt}

\noindent The first term of the r.h.s. of \eref{om.cal_var11} is again $S_{\rm p}$. The second term is
\begin{equation}
\label{om.cal_var12}
\left[ \frac {\partial L} {\partial \dot{x}} \delta x[t] \right]_{t_0}^{t_f} = \left[ \frac {\partial L} {\partial \dot{x}} \right]_{x = x[t_f]}^{\dot{x} = \dot{x}[t_f]} \delta x_f = p[t_f] \delta x_f
\end{equation}
where we have used \eref{om.cal_var10}. The third term is zero. We can conclude that ${\partial S}/{\partial x} = \lim_{\delta x \rightarrow 0}(S_{{\rm p} - \delta x} - S_{\rm p})/\delta x$ is equal to:
\begin{equation}
\label{om.HJ2}
\frac{\partial S(x_{\rm p}[t];x_0,t_0;x,t)} {\partial x} = m \dot{x}_{\rm p}[t] = p_{\rm p}[t]
\end{equation}
In summary, the variation of the stationary value of the action
function when we slightly modify the final position of a physical
trajectory is equal to the momentum of the trajectory at the final
time.\footnote{We have recovered the subindex ``p'' to emphasize that
\eref{om.HJ2} is only valid for physical trajectories.} To check
\eref{om.HJ2} for a particular case, see problem \ref{om.p2}.

\subsubsection{Classical Hamilton--Jacobi equation}

Now, using \eref{om.HJ1} and substituting the velocity (momentum) into
the Hamiltonian by \eref{om.HJ2}, we  obtain the well-known
Hamilton--Jacobi equation:
\begin{equation}
\label{om.hamilton_jacobi1D}
H\left(x,\frac{\partial S(x,t)}{\partial x},t \right) + \frac{\partial S(x,t)}{\partial t} = 0
\end{equation}
We want to use the Hamilton--Jacobi equation,
\eref{om.hamilton_jacobi1D}, to find all the physical trajectories.
For this reason,  we have eliminated
the dependence of $S$ on the trajectory $x[t]$. In addition, we have
also eliminated $t_0$ and $x_0$ because we consider that $x$ and $t$
are not parameters but variables. Notice that
\eref{om.hamilton_jacobi1D} is valid for physical trajectories so
that once we know $S(x,t)$, we are able to directly compute all
physical trajectories for all possible initial conditions.
Therefore, \eref{om.hamilton_jacobi1D} defines an (infinite)
ensemble of trajectories rather than just a single trajectory.
However, in most practical cases, the direct solution of the
Hamilton--Jacobi equation is much more difficult than using the Newton or
Lagrange formulation of classical mechanics. It has limited
practical interest. However, it provides a direct theoretical
connection with a classical wave equation similar to the
Schr\"odinger equation. See problem \ref{om.p3} for a particular
solution of \eref{om.hamilton_jacobi1D}.

\subsubsection{Local continuity equation for an (infinite) ensemble of classical particles}

\Eref{om.hamilton_jacobi1D} can be interesting when dealing with an ensemble of trajectories. For example, it can be used when we have a classical (single-particle) experiment with some practical difficulty in specifying the initial position and velocity of the particle, such that different experimental realizations can have slightly different initial conditions. The ensemble of trajectories solution of \eref{om.hamilton_jacobi1D} will provide a statistical (probabilistic) description of the classical (single-particle) experiment that accounts for the variability in the initial conditions.

For such an experiment, we can reasonably assume that the
variability of the initial position will be limited to a particular
spatial region. At the initial time $t_0 = 0$, we can define some
distribution of the initial position of the particles
$R^2(x,0)\geq0$.\footnote{Such a distribution is positive (or zero),
but it is not necessary to impose that $R^2(x,0)$ is normalized to
unity.} Such an ensemble of trajectories will evolve in time
according to \eref{om.hamilton_jacobi1D} so that we will obtain a
function $R^2(x,t)$ that describes the particular distribution of
particles at any time. The function $R^2(x,t)$ is constructed by
``counting'' the number of trajectories present inside the interval
$(x, x + dx)$ at time $t$. From \eref{om.HJ2}, we define the
particle velocity as $\dot x[t] = (1/m){\partial S(x,t)}/{\partial
x}$. Finally, we know that all these classical particles will move,
in a \textit{continuous} way, from one unit of volume to another.
Therefore, as we have extensively discussed in
\sref{om.sec_intro.1}, we can ensure that the ensemble of
trajectories accomplishes the following \textit{local} conservation
law:
\begin{equation}
\label{om.conservation_law1D}
\frac{\partial R^2(x,t)}{\partial t} + \frac {\partial } {\partial x} \left(\frac {1} {m} \frac {\partial S(x,t)}{\partial x} R^2(x,t) \right) = 0
\end{equation}
In summary, \eref{om.conservation_law1D} just certifies that if a
classical particle goes from one point to another, it has to go
through all the trajectory positions between these two places.

\subsubsection{Classical wave equation}

Now, we have all the ingredients to develop a wave equation for classical mechanics. In previous paragraphs we have been dealing with two (real) functions, $S(x,t)$ and $R(x,t)$. The first, $S(x,t)$, is the action function and appears in the Hamilton--Jacobi equation, \eref{om.hamilton_jacobi1D}. In particular, its spatial derivative determines the velocity of the particles (see \eref{om.HJ2}). The second, $R(x,t)$, tells us how an ensemble of trajectories is distributed at each time $t$. It evolves according to the conservation law, \eref{om.conservation_law1D}. In this sense, we can assume that they are some kind of ``field'' or ``wave'' that guide classical particles.

We can construct the following classical\footnote{For simplicity, we assume that $R(x,t)$ and $S(x,t)$ are single valued. This condition is equivalent to assigning, at each time, a unique velocity to each position of physical space. If this is not the case, then we will need different wave functions, that is, a mixed state, to describe the different velocity fields of the classical system.} (complex) wave function $\psi_{cl}(x,t) = R(x,t) \exp(i S(x,t)/\hbar)$.
We divide the action function inside the exponential by $\hbar$ (the reduced Planck's constant) in order to provide a dimensionless argument.
Then, it can be shown that the two previous (real) equations, Eqs. (\ref{om.hamilton_jacobi1D}) and (\ref{om.conservation_law1D}), for $S(x,t)$ and $R(x,t)$ are equivalent to the following (complex) classical wave equation for $\psi_{cl}(x,t)$:
\begin{eqnarray}
\label{om.classcho}
i \hbar \frac{ \partial \psi_{cl}(x,t)} {\partial t} &=& -\frac {\hbar^2}{2m} \frac{ {\partial}^2 \psi_{cl}(x,t)} {\partial x^2} + V(x,t) \psi_{cl}(x,t)+ \frac {\hbar^{2}} {2 m} \frac { {\partial}^2 |\psi_{cl}(x,t)|/ \partial x^2}{|\psi_{cl}(x,t)|} \psi_{cl}(x,t)\nonumber\\
\end{eqnarray}
The demonstration of this expression is left as an exercise to the
reader (see problem \ref{om.p3bis}). Additionally, see problem
\ref{om.p4} for a solution of the classical wave equation,
\eref{om.classcho}, for a free particle.

In conclusion, an ensemble of classical trajectories can be described with a wave function solution of a wave equation.
Thus, a common language for classical and quantum mechanics has been obtained by using an (infinite) ensemble of classical trajectories with different initial positions and velocities instead of just one single classical trajectory.

\subsection{Trajectories for quantum systems}\label{om.sec_single.3}

Before comparing the classical and quantum wave equations, let us first discuss in this section  whether trajectories can be also used to describe quantum systems.
We will introduce such trajectories in two different ways: first, as a direct consequence of the local conservation of particles extracted from the Schr\"odinger equation and, second, directly following the work presented by Bohm in his original paper \cite{om.bohm1952a}.

\subsubsection{Schr\"odinger equation}

As discussed in \sref{om.sec_intro.4}, the single-particle
Schr\"odinger equation in a 1D  quantum  system subjected to a scalar
time-dependent potential, $V(x,t)$, is:
\begin{equation}
\label{om.Schrodinger1D}
i \hbar \frac{\partial \psi(x,t)} {\partial t} = -\frac {\hbar^2} {2m} \frac{ {\partial}^2 \psi(x,t)} {\partial x^2} + V(x,t) \psi(x,t)
\end{equation}
It is important to emphasize that in the orthodox interpretation of
$\psi(x,t)$, \eref{om.Schrodinger1D} does not describe a single
experiment but an ensemble of identical (single-particle)
experiments. The orthodox meaning of the square modulus of the wave
function $|\psi(x,t)|^2$ is the probability density of finding a
particle at position $x$ at time $t$ when a measurement is
performed.  Such probabilities assume an infinite number of identical experiments.

\subsubsection{Local conservation law for an (infinite) ensemble of quantum trajectories}
\label{om.sec_conservation}

It is known that there is a \textit{local} continuity equation associated with \eref{om.Schrodinger1D}. Let us first mathematically derive it to later discuss its physical implications.

In order to find a local continuity equation, let us work with $\psi(x,t)$ and its complex conjugate $\psi^*(x,t)$. In particular, we can rewrite \eref{om.Schrodinger1D} as:
\begin{eqnarray}
\psi^*(x,t) i \hbar \frac{\partial \psi(x,t)} {\partial t} &=& -\psi^*(x,t)\frac {\hbar^2} {2m} \frac{ {\partial}^2 \psi(x,t)} {\partial x^2}+ \psi^*(x,t) V(x,t) \psi(x,t)
\label{om.Schrodinger1D_1}
\end{eqnarray}
and the complex conjugate of \eref{om.Schrodinger1D_1} as:
\begin{eqnarray}
-\psi(x,t)i \hbar \frac{\partial \psi^*(x,t)} {\partial t} &=& -\psi(x,t)\frac {\hbar^2} {2m} \frac{ {\partial}^2 \psi^*(x,t)} {\partial x^2}+ \psi(x,t)V(x,t) \psi^*(x,t)
\label{om.Schrodinger1D_2}
\end{eqnarray}
From Eqs. (\ref{om.Schrodinger1D_1}) and (\ref{om.Schrodinger1D_2}), we obtain:
\begin{equation}
\label{om.Schrodinger1D_4}
 \frac{\partial |\psi(x,t)|^2} {\partial t} = i\frac {\hbar} {2m} \frac {\partial} {\partial x} \left(\psi^*(x,t) \frac{ {\partial} \psi(x,t)} {\partial x} - \psi(x,t) \frac{ {\partial} \psi^*(x,t)} {\partial x} \right)
\end{equation}
We can easily identify \eref{om.Schrodinger1D_4} as the local conservation of particles discussed in \eref{om.difcurrent_density} where $\rho(x,t) = |\psi(x,t)|^2$ and the current density, $J(x,t)$, is defined as:
\begin{equation}
\label{om.current}
J(x,t) = i \frac {\hbar} {2 m} \left(\psi(x,t) \frac {\partial \psi^{*}(x,t)} {\partial x}- \psi^{*}(x,t) \frac {\partial \psi(x,t)} {\partial x} \right)
\end{equation}

Unlike other  wave equations, the Schr\"odinger equation is compatible
with a local conservation of particles due to the fact that $V(x,t)$
is a real function. We have noticed above that we can interpret
$\rho(x,t) = |\psi(x,t)|^2$ as a spatial distribution of an ensemble
of trajectories. Each trajectory correspond to a different experiment of the single particle system. Then, in spite of dealing with a single particle system, from a statistical point of view, such (very large) ensemble of trajectories can be interpreted as a (very large) ensemble of particles describing simultaneously all possible experiments. The presence of such local
conservation of particles is very relevant for us because it
justifies our aim to look for an ensemble of continuous trajectories
describing $\rho(x,t) = |\psi(x,t)|^2$.

\subsubsection{Velocity of Bohmian particles}

In Bohmian mechanics, the particle velocity is defined as:
\begin{equation}
\label{om.velocity}
v(x,t) = \frac{J(x,t)}{|\psi(x,t)|^2}
\end{equation}
where $J(x,t)$ is given by \eref{om.current}.
Taking into account that $|\psi(x,t)|^2$ is the distribution of the ensemble of particles, it is easy to show that this velocity definition is compatible with the local continuity equation, \eref{om.Schrodinger1D_4}, and that an ensemble of well-defined trajectories whose initial positions are all selected according to the distribution $|\psi(x,t_0)|^2$ will reproduce $|\psi(x,t)|^2$ at all times (see problem \ref{om.ppre6}).

Notice that one could add a divergence-free term to this velocity, and it would still fulfill \eref{om.Schrodinger1D_4}. Even though this would change the path followed by each individual trajectory, the trajectory distribution and, therefore, the measurement outcomes would remain unchanged.
However, if one develops the nonrelativistic limit of a relativistic treatment of Bohmian mechanics, one finds that the divergence-free  term does not appear for spin-0-particles and that the velocity is unambiguously defined by \eref{om.velocity}. In addition, it has also been demonstrated that possible Bohmian paths are naively observable from a large enough ensemble of weak values \cite{om.wiseman2007,om.4marian,om.7colomes} giving full support to the definition in \eref{om.velocity}. For other values of spin, the situation is a bit more complex \cite{om.ward2010,om.Holland1999,om.Holland2003} and far from the scope of this book.

\subsubsection{Quantum Hamilton--Jacobi equation}\label{sec.quantum_HJ.om}

Following the path described by Bohm in his original paper  [2], we
will now start from the quantum wave equation, that is, the
Schr\"odinger equation, to arrive at a ``quantum'' Hamilton--Jacobi
equation that describes the motion of quantum trajectories.

The first step is to write the quantum (complex) wave function, $\psi(x,t) = \psi_r(x,t) + i \psi_i(x,t)$, in polar form:
\begin{eqnarray}
R^2(x,t) = \psi_{r}^2(x,t) + \psi_{i}^2(x,t) \label{om.polar1D_1} \\
S(x,t) = \hbar \arctan \left(\frac {\psi_{i}(x,t)} {\psi_{r}(x,t)} \right)
\label{om.polar1D_2}
\end{eqnarray}

In principle, $S(x,t)$, the so-called quantum action, is not well defined when $\psi_r(x,t) =
\psi_i(x,t) = 0$, that is  at those points where $R(x,t) = 0$, meaning that no
particles will reach them.\footnote{We assume that the wave function
is single valued so that $R(x,t)$ is also single valued. However,
the definition of $S(x,t)$ has some practical difficulties. In
principle, $S(x,t)$ is a multivalued function because the function
$\arctan(x)$ itself is a multivalued function. If we want to use
Eqs. (\ref{om.polar1D_1}) and (\ref{om.polar1D_2}) to reconstruct
the wave function, then the multivalued problem can be eliminated by
imposing an  additional restriction on the definition of
$S$ \cite{om.Holand1993,om.Bohm1993,om.Durrllibre,om.reviewabc,om.llibreph}.}

The quantum Hamilton--Jacobi equation can be found by introducing $\psi(x,t) = R(x,t) \exp(i S(x,t)/\hbar)$ into \eref{om.Schrodinger1D}. On the one hand, the imaginary part of the resulting equation gives the \textit{local} conservation law identical to the one shown in \eref{om.conservation_law1D}, which we rewrite here for convenience:
\begin{equation}
\frac{\partial R^2(x,t)}{\partial t} + \frac {\partial } {\partial x} \left(\frac {1} {m} \frac {\partial S(x,t)}{\partial x} R^2(x,t) \right) = 0
\end{equation}
On the other hand, the real part gives a quantum Hamilton--Jacobi equation:
\begin{equation}
\label{om.hamilton_jacobi1D_des}
\frac{\partial S(x,t)}{\partial t} + \frac {1} {2 m} \left(\frac{ \partial S(x,t)} {\partial x} \right)^2 + V(x,t) + Q(x,t) = 0
\end{equation}

Since the last term in the r.h.s. of the classical wave equation, \eref{om.classcho}, is not present in the Schr\"odinger equation, \eref{om.Schrodinger1D}, an additional term appears in the quantum Hamilton--Jacobi equation, the so-called quantum potential, defined as:
\begin{equation}
\label{om.quantum_potential1D}
Q(x,t) = -\frac{\hbar^2} {2 m} \frac{{\partial}^2 R(x,t)/ \partial x^2} {R(x,t)}
\end{equation}

In conclusion, identical to the classical description of a system obtained from \eref{om.hamilton_jacobi1D_des}, we obtain an interpretation of the wave function solution of the Schr\"odinger equation as an ensemble of \textit{quantum} trajectories with different initial positions and velocities.
The velocity of each trajectory $x[t]$ is defined as:
\begin{equation}
\label{om.velocity_dSdx}
v[t] = \left[\frac{1}{m} \frac{\partial S(x,t)}{\partial x}\right]_{x = x[t]}
\end{equation}

Interestingly, it can be easily shown that this new expression for
the quantum velocity is identical to that mentioned in
\eref{om.velocity}. See problem \ref{om.p5} to show that both
definitions of the velocity of Bohmian trajectories are identical:
\begin{equation}
v(x,t) = \frac {1} {m} \frac {\partial S(x,t)} {\partial x} = \frac{J(x,t)} {|\psi(x,t)|^2}
\label{om.velocity2}
\end{equation}
where $J(x,t)$ is defined by \eref{om.current}.

\subsubsection{A quantum Newton-like equation}

If we compute the time derivative of the Bohmian velocity defined in \eref{om.velocity_dSdx}, we find a quantum Newton-like equation:
\begin{equation}
\label{om.Newton_law_quantic1}
m\frac {d} {dt} v(x[t],t)\! =\! m \frac {d} {dt} \left[ \frac{1}{m} \frac {\partial S} {\partial x} \right]_{x = x[t]} \!=\! \left[ \frac {\partial^2 S} {\partial x^2}\right]_{x = x[t]} \dot{x}[t] + \left[\frac {\partial} {\partial x} \frac {\partial S} {\partial t} \right]_{x = x[t]}
\end{equation}
We can rewrite \eref{om.Newton_law_quantic1} as:
\begin{equation}
\label{om.Newton_law_quantic2}
m\frac {d} {dt} v(x[t],t) = \left[ \frac {\partial} {\partial x} \left(\frac {1} {2m} \left(\frac {\partial S} {\partial x} \right)^{2} + \frac {\partial S} {\partial t} \right) \right]_{x = x[t]}
\end{equation}
Finally, using \eref{om.hamilton_jacobi1D_des} we get:
\begin{equation}
\label{om.Newton_law_quantic3}
m\frac {d} {dt} v(x[t],t) = \left[ -\frac {\partial} {\partial x} \left(V(x,t) + Q(x,t) \right) \right]_{x = x[t]}
\end{equation}

We conclude here the second route for finding a common language for
classical and quantum theories. The quantum (complex)
single-particle wave function can be interpreted as an ensemble of
trajectories that are all solutions of the same single-particle
experiment but with different initial conditions. The quantum
trajectories are not solutions of the classical Newton second  law with a
classical potential but solutions of the quantum Newton second law,
\eref{om.Newton_law_quantic3}, where a quantum potential (that
accounts for all nonclassical effects) is added to the classical
potential.

\subsection{Similarities and differences between classical and quantum mechanics}\label{om.sec_single.4}

In previous sections, we have provided a common language for
classical and quantum theories, in terms of either wave functions or
trajectories, to fairly compare differences and similarities between
both theories. Here we emphasize that an attempt to establish similarities and differences between classical and quantum mechanics have to be done by comparing either
classical and quantum wave functions or by comparing classical and quantum
ensembles of trajectories (not by comparing a single classical trajectory with a
quantum wave function). The main difference between the mathematical
description of the two theories is that the term \textit{$Q(x,t)$}
that appears in the quantum Hamilton--Jacobi equation,
\eref{om.hamilton_jacobi1D_des}, is exactly the same term that
appears in the classical wave equation, \eref{om.classcho}, but with a
change of sign. The term \textit{$Q(x,t)$} explains the
\textit{exotic} properties of quantum systems that are missing in
their classical counterparts:

\begin{figure}[b]
\centering
\includegraphics[width=14cm]{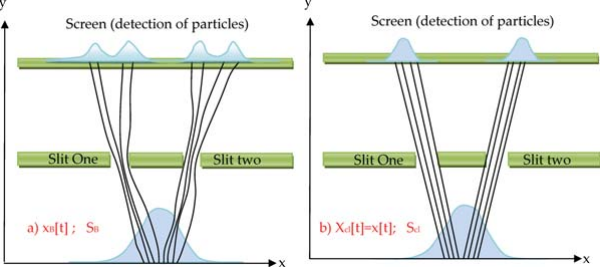}
\caption{A particle is sent toward a thin plate with two slits cut in it. This single-particle
experiment is repeated many times with some uncertainty on the
initial conditions of the particle positions. The distribution of
detected particles on the screen as a function of the position is
different if we deal with a quantum (a) or a classical (b) system.
In both cases, we use an identical description of the initial
ensemble of particles, that is, identical $R(x,t_0)$ and $S(x,t_0)$.
However, the classical evolution of the trajectories $x_{cl}[t]$ is
described by \eref{om.hamilton_jacobi1D}, which is independent of
$R(x,t)$, while quantum (Bohmian) trajectories $x_{B}[t]$ are
determined by \eref{om.hamilton_jacobi1D_des}, which depends
explicitly on the other particles of the ensemble through $Q(x,t)$ that depends on  $R(x,t)$.
These differences can also be understood by observing that the
evolution of a classical wave function is described by a nonlinear
equation, \eref{om.classcho}, while a quantum wave function evolves
with a linear one, \eref{om.Schrodinger1D}.}
\label{om_fig_dobleslit}
\end{figure}
\begin{enumerate}
\item \textbf{Differences}
\begin{enumerate}
\item \textbf{Quantum superposition} \\ One can apply the superposition principle to find solutions of the Schr\"odinger equation, \eref{om.Schrodinger1D}, but it cannot be applied to the classical wave equation, \eref{om.classcho}, since its last term is nonlinear. In fact, if $\psi_1(x,t)$ and $\psi_2(x,t)$ are solutions of the Schr\"odinger equation, then $a\psi_1(x,t) + b\psi_2(x,t)$ with arbitrary complex $a$ and $b$ is also a solution. Quantum mechanics will keep the (complex) amplitudes $a$ and $b$ constant at any time because of its linearity, which is not the case for classical mechanics.

{\quad}This is a fundamental difference between classical and quantum systems, which has deep consequences. Quantum mechanics can be developed in a linear vector space, while classical mechanics cannot.

\item \textbf{Quantum wholeness} \\ The presence of $Q(x,t)$ in the quantum Hamilton--Jacobi equation, \eref{om.hamilton_jacobi1D_des}, implies that Bohmian trajectories depend not only on the classical potential $V(x,t)$ but also on the quantum potential $Q(x,t)$, which is a function of the type of distribution of trajectories associated to different repetitions of the single-particle experiment, $R(x,t)$. In fact, it is the \textit{shape} and not the absolute value of $R(x,t)$ that acts on each individual quantum trajectory. On the contrary, each classical trajectory can be computed independently of the shape of the ensemble. The fact that the dynamics of one quantum trajectory in one particular experiment depends on the ensemble of other trajectories build from other identical experiments is highly counter-intuitive for our classical mind. 

{\quad}This surprising result for quantum mechanics can  be
illustrated with the double-slit experiment \cite{om.Dewdney}. We  assume that the initial wave functions
for the classical and quantum ensembles are identical at $t_0 = 0$.
Their difference appears in the time evolution of the trajectories.
For the quantum trajectories of the ensemble, the shape of the
ensemble (whether particles are stopped or not by the double-slit
screen) determines the shape of $R(x,t)$, which will affect the
dynamics of each trajectory, even those trajectories
that are far from the slit (see \fref{om_fig_dobleslit}a). On the
contrary, the classical Hamilton--Jacobi equation,
\eref{om.hamilton_jacobi1D}, is totally independent of $R(x,t)$, so
a single trajectory is completely independent from the rest of
trajectories of other experiments as seen in \fref{om_fig_dobleslit}b.
\end{enumerate}
\item \textbf{Similarities}
\begin{enumerate}
\item \textbf{Uncertainty} \\ As far as we deal with an ensemble of (classical or quantum) trajectories, there is an uncertainty in the exact value of any magnitude that can be measured from the ensemble. For example, one can compute the mean value and the standard deviation of the position of the classical ensemble. The classical and quantum uncertainties can have different origins, but both ensembles have uncertainties. The classical uncertainty is due to the technical difficulties in exactly repeating the initial conditions of a particular experiments; the quantum uncertainty has a more intrinsic origin  \cite{om.extra2,om.extra9}. See Ref. \cite{om.randomness} for a discussion on how Bohmian mechanics, although being a deterministic theory at the ontological level, provides a quite simple explanation on the origin of the quantum uncertainty at the empirical level. 

{\quad}\looseness-1Even assuming their different physical origin, what we want to
emphasize here is  that it is not licit to compare the uncertainty
of a quantum wave function with the uncertainty of a single
classical trajectory, saying that the classical trajectory has no
uncertainty and the quantum wave function has. As we have previously discussed,
we have to compare the uncertainty of classical and quantum wave
functions (or classical and quantum ensembles of trajectories).

\item \textbf{Initial conditions for the (classical or quantum)  Hamilton--Jacobi equation} \\ The classical Newton equation is a second-order differential equation, where both the initial position and the velocity have to be fixed. In the classical or quantum Hamilton--Jacobi equation, it seems that only an initial position is needed because the initial velocity is directly determined by the spatial derivative of the action. However, one can argue that in the Hamilton--Jacobi equation, two initial conditions are fixed, the initial position and the initial wave function (or initial action) that fixes the initial velocity. In other words, for one particular Hamiltonian, even if we fix one particular initial position, then it is possible to obtain different initial velocities if we select different initial wave functions. In conclusion, the discussion that classical trajectories are solved from second-order equations, while Bohmian trajectories from first-order equations, is somehow artificial if one recalls the quantum equilibrium hypothesis.\footnote{The quantum equilibrium hypothesis assumes that the initial positions and velocities of Bohmian trajectories are defined distributed according to the initial wave function \cite{om.extra2,om.extra9,om.Valentini2006,om.nonequilibrium}. This topic will be further discussed at the end of \sref{om.sec_many.5}.}

\item \textbf{Single-valued wave function and multivalued action function} \\ In principle, the action function solution of the (classical or quantum) Hamilton--Jacobi equation can be multivalued yielding different velocities in a particular $x$ and $t$ point. However, the classical wave or the Schr\"odinger equations deal with single-valued wave functions. If we want to model (classical or quantum) scenarios with multivalued velocities, then we have to consider several wave functions, one wave function for each possible velocity at $x$ and $t$.\footnote{In the quantum language, this means working with a density matrix, that is, with mixed states rather than with pure states.} On the contrary, if we only work with a single-valued (classical or quantum) wave function, then, the velocity itself is single-valued everywhere. Therefore, two (classical or quantum) trajectories that coincide in one configuration point will have identical velocities. This means that they will follow identical trajectories for any future time. This has the important consequence that all Bohmian trajectories (or classical trajectories) associated with a single-valued wave function cannot cross in the configuration space.
\end{enumerate}

\end{enumerate}

In conclusion, the differences between quantum and classical
ensembles of trajectories is not a difference between waves and
particles, because both waves and particles can be used to study
classical or quantum systems. On the contrary, the difference
resides between a linear wave equation (for quantum mechanics),
\eref{om.Schrodinger1D}, and a nonlinear wave equation (for
classical mechanics), \eref{om.classcho}. One of the most important
consequences of such difference is that quantum (Bohmian)
trajectories depend on the shape of the ensemble (i.e., quantum
wholeness), as seen in \fref{om_fig_dobleslit}a, while classical
trajectories are independent of the shape of the ensemble, as seen
in \fref{om_fig_dobleslit}b. This difference between classical and
quantum systems has important consequences at a computational level.
One can compute a unique classical trajectory. However, because of the quantum wholeness, somehow, one needs, in principle, to deal somehow with the whole ensemble of quantum
trajectories to know the dynamics of a unique quantum trajectory.

\subsection{Feynman paths}\label{om.sec_single.5}

\begin{figure}
\centering
\includegraphics{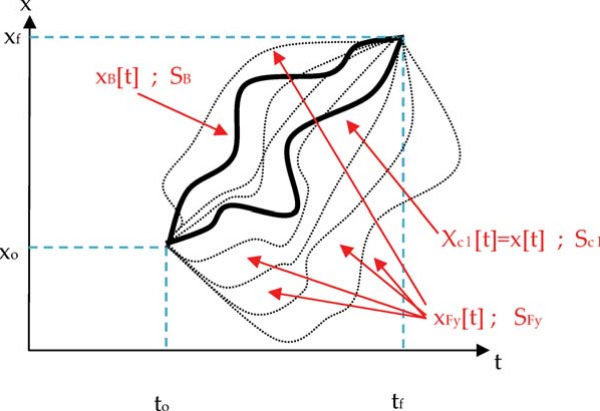}
\caption{Schematic representation of physical (solid lines) and nonphysical (dashed
lines) trajectories in the configuration space $(x,t)$ starting at
$(x_0,t_0)$ and finishing at $(x\!_f,t\!_f)$. In general, Feynman paths
$x_\textit{Fy}[t]$ do not provide stationary values of the action function,
while the quantum (Bohmian) trajectory $x\!_{B}[t]$ provides a
stationary value of the quantum action function, $S\!_{B}$ in
\eref{om.action_function_quan}, and the classical trajectory
$x_{cl}[t]$ a stationary value of the classical action function,
$S_{cl}$ in \eref{om.action_function_clas}.} \label{om_fig_Feynman}
\end{figure}

\looseness-1In the introduction of this chapter we mentioned that quantum mechanics can be described either in the matrix formulation proposed by Heisenberg and coworkers or in the wave equation formalism developed by Schr\"odinger. There are alternative representations of quantum phenomena. For example, the Feynman path  (see, for instance, Ref. \cite{om.feynmann1965}). In this context, the time evolution of a wave function can be written using the Green function (or propagator or transition amplitude) as:
\begin{equation}
\label{om.Feynman1}
\psi(x,t) = \int_{-\infty}^{\infty} G(x_0,t_0;x,t) \psi(x_0,t_0) dx_0
\end{equation}
Feynman provided an original technique for computing 
$G(x_0,t_0;x,t)$ from the classical Lagrangian of
\eref{om.lagrangian}. One considers all (physical and nonphysical)
paths $x_{Fy}[t]$ that may link the two points $(x,t)$ and
$(x_0,t_0)$. See dashed curves in \fref{om_fig_Feynman}. Feynman
associated to each path (physical or nonphysical) a complex
amplitude $\exp(iS(x[t];x_0,t_0;x,t)/\hbar)$, where
$S(x[t];x_0,t_0;x,t)$ is defined by \eref{om.action_function1D}. We
emphasize that we are using not only the trajectory that provides an
stationary value of $S(x[t];x_0,t_0;x,t)$ but all trajectories.
Then, Feynman defines the Green function as:
\begin{equation}
\label{om.Feynman2}
G(x_0,t_0;x,t) = C\sum_{\rm all\;paths} e^{{i S(x[t];x_0,t_0;x,t)}/{\hbar}}
\end{equation}

\noindent where $C$ is a normalization constant. It can be
demonstrated that the wave function constructed from
\eref{om.Feynman1} with \eref{om.Feynman2} reproduces the
Schr\"odinger equation \cite{om.feynmann1965}. Certainly, Feynamn
paths and Bohmian trajectories are completely different. In
particular, there is only one Bohmian trajectory $x\!_{B}[t]$ solution
of \eref{om.hamilton_jacobi1D_des} that goes from point $(x_0,t_0)$
to point $(x\!_f,t\!_f)$, as seen in \fref{om_fig_Feynman}, while there
are infinite Feynman paths $x_\textit{Fy}[t]$ that connect both points. In
particular, one of the Feynman paths is certainly the Bohmian trajectory  $x\!_{B}[t]$ and
another is the classical trajectory $x_{cl}[t]$. In general,
$x\!_{B}[t]$ is different from $x_{cl}[t]$. The later is the
trajectory that minimizes the classical action function:
\begin{equation}
\label{om.action_function_clas}
 S(x_{cl}[t];x_0,t_0;x_f,t_f) = \int_{t_0}^{t_f} \left[ \frac {1} {2m} (\dot{x}[t])^2 -V(x[t],t) \right] dt
\end{equation}
while the former is the trajectory that minimizes the quantum action function:
\begin{equation}
\label{om.action_function_quan}
 S(x_{B}[t];x_0,t_0;x_f,t_f) =\! \int_{t_0}^{t_f} \left[ \frac {1} {2m} (\dot{x}[t])^2 -V(x[t],t) -Q(x[t],t) \right] dt
\end{equation}
The derivation of the last expression can be straightforwardly obtained by repeating the steps done from the classical action function to the classical Hamilton--Jacobi equation but using now the quantum Hamilton--Jacobi equation. We conclude that Bohmian trajectories are quite different from Feynman paths. The Bohmian trajectories are part of the basic ontology of the theory, while the Feynman paths are basically a mathematical tool to compute the probabilities.

On the other hand, one could be interested in discussing whether the Feynman approach could be a more useful computational tool than Bohmian mechanics for solving quantum systems. In principle, it seems that Feynman paths would be less practical because we would have to compute an infinite number of trajectories from one particular initial position and then repeat the procedure for all initial positions. On the contrary, in Bohmian mechanics we only need to compute one quantum trajectory for each initial position. See problems \ref{om.P9}, \ref{om.P10} and \ref{om.P11}.

\subsection{Basic postulates  for a single-particle}\label{om.sec_single.6}

After introducing the reader to the language of trajectories for
quantum mechanics (and wave functions for classical mechanics), we
can state the basic postulates of the Bohmian theory  for a single particle. hey summarize in a few sentences all the discussions held so far. In general, the
postulates of any physical theory can be presented in different
compatible ways. For example, classical mechanics can be postulated
from Newton's laws or Hamilton's principle. We follow here the
standard presentation of Bohmian mechanics that involves a smaller
number of ingredients
\cite{om.Holand1993,om.Valentini2006,om.Bohmian1996,om.Durrllibre,om.reviewabc,om.llibreph}.\\

\noindent\textbf{FIRST POSTULATE}: \textit{The dynamics of a single particle in a single experiment in a quantum system is defined by a trajectory $x[t]$ that moves continuously under the guidance of a wave function $\psi(x,t)$.
The wave function $\psi(x,t)$ is a solution of the Schr\"odinger equation:}
\begin{equation*}
i \hbar \frac{\partial \psi(x,t)} {\partial t} = -\frac {\hbar^2} {2m} \frac{ {\partial}^2 \psi(x,t)} {\partial x^2} + V(x,t) \;\psi(x,t)
\end{equation*}
\textit{The trajectory $x[t]$ is obtained by time-integrating the particle velocity $v[t]$ defined as:}
\begin{equation*}
v(x,t) = \frac{J(x,t)} {|\psi(x,t)|^2}
\end{equation*}
\textit{where $J(x,t)$ is the (ensemble) current density given by:}
\begin{equation*}
J(x,t) = i \frac {\hbar} {2 m} \left(\psi(x,t) \frac {\partial \psi^{*}(x,t)} {\partial x}- \psi^{*}(x,t) \frac {\partial \psi(x,t)} {\partial x} \right)
\end{equation*}
\textit{The initial position $x[t_0]$ and velocity $v[t_0]$ must be specified to completely determine the trajectory $x[t]$.} \\

\noindent\textbf{SECOND POSTULATE} (quantum equilibrium hypothesis): \textit{The initial position and velocity of a particular trajectory cannot be known with certainty. When the experiment is repeated many times}  (\textit{\,j $=$ 1$,\ldots,$ M}), \textit{the initial positions $\{ x^j[t_0] \}$ of an ensemble of trajectories $\{x^j[t] \}$, associated to the same $\psi(x,t)$, have to satisfy $R^2(x,t_0) = |\psi(x,t_0)|^2$, that is the number of trajectories of the ensemble between $x$ and $x + dx$ at the initial time $t_0$ is proportional to $R^2(x,t_0) = |\psi(x,t_0)|^2$. The initial velocity of each trajectory is determined by $v^j[t_0] = J(x^j[t_0],t_0)/|\psi(x^j[t_0],t_0)|^2$.}\\

The condition on the distribution of the initial position in different experiments can be written mathematically as:
\begin{equation}
\label{om.sum_0f_particles_to}
R^2(x,t_0) = \lim_{M\rightarrow\infty} \frac {1} {M} \sum_{j = 1}^{M} \delta(x-x^j[t_0])
\end{equation}
where $j = 1,\ldots,M$ is the index of the different trajectories belonging to different experiments. If the set of $M$ positions of different experiments follows the distribution in \eref{om.sum_0f_particles_to} at the initial time $t=0$, it is easy to demonstrate that the distribution of positions satisfies $R(x,t)^2 = |\psi(x,t)|^2$ at any time other time $t$:
\begin{equation}
\label{om.equivariance}
R^2(x,t) = \lim_{M\rightarrow\infty} \frac {1} {M} \sum_{j = 1}^{M} \delta(x-x^j[t])
\end{equation}
provided that the many-particle wave function evolves according to the Schr\"odinger equation and that the particles move according to the corresponding Bohmian velocity, see problem \ref{om.ppre6}. This property is known as equivariance~\cite{om.extra9,om.llibreph} and it is key for the empirical equivalence between Bohmian mechanics and other quantum theories.
\eref{om.equivariance} says that Born's law is always satisfied by counting particles~\cite{om.Holand1993,om.Durrllibre,om.reviewabc,om.llibreph}. 
\eref{om.equivariance} also explains 
why quantum results are unpredictable at the empirical arena \cite{om.randomness}. 

The second postulate needs some additional remarks. It is argued by some authors that the second postulate, the quantum
equilibrium hypothesis, is not a necessary postulate of the Bohmian theory.  We will briefly explain the reasons in \sref{om.sec_many.5} when dealing with the postulates of a many particle system.

These postulates represent a minimalist explanation of the Bohmian
interpretation of quantum mechanics without mentioning either the Hamilton--Jacobi
equation, \eref{om.hamilton_jacobi1D_des} or the quantum potential,
\eref{om.quantum_potential1D}. Certainly, we can formulate Bohmian
mechanics without the use of the Hamilton--Jacobi formalism;
however, in the authors' opinion, the quantum Hamilton--Jacobi and
the quantum potential allows us to improve our understanding of
Bohmian mechanics and provides clear arguments for discussing the
similarities and differences between classical and quantum theories. Those who dislike the use of the quantum potential (and the quantum Hamilton--Jacobi equation) argue that it \emph{naturally} appears in the Schr\"odinger equation, but it is not present in other wave equations such as Dirac's one. On the contrary, they state that the concept of Bohmian velocity, as defined in the first postulate, is much more general because it is always present in any wave equation, as far as a local continuity equation for the probability density can be established. This is true. However, in spite of this important limitation, we will show several times in this book that the quantum potential is a quite useful tool. 

Finally, let us emphasize that no postulate about measurement is
 formulated here because, in Bohmian mechanics, no postulate about measurement is
needed. For a more detailed explanation, please refer to Refs.
\cite{om.Holand1993,om.Valentini2006,om.Bohmian1996,om.Durrllibre,om.reviewabc,om.llibreph} and
\sref{om.sec_measurement}. In next section, we will explain how
these postulates for single-particle systems are
generalized to many-particle systems.

\section{Bohmian Mechanics for Many-Particle Systems}\label{om.sec_many}

\subsection{Preliminary discussions: The many body problem}\label{om.sec_many.1}

Up to now we have only studied a simple system composed of just one
particle. However, a single-particle quantum system is some kind of
crude idealization of any macroscopic or microscopic system found in nature, which is usually formed by a very large number of interacting particles. Strictly speaking, a single-particle system cannot be measured  (one needs to consider the particles of the measuring apparatus that interact with the single-particle system). In this section we will study nonrelativistic
many-particle quantum systems with Bohmian trajectories. First of
all, let us clarify the meaning of the term ``many-particle'' in
Bohmian mechanics, as it can easily be misinterpreted. As we have
explained in the previous section, we need an infinite ensemble of
trajectories, $M\rightarrow\infty$, to describe the
statistics of a single-particle quantum system. These different
trajectories should not be confused with different physical particles of the system, since they all refer to different realizations (experiments) of the same
single-particle quantum system.

Let us consider  now a quantum system with $N$ degrees of freedom,
that is, an $N$ body quantum system. We will use a particular
variable $x_k$ for each degree of freedom $k = 1,\ldots,N$. The wave
function will thus be a function of all $x_{1},\ldots,x_{N}$
variables. Now, a many-particle Bohmian trajectory will involve $N$
interacting physical particles $x_{1}[t],\ldots,x_{N}[t]$. The relevant point
that allows us to use the adjective ``many-particle'' is that the
$N$ particles interact between them, that is, the interacting
potential depends on all possible particle positions,
$V(x_{1},\ldots,x_{N})$, in a non-trivial way. Along this section, in order to simplify our
notation, we will use $\vec{x}[t] = (x_{1}[t],\ldots,x_{N}[t])$ or
$\vec{x} = (x_{1},\ldots,x_{N})$.

Now, an ensemble of identical experiments  will be composed of $M\rightarrow\infty$
Bohmian trajectories $\vec{x}^j[t]$, with $j = 1,\ldots,M$. The superscript
$j$ refers to the statistical index (counting experiments), while
the subscript index will refer to each of the $N$ interacting particles present in each experiment,
that is, $\vec{x}^j[t] = (x^j_{1}[t],\ldots,x^j_{N}[t] )$. In the rest of chapter, we will use $M$ to refer to the number of experiments and $N$ to the number of physical particles of the quantum system in each experiment. 

The first step to obtain the many-particle quantum trajectories is solving the following many-particle Schr\"{o}dinger equation:
\begin{equation}
\label{om.schordingerND}
i \hbar \frac{\partial \psi(\vec{x},t)}{\partial t} = \left(\sum_{k = 1}^N -\frac{\hbar^2}{2m}\frac {\partial^2} {\partial x_k^2} + V(\vec{x},t) \right) \psi(\vec{x},t)
\end{equation}
The solution $\psi(\vec{x},t)$ of this equation is the so-called
\textit{many-particle wave function}, that is defined in a
$N$-dimensional space (plus time). The problem of $N$ particles in a
1D space is formally equivalent to the one of a single particle in
an $N$-dimensional space. Here, we will use the term ``many-particle''
in a wide sense to include the single-particle $N\geq2$
dimensional case.\footnote{The variable $N$ can be defined as the
number of particles in a 1D space, or it can be related to the
number of particles in a 3D space. In
simple words, $\psi(x_1,x_2,x_3,t)$ can be interpreted as three
particles in a 1D space or just one particle in a 3D space. From a
physical point of view, one particle in a 3D space is a
``single-particle'' system. However, from the computational point of
view, it is equivalent to a three-particle system in 1D.}

\Eref{om.schordingerND} is analitically unsolvable most of the times, and its numerical integration is out of today's present computer capabilities, even for systems with $N \gtrsim 5$, since we need to compute the wave function $\psi(\vec{x},t)$ in the $N$-dimensional configuration space.
Let us roughly estimate the hard disk space that we would require to store $\psi(\vec{x},t)$.
Considering, for instance, a system with $N = 10$ particles confined in a 1D region of 10 nm, which we discretize with a spatial step of $\Delta x = 0.1$ nm, we have a grid of 100 points for each dimension $x_k$.
Then, the total number of points in the configuration space for the 10 particles is $100^{10} = 10^{20}$.
Using 4 bytes (32 bits) to store the complex value of the wave function at each grid point, the information contained in a 10-particle wave function would require more than $3 \times 10^8$ Terabytes (TB) (and more than $3\times 10^{28}$ TB for $20$ particles).
This practical limitation is the main reason why our knowledge of many-particle quantum systems is so poor. This difficulty is the so-called many body problem. 
In 1929, Dirac wrote the following \cite{om.dirac1929bis}:\\\vspace*{-6pt}

\begin{quote}
The general theory of quantum mechanics is now almost complete. The underlying physical laws necessary for the mathematical theory of a large part of physics and the entire chemistry are thus completely known, and the difficulty is only that the exact application of these laws leads to equations much too complicated to be soluble.\vspace*{-6pt}\\
\end{quote}

Three decades later, Born rephrased the issue \cite{om.born1960bis}:\\\vspace*{-6pt}

\begin{quote}
It would indeed be remarkable if Nature fortified herself against further advances in knowledge behind the analytical difficulties of the many body problem.\vspace*{-6pt}\\
\end{quote}

The adjective ``many-particle'' can certainly be used with classical
mechanics for $N$ particles interacting through $V(\vec{x},t)$. Now,
it would be interesting to address the previous problem of storing
the state of a 10-particle system but now from the classical
mechanics perspective. Solving the problem means finding the
many-particle trajectory $\vec x^{\rm cl}[t]$. As discussed at the
end of subsection 1.2.4, a classical many-particle trajectory can be
computed alone, without knowing the rest of the trajectories of the
ensemble (with different initial positions). At a computational
level, this feature of classical mechanics implies a dramatic
simplification. Basically, at each time step of the simulation, if
the analytical expression for the potential is known, we just need
to save the position and velocity of each particle. This means 20
real values for the 10-particle systems mentioned above. The
classical trajectory of each particle, $x^{\rm cl}_k[t]$, can be
obtained by solving a coupled system of Newton's  second laws:
\begin{equation}\label{om.NewtonMP}
m_k\frac{d^2 x^{\rm cl}_k[t]} {dt^2} = \left [-\frac{\partial V(\vec{x},t)}{\partial x_k} \right ]_{\vec x = \vec x^{\rm cl}[t]}
\end{equation}
Fortunately, to compute a single classical trajectory $\vec x^{\rm
cl}[t]$, we only need to evaluate the derivative of the
many-particle potential $V(\vec{x},t)$ along the trajectory. In the
quantum case, even finding a single trajectory means solving for the
entire wave function $\psi(\vec{x},t)$ that \emph{encapsulates} the information
of the entire ensemble of trajectories of all other experiments (the quantum wholeness). 

In the scientific literature, there are many attempts to provide
reasonable approximations to the many-body quantum problem. Density
functional theory \cite{om.kohn1964,om.kohn1965} and the
Hartree--Fock approximation \cite{om.Hartree,om.Hartree2,om.Fock} are
some of the most popular techniques among the scientific community
for dealing with the many body problem. In \sref{om.sec_many.1} we
will discuss whether Bohmian trajectories can help in solving the
many body problem. In what follows, we present the basic theory of
the many-particle Bohmian trajectories.

\subsection{Many-particle quantum trajectories}\label{om.sec_many.2}

\looseness-1The efforts done in the single-particle section to compare classical
and quantum mechanics are valid here. In particular, we could write
a classical (many-particle) wave equation or develop (many-particle)
trajectories for quantum systems. In this section, we will only
explicitly develop the latter. We start by considering
(nonrelativistic) spinless particles. In \sref{om.sec_many.4} we
will introduce spin.

\subsubsection{Many-particle continuity equation}

Following a similar development as we did for the single-particle
case, see \sref{om.sec_single.3}, we can derive the following
\textit{local} continuity equation associated to the many-particle
Schr\"odinger equation, \eref{om.schordingerND}:
\begin{equation}
\label{om.continuityND}
 \frac{\partial |\psi(\vec{x},t)|^2} {\partial t} + \sum_{k = 1}^{N} \frac { \partial}  {\partial x_k}  J_k(\vec{x},t) = 0  
\end{equation}
where we have defined
\begin{equation}
\label{om.currentND}
J_k(\vec{x},t) = i \frac {\hbar} {2 m} \left(\psi(\vec{x},t) \frac {\partial \psi^{*}(\vec{x},t)} {\partial x_k}- \psi^{*}(\vec{x},t) \frac {\partial \psi(\vec{x},t)} {\partial x_k} \right)
\end{equation}
as the $k$-th component of the current density, see problem \ref{om.ppre6}.

Since the many-particle Schr\"odinger equation is also compatible with a local conservation of particles, we can interpret $|\psi(\vec{x},t)|^2$ as the spatial distribution of an ensemble of many-particle trajectories assigned to an ensemble of different experiments. The Bohmian velocity of the $k$-th trajectory is:
\begin{equation}
\label{om.velocityND}
v_k(\vec{x},t) = \frac{J_k(\vec{x},t)} {|\psi(\vec{x},t)|^2}
\end{equation}
In fact, the strategy followed here to develop Bohmian mechanics can be extended to any quantum equation of motion where a continuity equations holds: first, look for a continuity equation for the probability density and, then, define a velocity for the Bohmian trajectories as the current density divided by the probability density.
A particular example will be developed for particles with spin in \sref{om.sec_many.4} and chapter 8.

\subsubsection{Many-particle quantum Hamilton--Jacobi equation}

Alternatively, we can obtain Bohmian mechanics from a quantum Hamilton--Jacobi equation. We start by introducing the polar form of the many-particle wave function $\psi(\vec{x},t) = R(\vec{x},t) e^{i S(\vec{x},t)/\hbar}$ into the many-particle Schr\"odinger equation, \eref{om.schordingerND}. Then, after a quite simple manipulation, one obtains from the imaginary part:
\begin{equation}
\label{om.charge_conservationND}
\frac{\partial R^{2}(\vec{x},t)}{\partial t} + \sum_{k = 1}^{N} \frac {\partial } {\partial x_k} \left(\frac {1} {m} \frac {\partial S(\vec{x},t)}{\partial x_k} R^2(\vec{x},t) \right) = 0
\end{equation}
where we recognize the velocity of the $k$-th particle as:
\begin{equation}
\label{om.velocityND_bis}
v_k(\vec{x},t) = \frac {1} {m} \frac {\partial S(\vec{x},t)}{\partial x_k}
\end{equation}
Equations (\ref{om.velocityND}) and (\ref{om.velocityND_bis}) are
identical. This is shown in problem \ref{om.p5} for a 1D system, but
it can straightforwardly be generalized to $N$ dimensions. The real
part of the Schr\"odinger equation leads to a many-particle version
of the quantum Hamilton--Jacobi equation:
\begin{equation}
\label{om.Hamilton_JacobiND}
\frac{\partial S(\vec{x},t)}{\partial t} + \sum_{k = 1}^{N} \frac {1} {2 m} \left(\frac{{\partial} S(\vec{x},t)} {\partial x_k} \right)^2 + V(\vec{x},t) + Q(\vec{x},t) = 0 
\end{equation}
where we have defined the quantum potential as:
\begin{equation}
\label{om.quantum_potentialND_total}
Q(\vec{x},t) = \sum_{k = 1}^{N} Q_k(\vec{x},t)
\end{equation}
with:
\begin{equation}
\label{om.quantum_potentialND_parcial}
Q_k(\vec{x},t) = -\frac{\hbar^2} {2 m} \frac{{\partial}^2 R(\vec{x},t)/ \partial x^2_k} {R(\vec{x},t)}
\end{equation}
Again, we can obtain a system of coupled Newton-like equations (one for each component of $\vec{x}$) for the many-particle Bohmian trajectories by computing the time-derivative of the Bohmian velocity of \eref{om.velocityND}:
\begin{equation}
\label{om.NewtonlikeMP}
m\frac{d^2x_k[t]} {dt^2} = \left [-\frac{\partial} {\partial x_k} \left( V(\vec{x},t) + Q(\vec{x},t)\right)  \right ] _{\vec x = \vec x[t]}
\end{equation}
Both the potential $V(\vec{x},t)$ and the quantum potential
$Q(\vec{x},t)$ introduce correlations between particles. All
physical interactions are essentially expressed as correlations (relations)
between the degrees of freedom. In the following, we will discuss
some important differences between classical and quantum
correlations.

\subsection{Factorizability, entanglement, and correlations}\label{om.sec_many.3}

There are important differences between the correlations introduced by the classical potential $V(\vec{x},t)$ and the quantum potential $Q(\vec{x},t)$:
\begin{enumerate}
\item \looseness-1In general, the term \textit{$V(\vec{x},t)$} decreases with the distance between particles.
A simple example is the Coulomb interaction.
However, $Q(\vec{x},t)$ depends only on the shape of the wave function, not on its value (see \eref{om.quantum_potentialND_parcial}).
Thus, the quantum potential that we would obtain from \eref{om.quantum_potentialND_parcial} using either $R(\vec{x},t)$ or $a R(\vec{x},t)$ would be exactly the same, even when $a\rightarrow 0$.
Thus, the quantum potential can produce a significative interaction between two particles, even if they are very far apart.\footnote{These highly nonclassical features of the quantum potential is what led Bohm to argue that the quantum potential interchanges information between systems \cite{om.infinite_potential,om.Bohm1993}.}

\item In general, the term \textit{$V(\vec{x},t)$} produces classical correlations between different particles. The particular dependence of $V(\vec{x},t)$ in all variables $x_1,\ldots,x_N$ imposes a restriction on the speed of the interaction. The variations of $x_i$ can only affect $x_j$ after a time larger than $|x_i-x_j|/c$, $c$ being the speed of light.
For example, the relation between the particles positions in the
electromagnetic interactions ensures that there is no superluminal
influence between particles. However, such a restriction is not
present in the quantum potential. Thus, very far particles have an
\textit{instantaneous} (nonlocal) interaction between them.
\end{enumerate}

The quantum potential is at the origin of all quantum correlations,
that is, entanglement, that can imply (nonlocal) faster-than-light
interactions when two distant particles are involved. As mentioned
in \sref{om.sec_intro.6}, this ``spooky action at a distance'' is
what bothered Einstein about Bohmian mechanics (and quantum
mechanics, in general). In 1964, Bell elaborated his famous theorem
that established clear experimentally testable mathematical
inequalities that would be fulfilled by \textit{local} theories but
would be violated by \textit{nonlocal} ones
\cite{om.Bell1964}. All experimental results obtained so far confirm
that Bell's inequalities are violated. Therefore, contrarily to
Einstein's belief, we have to accept the real existence, in nature,
of faster-than-light causation.\footnote{We insist that the
experimental violation of Bell's inequalities gives direct support
not only to the Copenhagen interpretation but also to the Bohmian
one, since the latter is also a nonlocal theory because of the presence of
$Q(\vec{x},t)$.} Entanglement is an intrinsic correlation in
quantum mechanics (whose complexity and potentialities eventually
come from the fact that a $N$-particle wave function lives in a
$N$-dimensional configuration space) and is at the core of quantum
information science, which makes teleportation, quantum
communication, quantum cryptography, and quantum computing possible.

To improve our understanding of correlations, let us discuss under
which conditions we cannot expect correlations between $N$ particles. Let us focus our attention on a
wave function $\psi(\vec{x},t)$ that can be written as a product of
single-particle wave functions associated with each of the
particles:
\begin{equation}
\label{om.many_function_factorizable}
\psi(\vec{x},t) = \prod_{k = 1}^N \psi_k(x_k,t)
\end{equation}
We call such a wave function ``factorizable" or ``separable". \Eref{om.many_function_factorizable} expresses the physical independence of the $N$ particles (even though the wave functions $\psi_k(x_k,t)$ may overlap). According to Born's statistical interpretation of wave functions, the squared modulus of a wave function is the probability density of the quantum particle; thus the quantum wave function, \eref{om.many_function_factorizable}, corresponds to a system without (classical or quantum) correlations between particles. It is well known that such a solution occurs when the potential in \eref{om.schordingerND} can be written as $V(\vec{x}) = \sum_k V_k(x_k)$. Next, let us reformulate this result with Bohmian trajectories.

We realize that the phase and modulus of \eref{om.many_function_factorizable} are given by:
\begin{equation}
\label{om.many_S_factorizable}
S(\vec{x},t) = \sum_{k = 1}^N S_k(x_k,t)
\end{equation}
and
\begin{equation}
\label{om.many_R_factorizable}
R(\vec{x},t) = \prod_{k = 1}^N R_k(x_k,t)
\end{equation}
by defining
\begin{equation}
\psi_k(x_k,t) = R_k(x_k,t) e^{i S_k(x_k,t)/\hbar}
\end{equation}

In this case, the many-particle quantum potential can be written as:
\begin{equation}
\label{om.many_Q_factorizable}
Q_k(\vec{x},t) = -\frac{\hbar^2} {2 m} \frac{{\partial}^2 R_k(x_k,t)/ \partial x^2_k} {R_k(x_k,t)}
\end{equation}
Then, we can easily deduce from \eref{om.Hamilton_JacobiND} that each $\psi_k(x_k,t)$ is a solution of the following single-particle quantum Hamilton--Jacobi equation:
\begin{equation}
\label{om.Hamilton_JacobiND_factorizable} \frac{\partial
S_k(x_k,t)}{\partial t} + \frac {1} {2 m} \left(\frac{{\partial}
S_k(x_k,t)} {\partial x_k}\right)^2+ V(x_k,t) + Q_k(x_k,t) = 0
\end{equation}
In addition, each $\psi_k(x_k,t)$ satisfies a conservation law:
\begin{equation}
\label{om.charge_conservationND_factorizable}
\frac{\partial R_k^{2}(x_k,t)}{\partial t} + \frac {\partial } {\partial x_k} \left(\frac {1} {m} \frac {\partial S_k(x_k,t)}{\partial x_k} R_k^2(x_k,t) \right) = 0
\end{equation}
From these equations we can easily deduce an independent guiding equation for each $k$-particle:
\begin{equation}
\label{om.many_guide_equation_factorizable}
m \frac {d x_k[t]} {dt} = \left(\frac {\partial S_k(x_k,t)} {\partial x_k} \right)_{x_k = x_k[t]}
\end{equation}
showing the absence of correlations between (the different components of the many-particle) Bohmian trajectory.

\subsection{Spin and identical particles}\label{om.sec_many.4}

Elementary particles, such as electrons or quarks, have spin, an internal (discrete) degree of freedom that can influence their quantum dynamics in a nontrivial manner.
In this section we will briefly explain how to extend (nonrelativistic) Bohmian mechanics to include spin.

\subsubsection{Single-particle with \textit{s} $=$ 1/2}

In the orthodox formulation of quantum mechanics, the state of a single particle with spin $s$ is described by a $2s + 1$ component vector of wave functions with:
\begin{equation}
\vec \Psi(\vec r,t) = \left(
\begin{array}{c}
\Psi_{1}(\vec r,t) \\ \vdots \\ \Psi_{2s + 1}(\vec r,t)
\end{array} \right)
\end{equation}
where $\vec r = (x_1,x_2,x_3)$ represents the 3D position of the particle.
The time evolution of this state is no longer governed by the Schr\"odinger equation but by more involved wave equations such as the Pauli equation for $s = {1}/{2}$ \cite{om.landaulif,om.ward,om.colijn2002}.

The strategy to find Bohmian trajectories for spin particles will be the following: we will look for a continuity equation of the probability density and define the Bohmian velocity as the current density divided by the probability density.
The idea is quite simple, but the mathematical development can be much more complicated. Therefore, we will focus on a particular example for a spin $s = 1/2$ charged particle, whose vectorial wave function,
\begin{equation}
\vec \Psi(\vec r,t) = \left(
\begin{array}{c}
\Psi_{\uparrow}(\vec r,t) \\ \Psi_{\downarrow}(\vec r,t)
\end{array}
\right)
\end{equation}
has two components and it is called a \textit{spinor}. In this particular case, the spin along a particular direction takes only two possible values, referred to as spin-up ($\uparrow$) and spin-down ($\downarrow$) states.
The time evolution is given by the Pauli equation \cite{om.ward}:
\begin{equation}
\label{om.pauli}
i \hbar \frac{ \partial}{\partial t} \vec \Psi(\vec r,t) =
\left[ \frac{1}{2m} \vec{\sigma} \cdot \left(- i \hbar \vec{\nabla} - q \vec{A} (\vec r,t) \right) ^2 + V(\vec r,t) \right]
\vec \Psi(\vec r,t)
\end{equation}
where $\vec{\sigma} = (\sigma_1, \sigma_2, \sigma_3)$ is a vector containing the Pauli matrices \cite{om.landaulif}:
\begin{eqnarray}
\sigma_1 & = & \begin{pmatrix} 0&1 \\ 1&0 \end{pmatrix}  \\
\sigma_2 & = & \begin{pmatrix} 0&-i \\ i&0 \end{pmatrix} \\
\sigma_3 & = & \begin{pmatrix} 1&0 \\ 0&-1 \end{pmatrix}
\end{eqnarray}
and $V(\vec r,t)/q$ and $\vec{A}(\vec r,t) = (A_1(\vec r,t),A_2(\vec r,t),A_3(\vec r,t))$ are, respectively, the electromagnetic scalar and vector potential.

From the Pauli equation we obtain a continuity equation:
\begin{equation}
\label{om.difcurrent_density_pauli}
\frac{\partial\rho(\vec{r},t)}{\partial t} + \vec{\nabla} \vec{J}(\vec{r},t) = 0
\end{equation}
with the probability and current densities defined as:
\begin{eqnarray}
\label{om.rhospin}
\rho(\vec r,t) & = & \vec \Psi^{\dagger}(\vec{r},t)\cdot\vec \Psi(\vec{r},t) \\
\label{om.currentspin}
\vec J(\vec{r},t) & = & i \frac {\hbar} {2 m}
\left(\vec\Psi(\vec{r},t)\cdot {\vec{\nabla} \vec
\Psi^{\dagger}(\vec{r},t)} - \vec\Psi^{\dagger}(\vec{r},t) \cdot
{\vec{\nabla} \vec \Psi(\vec{x},t)} \right)-\frac {\vec A(\vec r,t)} {mc} \rho(\vec r,t),
\end{eqnarray}
where $\vec\Psi^{\dagger}$ denotes the conjugate transpose of $\vec\Psi$.

Defining the Bohmian velocity as:
\begin{equation}
\vec v(\vec r,t) = \frac{ \vec{J}(\vec{r},t)}{\rho(\vec r,t)}
\end{equation}
one can develop Bohmian trajectories for spin particles. Notice that the spin is basically a property defined through the wave function of the particle, not through its position. In any case, since the Bohmian velocity of the particle is affected by the wave function, the spin can have a direct effect on the trajectory. \footnote{Alternatively, one could also consider the degree of freedom of spin as an additional three-angle variable $\{\alpha,\beta,\gamma\}$ of the wave function (such that $R(\vec r,\alpha,\beta,\gamma,t)$ and $S(\vec r,\alpha,\beta,\gamma,t)$) and look for the equations of motion of the trajectories of the positions and also of the trajectories of the angles (see chapter 10 in Ref. \cite{om.Holand1993}).} It is out of the scope of this book to further develop Bohmian mechanics for the Pauli equation.
In any case, practical examples of the particle trajectories for spin-1/2 particles can be found in Refs. \cite{om.Holland2003,om.colijn2002}.

The Pauli equation can be rewritten as:
\begin{eqnarray}
\label{om.pauli_SG}
i \hbar \frac{ \partial}{\partial t} \vec \Psi(\vec r,t) &=&
\left[ \frac{1}{2m} \left(- i \hbar \vec{\nabla} - q \vec{A}(\vec r,t)\right) ^2 -\frac{\hbar q}{2m} \vec{\sigma}\cdot\vec{B}(\vec r,t)\right.+ V(\vec r,t) \bigg]\vec \Psi(\vec r,t)
\end{eqnarray}
$\vec{B}(\vec r,t) = \vec{\nabla} \times \vec{A}(\vec r,t)$ being
the magnetic field. Note that the only term that can transfer
population between spin states is the one with $\vec{B}$ (the
so-called Stern--Gerlach term). If this term can be neglected then
each spin component of \eref{om.pauli_SG} reduces to the familiar
Schr\"odinger equation. 

On the other hand we can also consider the case where the Hamiltonian is separable into a part depending only on the particle position and a part depending only on the spin. Then if the initial state is:
\begin{equation}
\label{om.spin1Dinitialstate}
\vec \Psi(\vec r,0) = \psi(\vec r,0)\left(
\begin{array}{c}
\alpha_{\uparrow}(0) \\ \alpha_{\downarrow}(0)
\end{array}
\right)
\end{equation}
at later times one can write the spinor evolution in a simpler form as:
\begin{equation}
\label{om.spin1D}
\vec \Psi(\vec r,t) = \psi(\vec r,t)\left(
\begin{array}{c}
\alpha_{\uparrow}(t) \\ \alpha_{\downarrow}(t)
\end{array}
\right)\equiv\psi(\vec r,t) \vec \chi(t)
\end{equation}
where $\psi(\vec r,t)$ depends only on the coordinate of the particles and the function $\vec \chi(t)$ depends only on its spin. We call the former the \textit{coordinate} or \textit{orbital} wave function and the latter the \textit{spin} wave function.
In this case, the evolution of $\vec \chi(t)$ and $\psi(\vec r,t)$ is independent.
If we are not interested in the actual spin of the particles, we can just consider the dynamics of the coordinate function $\psi(\vec r,t)$, which will be determined by the Schr\"odinger equation.

\subsubsection{Many-particle system with \textit{s} $=$ 1/2 particles}

For a nonrelativistic system of many particles, where the orbital
and spin contributions of the Hamiltonian are initially separable,\footnote{For
example, in the absence of a magnetic field such that  the initial wave
function can be written as a product of an orbital part, $\psi(\vec
r_1,\vec r_2,\ldots\vec r_N,0)$, and a spin part, $\vec \chi(0)$.}
the many-particle wave function will be also separable at all times,
that is:
\begin{equation}
\label{om.spinND}
\Psi(\vec r_1,\ldots\vec r_N,t) = \psi(\vec r_1,\ldots\vec r_N,t) \vec \chi(t)
\end{equation}
where $\psi(\vec r_1,\ldots\vec r_N,t)$ is the time evolution of the orbital state, the solution of the many-particle Schr\"odinger equation, and $\vec \chi(t)$ is the time evolution of the spin part of the system state:
\begin{equation}
\vec \chi(t) = \left(
\begin{array}{c}
\alpha_{\uparrow_1 \uparrow_2\ldots\uparrow_N}(t) \\ \alpha_{\uparrow_1 \uparrow_2\ldots\downarrow_N}(t) \\ \vdots \\ \alpha_{\downarrow_1 \downarrow_2\ldots\downarrow_N}(t)
\end{array}
\right) = \left(
\begin{array}{c}
\alpha_{1}(t) \\ \alpha_{2}(t) \\ \vdots \\ \alpha_{W}(t)
\end{array}
\right)
\end{equation}
$W \equiv (2s + 1)^N$ is the number of possible combinations of spin
projections in one direction for all the particles in the
system.\footnote{Additionally, one can also
look for the projection of the total spin of the system. Both
procedures are connected by the Clebsch--Gordan coefficients
\cite{om.landaulif}.}

In many practical situations, we are only interested in the
evolution of the orbital wave function through the many-particle
Schr\"odinger equation. However, even in this case, there is a
peculiar dependence of the dynamics of the particles on the total
spin. As explained in many textbooks, there is a pure quantum
interaction between identical particles, named ``exchange
interaction'' \cite{om.sakurai94,om.landaulif}. This interaction is
not classical, and we cannot find a term in the potential energies
of the many-particle Hamiltonian of the Schr\"odinger equation that
accounts for it. Alternatively, this new interaction is introduced
in the ``shape'' of the global wave function, through the
requirement of a particular symmetry. We
say that a many-particle wave function is anti-symmetric when the interchange of
the position and spin degrees of freedom associated to two
identical fermions (e.g., electrons) results only in a change of the
sign of the global wave function, $\vec \Psi(\vec r_1,\vec
r_2,\ldots,\vec r_N,t)$. Analogously, we say that the
many-particle wave function is symmetric when it
remains unchanged after the interchange of the degrees of freedom of
two identical  bosons.

The crucial point why we cannot forget about spin when dealing with
many-particle Hamiltonians is that for separable wave
functions, for example, \eref{om.spinND}, the symmetry of the
orbital part depends on the symmetry of the spin part. For example,
since the total wave function must be antisymmetric, two electrons
can have a symmetric orbital part if the spin part is antisymmetric,
and vice versa. In general, note that the wave function is not
separable, and it makes no sense to talk about the symmetry of the
orbital and spin parts alone but only of the symmetry of the total
wave function. We see that the exchange interaction induces
correlations between particles by imposing symmetries (or shapes) to
their many-particle wave function.\enlargethispage{-1pc}

A standard claim in many quantum mechanics textbooks is that identical particles, for example, two electrons with an antisymmetric wave function, are indistinguishable. It is affirmed that if the particles would have trajectories, they would automatically be distinguishable. In Bohmian mechanics, even with the symmetrization postulate, the adjective ``indistinguishable'' is inappropriate because one can label one particle's trajectory $\vec r_1[t]$ and the other $\vec r_2[t]$ and thus distinguish them perfectly at the ontological staff.

Bohmian trajectories can actually help us improve our understanding of the symmetrization postulate. Let us assume a two-electron system with an antisymmetric orbital wave function $\psi(\vec r_1,\vec r_2,t)$. We assume that an electron labeled $1$ with the initial position $\vec {r_1}[0]$ evolves into $\vec {r_1}[t]$ and an another electron labeled $2$ evolves from $\vec {r_2}[0]$ to $\vec {r_2}[t]$. Then, it can be easily understood that $\vec {r_1}'[0] = \vec {r_2}[0]$ evolves into $\vec {r_1}'[t] = \vec {r_2}[t]$ and $\vec {r_2}'[0] = \vec {r_1}[0]$ evolves into $\vec {r_2}'[t] = \vec {r_1}[t]$. We use primes to notice that $\vec {r_1}[t]$ and $\vec {r_1}'[t]$ correspond to trajectories of the same particle with different initial positions ($\vec {r_1}[0]$ or $\vec {r_2}[0]$).
This result follows from the symmetry of the velocity, that is, the symmetry of the current density and modulus, when positions of the two electrons are interchanged. In summary, Bohmian trajectories of identical particles are clearly distinguishable in our computation, but the observable (ensemble) results obtained from them are indistinguishable when we interchange their initial positions. Hence in Bohmian mechanics a set of particles with exchange symmetry are perfectly distinguishable at the ontological plane, while they become indistinguishable at the empirical plane. 

\subsection{Basic postulates  for many-particle systems}\label{om.sec_many.5}

The basic postulates of Bohmian mechanics for many-particle systems are exactly the same as those we have developed for a single particle but adding the symmetrization postulate. As done in \sref{om.sec_single.6}, we also present a minimalist formulation of the postulates.

\enlargethispage{-1pc}
In order to simplify our description, we assume the simpler many-particle wave function written as:
\begin{equation}
\label{om.wavepostulate}
\vec \Psi(\vec r_1,\ldots,\vec r_N,t) = \psi(\vec r_1,\ldots,\vec r_N,t)\left(
\begin{array}{c}
\uparrow_1 \uparrow_2\ldots\uparrow_N(t) \\ 0 \\ \vdots \\ 0
\end{array}
\right)
\end{equation}
The generalization of the present postulates to include the more general wave functions of \eref{om.spinND} follows straightforwardly. \\

\noindent\textbf{FIRST POSTULATE}: \textit{The dynamics of a many-particle quantum  system comprises a wave function,
\eref{om.wavepostulate}, whose orbital  part $\psi(\vec r_1,\vec r_2,\ldots,\vec r_N,t)$ is defined in the configuration space 
$\{\vec r_1,\vec r_2,\ldots,\vec r_N\}$ plus time, and a many-particle trajectory $\{\vec r_1[t]$, $\vec r_2[t],\ldots,\vec r_N[t]\}$ that moves continuously under the guidance of the wave function}.

\textit{The orbital part of the wave function $\psi(\vec r_1,\vec r_2,\ldots,\vec r_N,t)$ is a solution of the many-particle Schr\"odinger equation}:
\begin{eqnarray*}
i \hbar \frac{\partial \psi(\vec r_{1},\ldots,\vec
r_{N},t)}{\partial t} &=& \left( \sum_{k = 1}^N -\frac{\hbar^2}{2m}
\nabla^2_{\vec r_k} \right.+\ \ V(\vec r_{1},\ldots,\vec r_{N},t) \Bigg)
\psi(\vec r_{1},\ldots,\vec r_{N},t)
\end{eqnarray*}

\textit{Each component $\vec r_k[t]$ of the many-particle trajectory  $\{\vec r_1[t],\ldots,\vec r_N[t]\}$ is obtained by time-integrating the particle velocity $\vec v_k[t] = \vec v_k(\vec r_1,\ldots,\vec r_N,t)$ defined from}:
\begin{equation}
\vec v_k(\vec r_1,\ldots,\vec r_N,t) = \frac{\vec J_k(\vec r_1,\ldots,\vec r_N,t)} {|\psi(\vec r_1,\ldots,\vec r_N,t)|^2}
\nonumber
\end{equation}
\textit{where $|\psi(\vec r_1,\ldots,\vec r_N,t)|^2 = \psi(\vec
r_1,\ldots,\vec r_N,t) \psi^*(\vec r_1,\ldots,\vec r_N,t)$ is the
square modulus of the wave function and $\vec J_k = \vec J_k(\vec
r_1,\ldots,\vec r_N,t)$ is the k-particle  current density}:
\begin{eqnarray*}
\vec J_k &=& i \frac {\hbar} {2m} \Big(\psi(\vec r_1,..,\vec r_N,t) \vec \nabla_{\vec{r}_k} {\psi^{*}(\vec r_1,..,\vec r_N,t)}-\psi^{*}(\vec r_1,..,\vec r_N,t) \vec \nabla_{\vec{r}_k} \psi(\vec r_1,..,\vec r_N,t) \Big)
\end{eqnarray*}
\textit{The initial positions $\{\vec r_{10},\vec r_{20},\ldots,\vec r_{N0}\}$ and
velocities $\{\vec v_{10}$, $\vec v_{20},\ldots$, $\vec v_{N0}\}$ have to be specified in order to completely determine the many-particle trajectory}.\\

\noindent\textbf{SECOND POSTULATE} (quantum equilibrium hypothesis): \textit{The initial positions $\{\vec r^j_{1}[t_0],\vec r^j_{2}[t_0],\ldots,\vec r^j_{N}[t_0] \}$
and velocities $\{\vec v^j_{1}[t_0],\vec v^j_{2}[t_0],\ldots$, $\vec r^j_{N}[t_0]\}$ of a particular many-particle $j$ trajectory cannot be known with certainty.
When the experiment is repeated many times, these initial positions of an ensemble of trajectories associated with the same $\psi(\vec r_1,\ldots,\vec r_N,t)$ satisfy that the number of trajectories of the ensemble between $(\vec r_1,\ldots,\vec r_N)$ and $(\vec r_1 + d\vec r_1,\ldots,\vec r_N + d \vec r_N)$ at the initial time $t_0$ is proportional to $R^2(\vec r^j_{1}[t_0],\ldots,\vec r^j_{N}[t_0]) = |\psi(\vec r^j_{1}[t_0],\ldots,\vec r^j_{N}[t_0],t_0)|^2$.
The initial velocity is determined by $\vec v_k^j[t_0] = \vec J_k(\vec r^j_{1}[t_0],\ldots,\vec r^j_{N}[t_0],  t_0)/|\psi(\vec r^j_{1}[t_0],\ldots,\vec r^j_{N}[t_0],t_0)|^2$ .}\\

The condition on the initial position can be written mathematically as:
\begin{equation}
\label{om.sum_0f_particlesNDo}
R^2(\vec r_1,\ldots,\vec r_N,t_0) = \lim_{M\rightarrow\infty} \frac {1} {M} \sum_{j = 1}^{M} \prod_{k = 1}^{N}\delta(\vec r-\vec r_{k}^j[t_0])  \; \; {\rm for}\; t = t_0
\end{equation}
Notice the presence of two indices, the $j = 1,\ldots,M$ for the infinite ensemble of identical experiments and the $k = 1,\ldots,N$ for the $N$ degrees of freedom.\\

\noindent\textbf{THIRD POSTULATE} (symmetrization postulate of quantum mechanics): \textit{If the variables $\vec r_i,\uparrow_i$ and $\vec r_j,\uparrow_j$ refer to two identical particles of the ensemble, then the wave function, \eref{om.wavepostulate}, is either symmetric:
\begin{equation}
\label{om.spinbos}
\psi(.,\vec r_i,.,\vec r_j,.,t)\!\!\left(
\begin{array}{c}
\uparrow_i\ldots \uparrow_j\ldots\\ 0 \\\vdots\\ 0
\end{array}
\right) = \psi(.,\vec r_j,.,\vec r_i,.,t)\left(
\begin{array}{c}
\uparrow_j\ldots \uparrow_i\ldots\\ 0 \\\vdots\\ 0
\end{array}
\right)
\end{equation}
if the particles are bosons (every particle with an integer spin 0, 1, 2, \ldots is a boson), or antisymmetric:
\begin{equation}
\label{om.spinfer} \psi(.,\vec r_i,.,\vec r_j,.,t)\!\!\left(
\begin{array}{c}
\uparrow_i\ldots \uparrow_j\ldots\\ 0 \\\vdots\\ 0
\end{array}
\right)\!\! = -\psi(.,\vec r_j,.,\vec r_i,.,t)\!\!\left(
\begin{array}{c}
\uparrow_j\ldots \uparrow_i\ldots\\ 0 \\ \vdots \\ 0
\end{array}
\right)
\end{equation}
if the particles are fermions (every particle with a half-odd spin 1/2, 3/2, \ldots is a fermion). In \eref{om.spinbos} and \eref{om.spinfer} it is understood that all other degrees of freedom of the other particles remain unchanged.\footnote{This simple spin vector wave function is clearly symmetric so that the orbital wave function has to be either symmetric or antisymmetric. For general wave functions, such as the one in \eref{om.spinND}, this postulate implies much more complicated restrictions on the possible orbital and spin wave functions.}} \\

As we  already did in the development of the postulates for a single-particle system, we remind that the second postulate needs some additional remarks since it has been argued by some authors that this postulate, the quantum equilibrium hypothesis, is not a necessary postulate of the Bohmian theory. 

First, it has been demonstrated that if one assumes that the many-particle wave function of the whole Universe, the so-called Universal wave function that includes all degrees of freedom of the Universe, satisfies some (typical) conditions then, it follows that each (conditional) wave function of each quantum (sub) system will satisfy its own \eref{om.sum_0f_particlesNDo}. Thus, the justification about the quantum equilibrium postulate has to be done only for the Universal wave function, not for each individual (sub)system.  

Probably, the most accepted view against taking the quantum equilibrium as a postulate comes from the seminal work by D\"urr, Goldstein, and Zangh\`i~\cite{om.extra9,om.llibreph}, where the equivariance in any system is discussed from the initial configurations of (Bohmian) particles in the Universe.
Using Bohmian mechanics to describe the wave function of the whole Universe, then the wave function associated to any (sub)system is an effective (conditional) wave function of the universal one. Using typicality arguments, they showed that the overwhelming majority of possible selections of initial positions of particles in the Universe will satisfy  \eref{om.sum_0f_particlesNDo} in a subsystem (or \eref{om.equivariance} for a subsystem with one degree of freedom)~\cite{om.extra9,om.llibreph}.
Other authors~\cite{om.towler2011} have attempted to dismiss \eref{om.sum_0f_particlesNDo} as a postulate by showing that any initial configuration of Bohmian particles will relax, after some time, to a distribution very close to \eref{om.equivariance} for a subsystem.
Such discussions about quantum equilibrium are far from the scope of this book. Therefore, from a practical point of view, one can postulate \eref{om.sum_0f_particlesNDo} (at some initial time) in the Bohmian theory in the same way that Born's law is a postulate in the orthodox theory.

Again, no postulate about measurement is needed, since in Bohmian
mechanics measurement is treated as a particular case of interaction between particles (see \sref{om.sec_measurement}).

\subsection{The conditional wave function: many-particle Bohmian trajectories without the many-particle wave function}\label{om.sec_many.6}

In section 1.2.4, when discussing the similarities and differences
between classical and quantum mechanics, we mentioned quantum
wholeness (the dependence of each individual Bohmian trajectory on
the rest of trajectories through the wave function) as a fundamental
difference. This concept acquires an even more dramatic meaning when
dealing with many-body systems. A particular Bohmian trajectory of
the $k$-particle $\vec r_k^j[t]$ depends on the rest of infinite $j = 1, \ldots, M$ possible
trajectories (with different initial position associated to different experiments) of all $k = 1,\ldots, N$
particles (corresponding to the $N$ degrees of freedom of the quantum system). This property is mathematically manifested by the fact
that the many-particle wave function is defined in an $N$-dimensional
configuration space rather than in the usual real 3D real space. In a 3D
real space one can easily define whether or not two particles are
far away. In general, the potential profile that determines the
interaction between these two particles decreases with the distance.
However, two distant particles that share a common region of the
$N$-dimensional configuration space where the wave function is
different from zero have an interaction (entanglement) independently
of their distance.

As discussed in section \sref{om.sec_many.1}, the need of computing the
many-particle wave function in the $N$-dimensional configuration
space is the origin of the quantum many-body problem. Among the
large list of approximations present in the literature to tackle
this problem, Bohmian mechanics provides a natural, original and
mainly unexplored solution through the use of the conditional wave
function \cite{om.norsen}. Due to the Bohmian dual description of a quantum
system as particles and waves, one can reduce the complexity of a
many-particle wave function through the substitution of  some of the
$N$ degrees of freedom by its corresponding Bohmian trajectories.
The new wave function with a reduced number of degrees of freedom is
named conditional wave function.  For a discussion about the
fundamental implications of working with conditional wave functions
instead of full many-body wave functions, see Refs \cite{om.norsen,om.norsen2}. We will
present now  some work on the computational abilities of such
conditional wave functions.

Hereafter, for simplicity, we return to a definition of the degrees
of freedom of the many-particle quantum system in terms of 
$\{x_1, x_2,\ldots ,x_N\}$ instead of $\{ \vec r_1,\vec r_2, \ldots
,\vec r_N \}$. The direct solution of  the many-particle
Schr\"odinger equation, \eref{om.schordingerND}, is intractable
numerically. On the contrary, differential classical
equations of motion deal with solutions in a much smaller configuration
space,  $\{x_1[t],\ldots,x_{i -
1}[t],x_a,x_{i1}[t],\ldots,x_N[t]\}$, where all other trajectories
are known parameters, except $x_a$. In other words, the Newton
solution of $x_a[t]$ just needs the spatial dependence of
$V(x_1[t],\ldots,x_a,\ldots,x_N[t])$ on the variable $x_a$. In this
section, we summarize the formalism developed by Oriols \cite{om.oriolsprl} to compute many-particle Bohmian trajectories
without knowing the many-particle wave function. It is a clear
example of how Bohmian mechanics can be a powerful computational 
tool.

The main idea behind the work developed in \cite{om.oriolsprl} is
that any Bohmian trajectory $x_a[t]$ that is computed from the
many-particle wave function $\Psi(\vec x,t)$ solution of
\eref{om.schordingerND} can be alternatively computed from a much
simpler single-particle conditional wave function $\phi_a(x_a,t) = \Psi(x_a,\vec
x_b[t],t)$. Here, we use the notation $\vec x = \{x_a,\vec x_b\}$
with $\vec x_b = \{x_1,\ldots,x_{a-1},x_{a + 1},\ldots,x_N\}$ for
particle positions and $\vec x_b[t] =
\{x_1[t],\ldots,x_{a-1}[t],x_{a + 1}[t],\ldots,x_N[t]\}$ for Bohmian
trajectories.

It is quite trivial to demonstrate the ability of $\phi_a(x_a,t)$ in
reproducing $x_a[t]$. By construction, when we use a polar form
$\phi_a(x_a,t) = r_a(x_a,t) \; e^{i \; s_a(x_a,t)/\hbar}$, the angle
$s_a(x_a,t)$ is identical to the angle $S(x_a,\vec x_b,t)$ of
$\Psi(x_a,\vec x_b,t)$ evaluated at $\{x_a,\vec x_b[t]\}$.
Therefore, since the velocity of the trajectory $x_a[t]$ is computed
from the spatial dependence of $S(x_a,\vec x_b,t)$ on $x_a$ when all
other positions are fixed at $\vec x_b = \vec x_b[t]$, the same
velocity will be obtained from the spatial dependence of
$s_a(x_a,t)$. Interestingly, $\phi_a(x_a,t)$ is solution of a
single-particle (pseudo) Schr\"odinger equation because it depends
only on time $t$ and position $x_a$. Next, our effort will be
focused on determining such single-particle  equations.

\subsubsection{Single-particle pseudo-Schr\"odinger equation for many-particle systems}

First of all, we show that any (totally arbitrary) single-valued complex function $\phi_a(x_a,t)$, which has a well-defined second-order spatial derivative and first-order temporal derivative, can be obtained from a Schr\"odinger-like equation when the following potential $W(x_{a},t)$ is used:\vspace*{-6pt}
\begin{equation}
\label{om.W_prl}
W(x_{a},t) = \frac {i\hbar\frac{\partial \phi_{a}(x_{a},t)}{\partial t} + \frac{\hbar^2}{2m}\frac{\partial^2\phi_{a}(x_{a},t) }{\partial_{x_a}^2}} {\phi_{a}(x_{a},t)}
\end{equation}
For an arbitrary (complex) function, the potential energy
$W(x_{a},t)$ can be complex, too. In fact, we are interested in rewriting  $W(x_{a},t)$ in terms of the polar form of the wave
function $\phi_a(x_a,t) = r_a(x_a,t) \; e^{i \; s_a(x_a,t)/\hbar}$.
We obtain for the real part:
\begin{eqnarray}
\label{om.ReW_prl}
{Real}[W(x_{a},t)] &=& - \Bigg( \frac {1} {2 \; m} \left (\frac {\partial s_a(x_a,t)} {\partial x_a} \right)^2-\frac{\hbar^2} {2 \; m \; r_a(x_a,t)}
\frac {\partial^2 r_a(x_a,t)} {\partial^2 x_a}
 + \frac {\partial s_a(x_a,t)} {\partial t} \Bigg)\qquad\nonumber\\
\end{eqnarray}
From \eref{om.W_prl}, we do also obtain  for the imaginary part:
\begin{eqnarray}
\label{om.ImW_prl}
\textit{Imag}[W(x_{a},t)] &=& \frac{\hbar}
{2r_a^2(x_a,t)}\!\! \left( \frac {\partial r_a^2(x_a,t)} {\partial
t}\right.\left. + \frac {\partial} {\partial x_a} \left( \frac {r_a^2(x_a,t)} {m}
\frac {\partial s_a(x_a,t)} {\partial x_a} \right) \right)
\end{eqnarray}
It can be easily verified that $\textit{Imag}[W(x_{a},t)] = 0$ when the single-particle wave function preserves the norm.

Finally, we are interested in using the above expressions when the single-particle wave function is the one mentioned in the introduction of this subsection, $\phi_a(x_a,t) = \Psi(x_a,\vec x_b[t],t)$. In particular, we use $r_a(x_a,t) = R(x_{a},\vec x_{b}[t],t)$ and $s_a(x_a,t) = S(x_{a},\vec x_{b}[t],t)$. Then, we realize that \eref{om.W_prl} transforms into:
\begin{eqnarray}
\label{om.pseudoscho_prl}
i\hbar\frac{\partial \phi_{a}(x_{a},t)}{\partial t} &=& \bigg( -\frac
{\hbar^2} {2m}\frac {\partial^2}{\partial^2 {x_a}} + U_{a}(x_{a},\vec
x_{b}[t],t)+ G_{a}(x_{a},\vec x_{b}[t],t) + i J_{a}(x_{a},\vec x_{b}[t],t) \bigg) \phi_{a}(x_{a},t)\nonumber\\
\end{eqnarray}
where we have defined:
\begin{eqnarray}
\label{om.G_prl}
G_{a}(\vec x,t) &=& U_b(\vec x_{b},t) + \sum_{k = 1;k\neq a}^{N} \Bigg( \frac {1} {2 \; m} \left (\frac {\partial S(\vec x,t)} {\partial x_k} \right)^2+ Q_k(\vec x,t) - \frac {\partial S(\vec x,t)} {\partial x_k} v_k(\vec x[t],t) \Bigg)\nonumber\\
\end{eqnarray}
with $Q_k(\vec x,t)$ as the quantum potential energy defined in \eref{om.quantum_potentialND_parcial}. The terms \textit{$U_{a}(x_{a},\vec x_{b}[t],t)$} and \textit{$U_b(\vec x_{b},t)$} are defined from the many-particle potential that appears in the original many-particle potential energy in \eref{om.schordingerND} as:
\begin{equation}
\label{om.U_prl}
V(\vec x,t) = U_{a}(x_{a},\vec x_{b},t) + U_b(\vec x_{b},t)
\end{equation}
In addition, we have defined:
\begin{eqnarray}
\label{om.J_prl}
J_{a}(\vec x,t) &=& \sum_{k = 1;k\neq a}^{N} \frac {\hbar} {2R^2(\vec
x,t)} \Bigg( \frac {\partial R^2(\vec x,t)} {\partial x_k} v_k(\vec
x[t],t) - \frac {\partial} {\partial x_k} \left( \frac {R^2(\vec x,t)} {m} \frac {\partial S(\vec x,t)} {\partial x_k} \right) \Bigg)\nonumber\\
\end{eqnarray}
In order to obtain these expressions, one has to carefully evaluate:
\begin{eqnarray}
\label{om.prlpas}
\frac {\partial S(x_a,\vec x_b[t],t)} {\partial t} &=& \left( \frac
{\partial S(x_a,\vec x_b,t)} {\partial t}\right)_{\vec x_b = \vec
x_b[t]}+ \sum_{k = 1; k \neq a}^{N}\frac {\partial S(x_a,\vec x_b[t],t)}{\partial x_k} \; v_k(\vec x[t],t)\nonumber\\
\end{eqnarray}
and use \eref{om.Hamilton_JacobiND} evaluated at $\{x_a,\vec x_b[t]\}$. Identically for  $\partial R^2(x_a,\vec x_b[t],t)/\partial t$ with \eref{om.charge_conservationND}.

The single-particle pseudo-Schr\"odinger equation,
\eref{om.pseudoscho_prl}, is the main result of this algorithm. Let
us discuss the role of each potential term in
\eref{om.pseudoscho_prl}:
\begin{enumerate}
\item The term \textit{$U_{a}(x_{a},\vec x_{b}[t],t)$} is a real-valued potential whose explicit dependence on the positions is known. It has to be evaluated from the particular Bohmian position of all particles except $x_a[t]$.

\item The term \textit{$G_{a}(x_{a},\vec x_{b}[t],t)$} is a real-valued potential whose explicit dependence on the positions is unknown (unless we know the many-particle wave function) and needs some educated guess. It takes into account, for example, the exchange interaction between particles.

\item The term \textit{$i J_{a}(x_{a},\vec x_{b}[t],t)$} is an imaginary-valued potential whose explicit dependence on the positions is also unknown and needs some educated guess. It takes into account that the norm of $\phi_{a}(x_{a},t)$ is not directly conserved (the norm of the many-particle wave function $\Psi(x_a,\vec x_b,t)$ is conserved in the entire configuration space, but this is not true for $\Psi(x_a,\vec x_b[t],t)$ in the $x_a$ space).
\end{enumerate}

By using \eref{om.pseudoscho_prl} for each particle, $x_a[t]$ for $a = 1,\ldots,N$, we obtain a  system of $N$ coupled single-particle pseudo-Schr\"odinger equations that is able to compute many-particle Bohmian trajectories without knowing the many-particle wave function. The great merit of \eref{om.pseudoscho_prl} is to demonstrate that such a single-particle solution of a many-particle problem exists, although we do not know exactly the values of the terms $G_{a}(x_{a}, \vec x_{b}[t],t)$ and $J_{a}(x_{a},\vec x_{b}[t],t)$. Our algorithms have similarities with the original work on density functional theory \cite{om.kohn1964,om.kohn1965}. The formidable computational simplification comes at the price that some terms of the potential of the corresponding single-particle Schr\"odinger equations are unknown, that is, the exchange-correlation functional in density functional theory \cite{om.kohn1964,om.kohn1965} and, here, the terms in Eqs. (\ref{om.G_prl}) and (\ref{om.J_prl}).

\subsubsection{Example: Application in factorizable many-particle systems}

Let's start by discussing which  will be the solution
$\phi_{a}(x_{a},t)$ of \eref{om.pseudoscho_prl} when the
many-particle wave function $\Psi(x_a,\vec x_b,t)$ is factorizable
(i.e., when it can we written as
\eref{om.many_function_factorizable}). Then,
\eref{om.Hamilton_JacobiND_factorizable} is valid for each summand
of $G_{a}(x_{a},\vec x_{b}[t],t)$ in
\eref{om.pseudoscho_prl}, so it can be written in a compact way as:
\begin{equation}
\label{om.prlpas_factorizable1}
G_{a}(x_{a},\vec x_{b}[t],t) = -\sum_{k = 1;k\neq a}^{N} \frac {d S_k(x_k[t],t)} {dt}
\end{equation}
Then, a (real) time-dependent term (without any spatial dependence)
appears in the potential of \eref{om.pseudoscho_prl}. It can be
easily demonstrated that such a term introduces the following
time-dependent contribution  \textit{$ \beta_a(t)$} into the phase of
$\phi_a(x_a,t)$:
\begin{equation}
\label{om.prlbeta}
\beta_a(t) = - \sum_{k = 1;k\neq a}^{N} \int_{t_0}^{t} \frac {d S_k(x_k[t'],t')} {dt'} dt' = -\sum_{k = 1;k\neq a}^{N} S_k(x_k[t],t)
\end{equation}

Identically, \eref{om.charge_conservationND_factorizable} is valid for each term of $J_{a}(x_{a},\vec x_{b}[t],t)$, so it can be written as:
\begin{equation}
\label{om.prlpas_factorizable2}
J_{a}(x_{a},\vec x_{b}[t],t) = -\sum_{k = 1;k\neq a}^{N} \frac {\hbar} {2} \frac {d} {dt} \ln \left( R_{k}^2(x_k[t]) \right)
\end{equation}
Using $\ln (a \; b) = \ln (a) + \ln (b)$, we obtain a contribution  $\alpha_a(t)$ into the phase of $\phi_a(x_a,t)$:
\begin{eqnarray}
\label{om.prlalpha} \alpha_a(t) &=& -\hbar \int_{t_0}^{t} \frac {d}
{dt'} \ln \left( \prod_{k = 1;k\neq a}^{N} R_k(x_k[t'],t') \right) \;
dt'\nonumber\\ &=& -\hbar \ln \left( \prod_{k = 1;k\neq a}^{N}
R_k(x_k[t],t) \right)
\end{eqnarray}
Finally, we use that a pure (real or imaginary) time-dependent
potential (without spatial dependence) in a Schr\"odinger-like
equation does only introduce a pure (imaginary or real)
time-dependent global phase. Thus, we obtain:
\begin{eqnarray}
\label{om.prlpas_factorizable3}
\phi_a(x_a,t) &=& \exp\left(\frac {i \; \beta_a(t)-\alpha_a(t)} {\hbar}\right) \;\psi_a(x_a,t)= \psi_1(x_1[t],t)\ldots\psi_a(x_a,t)\ldots\psi_N(x_N[t],t)\nonumber\\
\end{eqnarray}
which is, certainly, the expected result. Each term \textit{$\psi_k(x_k,t)$} is a single-particle wave function, whose evolution is found from \eref{om.Schrodinger1D} after appropriately defining the initial wave packet at time $t = 0$. Notice the difference between $\psi_a(x_a,t)$ and $\phi_a(x_a,t)$. The former is a single particle wave packet, while the latter has an additional time-dependent function multiplying $\psi_a(x_a,t)$.

\subsubsection{Example: Application in interacting many-particle systems without exchange interaction}

Up to here, we have demonstrated that a Bohmian solution to the
many-particle problem exists in terms of a system of coupled
single-particle Schr\"odinger equations (see
\eref{om.pseudoscho_prl}). The significant computational
simplification comes at the price that the terms
\textit{$G_{a}(x_{a},\vec x_{b}[t],t)$} and
\textit{$J_{a}(x_{a},\vec x_{b}[t],t)$} of the corresponding
single-particle Schr\"odinger equations are unknown. It is in this
sense that we mentioned that these algorithms have similarities with
density functional theory \cite{om.kohn1964,om.kohn1965}. In this
subsection we provide a simple approximation for a system of $N$
electrons with Coulomb interaction but without exchange interaction.
Later, we will include the exchange interaction.

As mentioned, the solution of \eref{om.pseudoscho_prl} needs
educated guesses for the terms \textit{$G_{a}(x_{a},\vec
x_{b}[t],t)$}, \eref{om.G_prl}, and \textit{$J_{a}(x_{a},\vec
x_{b}[t],t)$}, \eref{om.J_prl}. Since no exchange interaction is
considered, we assume that the origin of the correlations between the $a$ electron and the rest is
mainly contained in the term \textit{$U_{a}(x_{a},\vec
x_{b}[t],t)$}. We develop  a Taylor expansion of the other two terms, Eqs.
(\ref{om.G_prl}) and (\ref{om.J_prl}), in the variable $x_a$
around the point $x_a[t]$:
\begin{equation}
\label{om.G_prl_taylor}
G_{a}(x_a,t) = G_{a}(x_a[t],t) + \left[\frac {\partial G_{a}(x_a,t)} {\partial x_a} \right]_{x_a = x_a[t]}(x_a-x_a[t]) + \ldots
\end{equation}
and
\begin{equation}
\label{om.J_prl_taylor}
J_{a}(x_a,t) = J_{a}(x_a[t],t) + \left[\frac {\partial J_{a}(x_a,t)} {\partial x_a}\right]_{x_a = x_a[t]} (x_a-x_a[t]) + \ldots
\end{equation}
The simplest approximation is just a zero-order Taylor term \textit{$G_{a}(x_a,t) \approx G_{a}(x_a[t],t)$} and \textit{$J_{a}(x_a,t) \approx J_{a}(x_a[t],t)$}.

The conditional wave function $\phi_{a}(x_a,t)$ solution of \eref{om.pseudoscho_prl} can be constructed in two steps. First, by solving \eref{om.pseudoscho_prl} without considering the purely time-dependent potential terms, $G_{a}(x_a[t],t)$ and $J_{a}(x_a[t],t)$, to find $\tilde{\psi}_a(x_a,t)$:
\begin{equation}
\label{om.pseudoschobis_prl}
i\hbar\frac{\partial \tilde \psi_{a}(x_{a},t)}{\partial t} = \\ \left( -\frac{\hbar^2}{2m}\frac{\partial^2}{\partial {x_a^2}} + U_{a}(x_{a},\vec x_{b}[t],t) \right) \tilde\psi_{a}(x_{a},t)
\end{equation}
where the term \textit{$U_{a}(x_{a},\vec x_{b}[t],t)$} is defined in \eref{om.U_prl}. Second, by multiplying the wave function $\tilde{\psi}_a(x_a,t)$ by time-dependent (real or imaginary) values (without any spatial dependence) for the final solution:
\begin{equation}
{\phi}_a(x_a,t) \approx \tilde{\psi}_a(x_a,t) \exp (z_a(t))
\label{om.mpnocoulomb}
\end{equation}
with $z_a(t) = i\beta_a(t)/\hbar-\alpha_a(t) \hbar$ defined
according to \eref{om.prlpas_factorizable3}. Again, we have used the
well-known result that a pure (real or imaginary) time-dependent
potential term (without spatial dependence) added into the Hamiltonian of
the Schr\"odinger-like equation does only introduce a pure
(imaginary or real) time-dependent global phase, that is, $\exp (z_a(t))$.
Since the velocity of Bohmian trajectories does not depend on these
pure time-dependent terms \textit{$\exp (z_a(t))$}, we do not have
to compute $\beta_a(t)$ and $\alpha_a(t)$ explicitly.

\begin{figure}
\centering
\includegraphics{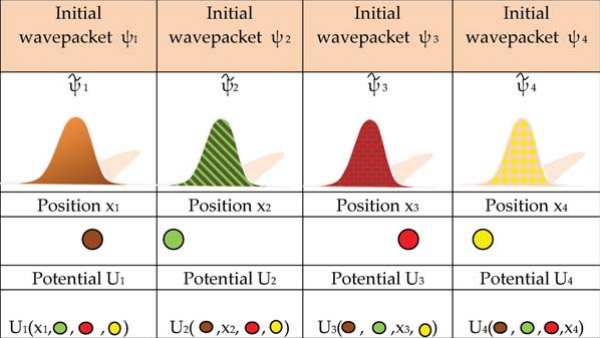}
\caption{For $4$ interacting particles without exchange interaction, the present algorithm needs
$N = 4$ single-particle wave functions, $\psi_a(x_a,t)$. The
subindex $a$ of the wave function is associated with the potential
$U_a(x_a,\vec x_b[t],t)$ and the initial wave packet. Each wave
function $\psi_a(x_a,t)$ determines the $a$-Bohmian trajectories
$x_a[t]$. See also Color Insert.}
\label{om_fig_manyprl1}
\end{figure}

For example, for a system of $N = 4$ interacting electrons, as
depicted in the scheme  of \fref{om_fig_manyprl1}, we need to solve
$N = 4$ Schr\"odinger-like  equations defined by  \eref{om.pseudoschobis_prl} to find $N =
4$ wave functions $\tilde{\psi}_a(x_a,t)$. Many-particle Bohmian
trajectories are computed from the Bohmian velocity
\eref{om.velocity} using $\tilde{\psi}_a(x_a,t)$ to compute the
current and the square modulus. The term \textit{$U_{a}(x_{a},\vec
x_{b}[t],t)$} can be solved from a Poisson equation (if we deal with
Coulomb interaction) and it introduces correlations between
particles. The initial wave function $\tilde{\psi}_a(x_a,t = 0)$ has
to be specified. For example, if we assume that the initial wave
function is defined in a region of the space where the many-particle
wave function is factorizable, we can use $\tilde{\psi}_a(x_a,0) =
{\psi}_a(x_a,0)$.

\begin{figure}
\centering
\includegraphics{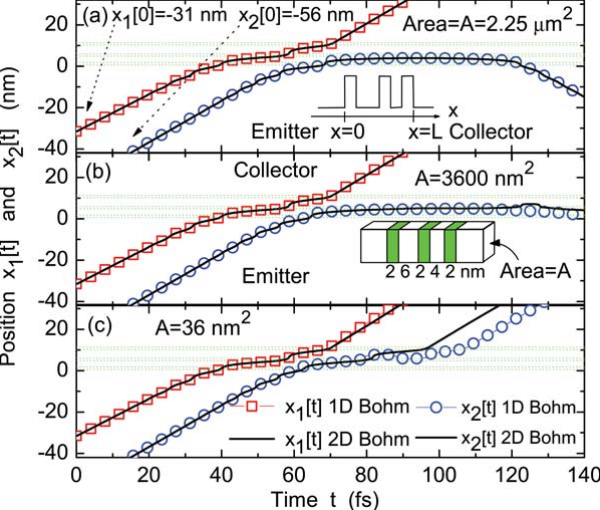}
\caption{Two-particle interacting Bohmian trajectories in a triple barrier tunneling scenario
computed from our 1D approach (symbols) and from exact 2D results
(solid line) for three different lateral areas that modify the
Coulomb interaction. Reprinted with permission from
\cite{om.oriolsprl}. Copyright 2007 American Physical Society.}
\label{om_fig_prl1}
\end{figure}

We will show now an example of the goodness of this simple
approximation. We control the strength of the Coulomb interaction
between two electrons by changing the lateral area of a 3D electron
device, as described in \cite{om.oriolsprl}. \Fref{om_fig_prl1}
shows the excellent agreement between exact Bohmian trajectories and
those computed with our algorithm. In \fref{om_fig_prl1}a, the
lateral area is so large that it makes the Coulomb interaction quite
negligible. The first electron is transmitted, while the second is
reflected. However, as shown in \fref{om_fig_prl1}c, the smaller lateral area
provides strong Coulomb interaction between the electrons, and the
second is finally transmitted because of the presence of the first
one in the barrier region.

Certainly, we have used the simplest approximation in Eqs. (\ref{om.G_prl_taylor}) and (\ref{om.J_prl_taylor}). Any possible improvement of this approximation will imply an even better agreement between this algorithm and the exact computation.

\subsubsection{Example: Application in interacting many-particle systems with exchange interaction}

Now, we generalize the previous result to an arbitrary system with
Coulomb and exchange interactions. For simplicity, we will consider
only the wave function defined in \eref{om.wavepostulate}, that is,
only the symmetry of the orbital wave function is considered. A
simple Taylor approximation for the terms \textit{$G_{a}(x_a,t)$}
and \textit{$J_{a}(x_a,t)$}, as we assume in \eref{om.G_prl_taylor}
and \eref{om.J_prl_taylor}, is not valid here. The reason can be
easily understood, for example, for a system of (identical)
electrons. Due to the Pauli exclusion principle, the modulus of the
wave function tends to zero, $R({{x}_{a}},{{\vec{x}}_{b}}[t],t)\to
0$, at every position where ${{x}_{a}}\to {{x}_{k}}[t]$. Thus, the term
\textit{${{G}_{a}}({{x}_{a}},{{\vec{x}}_{b}}[t],t)$} has asymptotes
at ${{x}_{a}}\to {{x}_{k}}[t]$ that repel other electrons, that is,
${G}_{a}({{x}_{a}},{{\vec{x}}_{b}}[t],t) \to \pm \infty $ at the
same particular positions $x_a$, invalidating the approximation
\eref{om.G_prl_taylor}. For the same reason,
${J}_{a}({{x}_{a}},{{\vec{x}}_{b}}[t],t) \to \pm \infty $.

Here, the strategy is assuming that an antisymmetric wave function,
$\Psi (\vec x,t)$, can be constructed from permutations of a
many-particle wave function without any symmetry $\Psi_\textrm{no-sym}(\vec
x,t)$:
\begin{equation}
\label{om.prl_sumexbis}
\Psi(\vec{x},{t}) = C \sum\limits_{l = 1}^{N!} \Psi_\textrm{no-sym}(x_{p(l)_1},x_{p(l)_2}, \ldots ,x_{p(l)_N},t) \; s\left( \vec{p}(l) \right)
\end{equation}
The subindex ``no-sym'' reminds that such wave functions have no
spatial symmetric or antisymmetric property. The constant $C$ is a
normalization constant that will become irrelevant for the
computation of the Bohmian velocity. The sum is over all $N!$
permutations $\vec{p}(l) = \left\{
p{{(l)}_{1}},p{{(l)}_{2}},\ldots,p{{(l)}_{N}} \right\}$, and
$s\left( \vec{p}(l) \right) = \pm 1$ is the sign of the
permutations. Then, each wave function
$\Psi_\textrm{no-sym}(x_{p(l)_1},x_{p(l)_2}, \ldots , x_{p(l)_N},t)$
evaluated at $\vec x = \{x_a,\vec x_b[t]\}$, can be computed
following the previous (no-exchange) algorithm, that is  Eqs.
(\ref{om.pseudoschobis_prl}) and (\ref{om.mpnocoulomb}). In
particular:
\begin{eqnarray}
\label{om.prl_def}
&&\Psi_\textrm{no-sym}(x_{p(l)_1}[t],\ldots , x_{p(l)_j},\ldots ,
x_{p(l)_N}[t],t) = \tilde{\psi}_{a,{p}(l)_j}(x_a,t) \exp { (z_{a,\vec{p}(l)}(t)) }\nonumber\\
\end{eqnarray}
where the permutation $\vec p(l)$ gives $x_{p(l)_j} = x_a$, that is, ${p(l)_j} = a$. Now, we have to use two labels in the subindex of $\tilde{\psi}_{a,h}(x_a,t)$ to specify the solution of \eref{om.pseudoschobis_prl}. The first label $a$ accounts for the degree of freedom, that is, the particular trajectory, that we are computing and it also fixes the potential energy $U_{a}(x_{a},\vec x_{b}[t],t)$ in \eref{om.pseudoschobis_prl}. The second label $h$ fixes the initial wave function that we will consider. If the initial many-particle wave function can be defined, in a region without interactions, as:
\begin{equation}
\label{om.prl_defini}
\Psi_\textrm{no-sym}(x_{1}, \ldots ,x_{N},0) = \psi_1(x_1,0) \ldots\psi_N(x_N,0)
\end{equation}
then $\tilde{\psi}_{a,{p}(l)_j}(x_a,0) = \psi_{p(l)_j}(x_a,0)$. In other words, identical initial wave functions $\Psi_{i,j}(x_{i},t_{0}) = \Psi_{k,j}(x_{k},t_{0})$ can evolve differently when using $U_{i}(x_{i},\vec{x}_{i}[t],t)$ or $U_{k}(x_{k},\vec{x}_{k}[t],t)$.

\begin{figure}
\centering
\includegraphics[width=12cm]{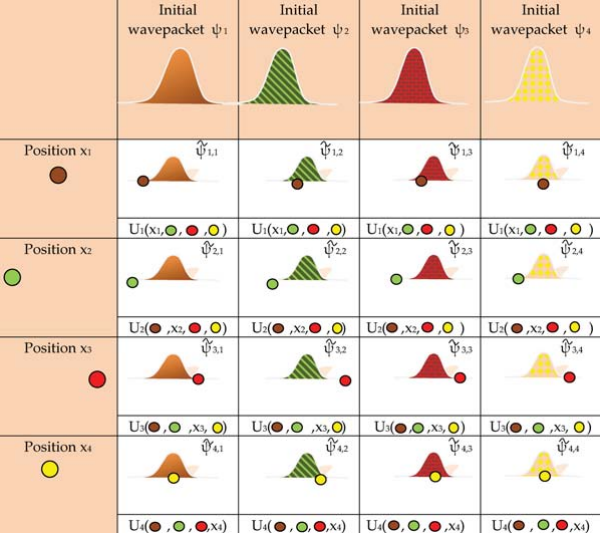}
\caption{When Coulomb plus exchange interactions are considered, a $4\cdot 4$ matrix of the
single-particle wave functions, $\psi_{a,h}(x_a,t)$ (with $a = 1,
\ldots,4$ for the potential energies and $j = 1, \ldots, 4$ for the
initial wave packets) is needed to compute (through a Slater
determinant) the wave function associated with each Bohmian
trajectory. See also Color Insert.}
\label{om_fig_manyprl2}
\end{figure}

Finally, using \eref{om.mpnocoulomb}, the many-particle wave function $\phi(x_a,t) = \Psi(x_a,\vec{x}_b[t],{t})$ can be written as:
\begin{equation}
\label{om.prl_manywave}
\phi(x_a,t) = \Psi(x_a\vec{x}_b[t],{t}) = C \sum_{l = 1}^{N!} \tilde{\psi}_{a,{p}(l)_j}(x_a,t) \exp { (z_{a,\vec{p}(l)}(t)) } \; s\left( \vec{p}(l) \right)
\end{equation}
Here, the angles $z_{a,\vec{p}(l)}(t) $ are relevant and cannot be
ignored (only a global phase can be ignored). We emphasize that $j$
is selected according to the condition ${p(l)_j} = a$, which depends
on the index $l$ of the permutations.

\begin{figure}
\centering
\includegraphics{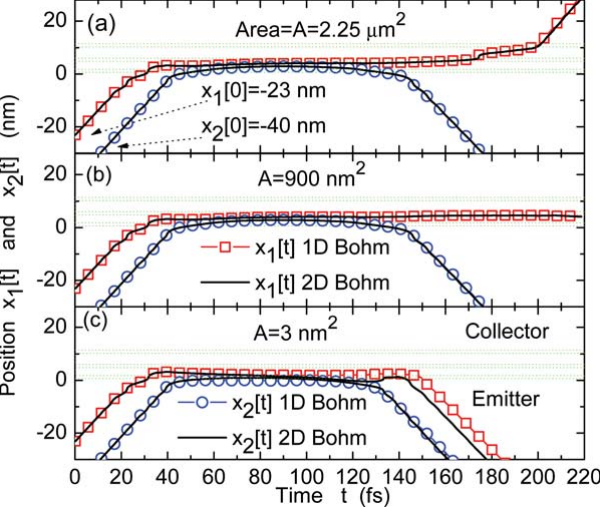}
\caption{Two-particle Bohmian trajectories with Coulomb and exchange interactions computed from our 1D approach (symbols) and from 2D exact results (solid line) for three lateral areas. Reprinted with permission from \cite{om.oriolsprl}. Copyright 2007 American Physical Society.}
\label{om_fig_prl2}
\end{figure}

In fact, we can guarantee the expected symmetry of the quantum wave
function by defining appropriately the phases. It is quite simple
to realize that the following definition of the angles:
\begin{equation}
\label{om.prl_fases}
 \exp { (z_{a,\vec{p}(l)}(t)) } = \prod^{N}_{k = 1,k\neq a} \tilde \Psi_{a,p(l)_k}(x_k[t],t)
\end{equation}
 accomplishes the symmetry requirements. Finally, putting \eref{om.prl_fases} into \eref{om.prl_manywave}, we obtain the final wave function:
\begin{eqnarray}
\phi(x_a,t) &=& C \sum\limits_{l = 1}^{N!}{{{\tilde{\Psi }}}_{1,
p{{(l)}}_1}}({{x}_{1}}[t],t)\ldots {{{\tilde{\Psi }}}_{a, p{{(l)}}_a}}({{x}_{a}},t)\ldots{{{\tilde{\Psi }}}_{N, p{{(l)}}_N}}{({{x}_{N}}[t],t) s(\vec{p}(l))}\nonumber\\
 \label{om.prl_fin}
 \end{eqnarray}
\looseness-1This Eq. (\ref{om.prl_fin}) can be simply computed from a (Slater) determinant of the $N \cdot N$ matrix, $\tilde{\psi}_{a,h}(x_a,t)$ with $a = 1, \ldots , 4$ and $j = 1, \ldots , 4$ for the initial wave packets, which are defined in \fref{om_fig_manyprl2}.

\begin{figure}
\centering
\includegraphics{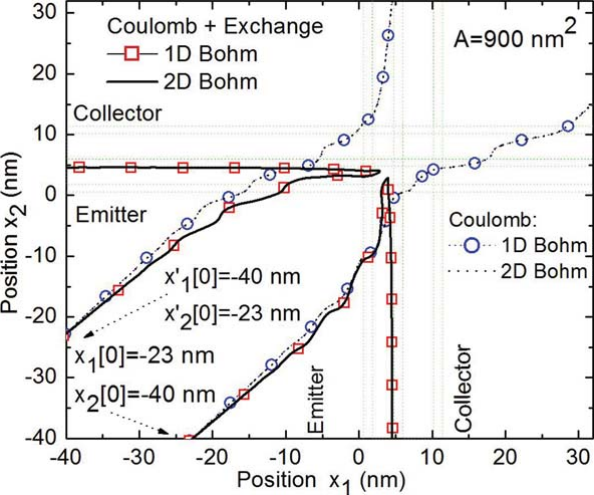}
\caption{Two-particle Bohmian trajectories with Coulomb and exchange interactions computed from
our 1D approach (symbols) and from 2D exact results (solid line)
when initial positions are interchanged. Reprinted with permission
from \cite{om.oriolsprl}. Copyright 2007 American Physical Society.
} \label{om_fig_prl3}
\end{figure}

The reader can wonder why we need $N^2$ different wave functions and not $N!$. The reason is because the term \textit{$U_{a}(x_{a},\vec x_{b}[t],t)$} in \eref{om.pseudoschobis_prl} does not change when we interchange the positions of two of the trajectories in $\vec x_{b}[t]$ (only the interchange between $x_a[t]$ and a trajectory of $\vec x_{b}[t]$ becomes relevant). \Fref{om_fig_prl2} shows the excellent agreement between the exact Bohmian trajectories and those computed from this second algorithm. See details in \cite{om.oriolsprl}.

In order to clarify the meaning of \textit{indistinguishable
particles} in the Bohmian language, in \fref{om_fig_prl3},
we  simulate Bohmian trajectories with the
initial positions and with interchanged initial positions. By our
own construction of the wave function of \eref{om.prl_fin}, the
velocity has symmetry when we interchange particles, as discussed in
\sref{om.sec_many.4}. Therefore, although we can distinguish each
electron by its own trajectories, one cannot discern between them in
the final observable results. For example, the number of events
where the first trajectory is transmitted and the other reflected is
exactly identical to the events where the first is reflected, while
the second is transmitted. On the contrary, as seen in
\fref{om_fig_prl3}, this \textit{symmetry} is broken without the
exchange interaction algorithm.

\section{Bohmian Explanation of the Measurement Process}\label{om.sec_measurement}

The Bohmian explanation of a quantum measurement process is, perhaps, the most attractive (and also ignored) feature of the Bohmian explanation of the quantum nature \cite{om.bell1990,om.Durrllibre,om.Goldsteinobserver,om.Bell1966,om.bomhhiley1993,om.reviewabc,om.llibreph}. Let us start by noting that Bell disliked the word ``measurement'' \cite{om.bell1990}. He preferred the word ``experiment'' because ``when it is said that something is measured it is difficult not to think of the result as referring to some pre-existing property of the object in question.'' On the contrary, in an experiment, it is natural to think that everything can change (time-evolve), because of the interactions. Once we admit that the system itself can be modified during the measurement, we can easily understand that the output of a measurement depends on the duration and strength of the measurement process, on the previous measurements, etc\footnote{Technically,  it is stated that any quantum theory compatible with experiments has to be contextual. By construction, Bohmian mechanics is contextual since the measurement is just another type of interaction.}.

\subsection{The measurement problem}\label{measpro}

We have previously talked about the so-called \emph{measurement problem}\cite{om.Dieter,om.maudlin,om.bohm66} but, what is in fact this measurement problem? It is a problem related to the application of the superposition law in quantum systems. Quantum states of a particle are represented by vectors in a Hilbert space such that linear combinations of them, for example a superposition of macroscopically distinguishable states, also correspond to valid states of the Hilbert space.
However, such superposition of states is not always compatible with measurements~\cite{om.bohm66,om.nikolic}. The measurement problem can be formulated as the impossibility for a physical quantum theory (in empirical agreement with experiments) to simultaneously satisfy the following three assumptions~\cite{om.maudlin}.
\begin{enumerate}
\item The wave function always evolves deterministically according to the linear and unitary Schr\"odinger equation.
\item A measurement always finds the physical system in a localized state, not in a superposition of macroscopically distinguishable states.
\item The wave function is a complete description of a quantum system.
\end{enumerate}
A theory that includes all three assumptions is not empirically compatible with the experimental results. Different physical theories are developed depending on which assumption is ignored~\cite{om.herbert}. The \emph{measurement problem} appears because none of the proposed solutions fully satisfies the whole scientific community.  

The first type of solutions argues that the unitary and linear evolution of the Schr\"odinger equation is not always valid (such solutions ignore the (1) assumption).
For instance, in the instantaneous collapse theories~\cite{om.bassi13} (like the GRW interpretation~\cite{om.ghirardi86}), a new stochastic equation is used 
that breaks the superposition principle at the macroscopic level, while still keeping it at the microscopic one~\cite{om.bassi13}.
Another possibility is substituting the linear Schr\"odinger equation by a non-linear collapse law only when a measurement is performed~\cite{om.bohr20}.
This is the well-known orthodox or Copenhagen solution. The dissatisfaction with the Copenhagen solution is that the theory does not clearly specify when, in which circumstances, the linear or non-linear equation has to be used. 

A second type of solution ignores the (2) assumption that a measurement always finds the physical system in a localized state.
One then concludes that there are different worlds where different states of the superposition are found.
This is the many worlds solution~\cite{om.Everett57,om.wallace12} in which the famous Schr\"odinger's cat
is found alive in one world and dead in another.

There is a final kind of solutions that assumes that the wave function alone does not provide a complete description of the quantum state, that is, it ignores the (3) assumption and includes additional elements (hidden variables) in the theory. The most spread of these approaches is Bohmian mechanics explained in this book.  In a spatial superposition of two disjoint states for a single-particle system, only the one whose support contains the position of the particle becomes  relevant for the dynamics. Notice that, in Bohmian mechanics, it is not mandatory to define which interactions are considered a measurement and which are not. All interactions (implying a measurement or not) are treated identically. 

Let us discuss in more detail the differences between an orthodox and a Bohmian measurement. As we have already explained, in Bohmian mechanics, the evolution of the wave function is determined by the Schr\"odinger equation (when the appropriate Hamiltonian is used). This is not true for the orthodox formulation of the measurement because one linear equation (the Schr\"odinger equation) is used for non-measuring interactions, while a different non-linear equation (the collapse law) is needed to tackle measuring interactions. 

\subsubsection{The orthodox measurement process}\label{meaort}

\begin{figure}
\centering
\includegraphics[width=10cm]{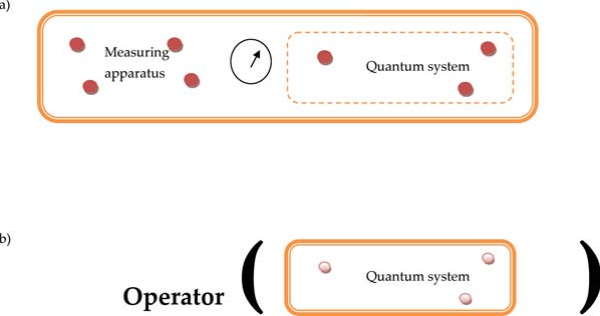}
\caption{(a) The Bohmian measurement is better explained by assuming that the quantum system and the measuring apparatus are two interacting systems. (b) The typical orthodox measurement is better explained by assuming that only the quantum system is explicitly simulated, while the measuring apparatus is substituted by a proper operator acting on the wave function of the system.}
\label{om_measure2}
\end{figure}

In any attempt to study a quantum system, one usually separates the quantum system itself and the rest. The separation is arbitrary, but mandatory in any practical computation since it is impossible to deal with all degrees of freedom of the system and the rest. This separation is also typical in the modeling of a quantum measurement. Let us notice that, in principle, the origin of the separation between the quantum system and the measuring apparatus is technical; it is not a direct consequence of the quantum theory that we select to describe the quantum world. However, it is true that, because of the ontology of the Copenhagen theory, the measurement process is explained by a proper separation between the system and the measuring apparatus (see \fref{om_measure2}b), while the Bohmian theory tends to consider system and apparatus altogether (see \fref{om_measure2}a). 

The typical orthodox prediction of some experimental property of the quantum system is described through the use of a proper operator $\hat{G}$ whose \textit{eigenvalues} give the possible outcomes of the measurement. When we measure a particular eigenvalue, the initial wave function is transformed into an \textit{eigenfunction} of the operator. This is the so-called von Neumman (or projective) measurement.
Thus, the time evolution of the wave function of a quantum system is governed by two (quite) different laws:
\begin{enumerate}
\item The first dynamical evolution is given by the Schr\"odinger equation. This dynamical law is deterministic in the sense that the final wave function of the quantum system is perfectly determined when we know the initial wave function and the Hamiltonian of the quantum system.

\item The second dynamical law is called the \textit{collapse} of the wave function. The collapse is a process that occurs when the wave function interacts with a measuring apparatus. The initial wave function before the measurement is substituted by one of the \textit{eigenstates} of the particular operator $\hat{G}$\!. Contrarily to the dynamical law given by the Schr\"odinger equation, the collapse is not deterministic, since the final wave function is randomly selected among the operator's eigenstates.
\end{enumerate}

The duality in the equation of motions (linear or non-linear) of a quantum system in the orthodox interpretation is certainly a persistent controversial issue. There are many scientist unsatisfied with this solution of the \textit{measurement problem} \cite{om.bell1990}. As repeatedly stressed by Bell, the orthodox theory is unprofessional because it does not explain with total accuracy which parts of the entire quantum system form the measuring apparatus and which the quantum system itself. It is even not clear if the measuring apparatus needs some kind of human activity (\textit{with a PhD?} \cite{om.bell1990}) to be accepted as a proper measuring apparatus. The problematical way in which measurement is treated in the orthodox formulation has been stressed by Bell \cite{om.Bell1987}:
\begin{quote}
The concept of ``measurement'' becomes so fuzzy on reflection that it is quite surprising to have it appearing in physical theory at the most fundamental level. Less surprising perhaps is that mathematicians, who need only simple axioms about otherwise undefined objects, have been able to write extensive works on quantum measurement theory-which experimental physicists do not find it necessary to read. [\ldots] Does not any analysis of measurement require concepts more fundamental than measurement? And should not the fundamental theory be about these more fundamental concepts?
\end{quote}
In short, Bell argued that the separation between the quantum
system and the measuring apparatus in \fref{om_measure2}b is
arbitrary. For practical computations, the \textit{encapsulation} of the rest of the world (except
the quantum system) into a mathematical entity called an operator,
$\hat{G}$\!, is a very clever \textit{trick} that allows for
straightforward calculations of the results of quantum measurements
\cite{om.Durrnaive,om.goldstein} without considering the rest of
the  world. However, conceptually, the role played by the operator $\hat{G}$ as a fundamental 
part of the orthodox theory is at the origin of the criticism against the orthodox solution of the measurement problem. 

The reader can argue that, since we have said that the separation between the system and the rest is always arbitrary, one can always enlarge what we consider as a system (see \fref{om_measure2}a) also in the orthodox formulation of the measurement. This would not solve the measurement problem in the orthodox formulation.
Certainly, we can enlarge the system in the orthodox quantum formulation of the measurement process by including as a system part of what was defined previously as an apparatus. The degrees of freedom of the new system will interact with the degrees of freedom of the old system through a linear equation of motion (Schr\"odinger equation). Therefore, we will always need \emph{something} external to the enlarged system (some new operator $\hat{G}$\!) to provide the \textit{collapse} of the wave function of the enlarged system.  

\subsubsection{The Bohmian measurement process}\label{meabom}

In Bohmian theory, the measurement process is treated just as any
other quantum process of interacting particles and the previous
measurement difficulties of the orthodox interpretation simply
disappear. At least, from a conceptual point of view. There is no need to introduce operators
\cite{om.Durrnaive,om.Durrllibre,om.goldstein,om.reviewabc,om.llibreph}. Here, the entire
quantum system is described by a many-particle trajectory plus a many-particle wave function
(rather than a wave function alone). The wave function and the
trajectory are both associated with the entire system, that is, the
quantum system plus the measuring apparatus. Then, there is one dynamical law for
the evolution of the wave function and another  for the
evolution of the trajectory:\enlargethispage{1pc}
\begin{enumerate}
\item The Schr\"odinger equation (with the appropriate Hamiltonian of the quantum system plus the measuring apparatus) determines the time evolution of the wave function independently of whether a measurement process takes place or not.

\item The time evolution of the particle is determined by the time integration of the Bohmian velocity independently of whether a measurement process takes place or not.
\end{enumerate}

For example, imagine that some kind of pointer indicates the
measured quantity; then, the particles (degrees of freedom) that
form the pointer must be present in the Hamiltonian.\footnote{In
modern electronic measuring devices, the pointer could be
represented by a seven-segment array of light-emitting diode (LED)
displays, each one with two possible states, ON and OFF. When
electrons are present inside the PN interface of one of the LEDs, a
radiative transition of the electrons from the conduction to the
valence band produces light corresponding to the ON state. The
absence of electrons is associated with an OFF state.} In this
sense, in Bohmian mechanics, a physical quantum system must be described by a
many-particle Hamiltonian. Once the Bohmian
trajectories associated with the positions of the pointer are known,
the value of the measurement is already predicted. We just need
knowledge of the positions of the pointer particles. Again, Bell
provided us with one of his didactic sentences \cite{om.Bell1987}:
\begin{quote}
In physics the only observations we must consider are position
observations, if only the positions of instrument pointers. It is a
great merit of the de Broglie--Bohm picture to force us to consider
this fact. If you make axioms, rather than definitions and theorems,
about the ``measurement'' of anything else, then you commit
redundancy and risk inconsistency.\enlargethispage{1pc}
\end{quote}

\begin{figure}
\centering
\includegraphics{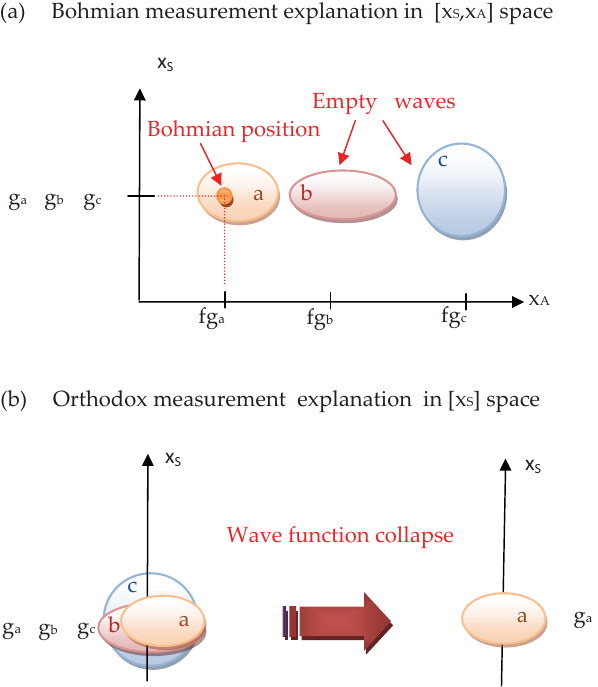}
\caption{(a) Bohmian measurement in the $[x_S,x_A]$ space: From the nonoverlapping
many-particle (system + apparatus) wave function, only the $g_a$
part of the wave function where the trajectory is present is needed
to compute the evolution of the Bohmian system. (b) Orthodox
measurement in the $[x_S]$ space: The (system) wave function
collapses into the $g_a$ part when the measurement takes place. See also Color Insert.}\vspace*{-6pt}
\label{om_measure}
\end{figure}

Therefore, a proper modeling of a Bohmian measurement just needs the explicit consideration of the degrees of freedom of the pointer in the many-particle wave function and many-particle Bohmian trajectories that define the entire system. See \fref{om_measure}. In particular, we have to introduce into the Hamiltonian the interaction of the particles of the pointer with the rest of the particles of the system. The back reaction of the measurement process on the wave function is trivially considered. Certainly, a Hamiltonian with or without the measuring apparatus will provide a different evolution of the quantum system wave function. For this reason, Bell preferred the word ``experiment'' instead of ``measurement.''

The Bohmian and the orthodox explanations of a measurement produce the same probabilistic predictions. However, the mathematical implementation of the equations of motion in each case is quite different. The orthodox quantum theory requires an operator to describe the effect of the measuring apparatus, but this operator is not needed in the Bohmian explanation.

To be fair, the simplicity of the Bohmian measurement is true at the conceptual level explained above. However, any practical attempt to directly implement the Bohmian protocol will have to face with the \emph{many body problem} discussed in \sref{om.sec_many.1}. A pointer in a measuring apparatus is a macroscopic object with a number of particles on the order of  Avogadro's number ($\approx 10^{23}$ particles). Can we simulate such large number of particles and look at the ``position of the instrument pointers'' to get the measured valued as suggested by Bell? From a practical point of view, the answer is no because of the computational burden. Approximations reducing the simulated degrees of freedom, implying the artificial separation of the whole Bohmian system (as depicted in \fref{om_measure2}b) into systems plus apparatus (as depicted in \fref{om_measure2}a), will be mandatory.

\subsection{Theory of the Bohmian measurement process}
\label{om.sec_measurement.1}

Now, we provide a mathematical demonstration of the equivalence between orthodox and Bohmian measurements for a typical ensemble over identical experiments. In the standard interpretation of quantum theory, the von
Neumann (projective) measurement process is defined in a particular
quantum region, which is defined through the degrees of freedom
$\vec x_S$. See \fref{om_measure}b. The state of the quantum system
in this particular region is determined through the wave function
$\psi_S(\vec x_S,t)$. The process of measuring a particular
magnitude is mathematically defined through an operator, for
example $\hat{G}$, acting on the wave function. The possible
outcomes of the measurement process correspond to one of the
possible eigenvalues $g$ of this operator that satisfy the equation
$\hat{G}\psi_g(\vec x_S) = g\psi_g(\vec x_S)$ with $\psi_g(\vec
x_S)$, an eigenvector of this operator. The set $\psi_g(\vec x_S)$
forms an orthonormal basis of the Hilbert space of the quantum
system so that the arbitrary initial wave function can be written
as:\enlargethispage{1pc}
\begin{equation}
\label{om.bmeasure1}
\psi_S(\vec x_S,t) = \sum_{g} c_g(t)\psi_g(\vec x_S)
\end{equation}
with $c_g(t)=\int \psi_S(\vec x_S,t)\psi_g^*(\vec x_S) $ a complex value with the only restriction $\sum_{g}
|c_g(t)|^2 = 1$ in order to ensure that $\psi_S(\vec x_S,t)$ is well
normalized. When measuring the eigenvalue $g_a$ the total wave
function $\psi_S(\vec x_S,t)$ collapses into $\psi_{g_a}(\vec x_S)$.
Then, the probability of obtaining $g_a$ in the measuring apparatus
is just $P_{g_a} = |c_{g_a}(t)|^2$. To avoid unnecessary
complications, hereafter, we have assumed that the basis $\psi_g(\vec x_S)$ has no
degeneracy.

In order to understand how this (von Neumann) measurement can be
described by means of Bohmian mechanics, we use the Schr\"odinger
equation plus a trajectory, defined both in the enlarged
configuration space that includes the quantum system and the
measuring apparatus. There is no privileged status for the
\emph{observer}, that is, the measuring apparatus, over the quantum
system. This is the reason why some authors describe Bohmian theory
under the title of ``quantum mechanics without observers''
\cite{om.Goldsteinobserver}.

Surprisingly, this idea was, somehow, anticipated in the book
\textit{Quantum Theory}, published in 1951 \cite{om.bohmbook}, by
Bohm. Although in this book Bohm followed an orthodox presentation
of quantum mechanics, he also provided a detailed and conceptually
intricate treatment of the measurement process in chapter 22. First,
Bohm demanded in unequivocal terms a quantum theoretical treatment
of the entire process of measurement within the Schr\"odinger
equation. In particular, he argued that a good measuring apparatus
will force the wave function to decouple into several nonoverlapping
wave packets. Each one of these wave packets can be associated to
one particular value of the measurement process. Second, he
supplemented his measurement process by adding some obscure
discussion about decoherence where only one of the several wave
packets survives (as we mentioned at the end of \sref{meaort}, even after enlarging the system, one has to invoke somehow the \emph{collapse} of the wave function). After 1951, when Bohm presented his ``hidden
variables'' theory \cite{om.bohm1952a,om.bohm1952b}, he kept the
first part of his previous explanation of the measurement process
and then provided a simple explanation about the final selection of
the available wave packets: the selected wave packet is the one that
contains the trajectory,\footnote{Very roughly speaking, Bohmian
mechanics avoids the need to select randomly the final wave packet
because the initial position of the trajectory has already selected the final wave packet at the beginning.} see Fig. 1.10(a).\enlargethispage{1pc}

Our mathematical demonstration on how Bohmian mechanics explains the (von Neumann) measurement follows 
\cite{om.Holand1993,om.bomhhiley1993,om.Durrnaive,om.Durrllibre,om.llibreph}. In
order to be able to define the measurement process in the Bohmian
formalism, apart from the degrees of freedom $\vec x_S$, we need the
degrees of freedom of the positions $\vec x_A$ of the pointer belonging to a measuring apparatus.\footnote{In fact, we would have
to include all others degrees of freedom of the ``environment,''
${\vec x_E}$, needed to be able to assume that the considered entire
system is a closed one described by a pure (not mixed)
state. In addition, as mentioned in \sref{meabom}, we emphasize that a pointer in a measuring apparatus is a macroscopic object with a number of particles on the order of Avogadro's number, not the single degree of freedom $\vec x_A[t]$ used here. A single degree of freedom cannot be considered a pointer because it is not a macroscopic object (we do not \emph{see} $\vec x_A[t]$). In any case and without loss of generality, we perform here a conceptual demonstration with a single degree of freedom $\vec x_A[t]$ for the pointer.} Thus, we define a total wave function
$\Phi_T(\vec x_S,\vec x_A,t)$ in a larger configuration space that
includes the quantum region plus the measuring region, $\{\vec
x_S,\vec x_A\}$. According to the Bohmian postulates, we do also
select a trajectory $\{\vec x_S[t],\vec x_A[t]\}$ in this larger
configuration space. The (Bohmian) time evolution of the total wave
function and the total trajectory is what we need to explain the (von Neumann) measurement process
described earlier .

There are some necessary conditions that the time-evolution of the 
entire Bohmian system has to satisfy to say that a measuring
apparatus is able to correctly determine the eigenvalues $g$. First,
the pointer positions $\vec x_A[t]$ of such an apparatus have to be
restricted to a particular region, $\vec x_A[t]\in S_{g_1}$, every
time that the quantum system is in the eigenstate $\psi_{g_1}(\vec
x_S)$. Let us define $\Phi_{g_1}(\vec x_S,\vec x_A,t)$ as the total
wave function that fits with the property that any experiment whose
quantum system is described by $\psi_{g_1}(\vec x_S)$ implies that
the pointer points in the particular region, $\vec x_A[t]\in
S_{g_1}$. A second necessary condition for a good measuring
apparatus of the eigenvalue $g$ is that $S_{g1} \cap S_{g2} = 0$
when we measure. We have defined the restricted region
allowed by the pointer positions associated with a second
eigenstate, $\psi_{g_2}(\vec x_S)$, as $\vec x_A[t]\in S_{g_2}$.
This implies that the states $\Phi_{g_1}(\vec x_S,\vec x_A,t)$ and
$\Phi_{g_2}(\vec x_S,\vec x_A,t)$ do not overlap in the larger
configuration space during the measurement.

Since the eigenstates $\psi_g(\vec x_S)$ form a complete basis, we can decomposed any function, and in particular each $\Phi_g(\vec x_S,\vec x_A,t)$, into the following sum:
\begin{equation}
\Phi_g(\vec x_S,\vec x_A,t) = \sum_{g'} f_{g'}(\vec x_A,t) \; \psi_{g'}(\vec x_S)
\label{om.bmeasure2bis},
\end{equation}
with $ f_g(\vec x_A,t) = \int \Phi_g(\vec x_S,\vec x_A,t)
\psi^*_g(\vec x_S) d\vec x_S$. However, from our previous discussion
about the properties of a ``good'' von Neumann measurement
apparatus, $\Phi_g(\vec x_S,\vec x_A,t)$ cannot be a sum over
different eigenstates $\psi_g(\vec x_S)$ in \eref{om.bmeasure2bis}
because then the measuring apparatus would provide the same pointer
position $\vec x_A[t]\in S_{g}$ for different eigenstates. Thus, during the time of the measurement, the
only good decomposition for $\Phi_g(\vec x_S,\vec x_A,t)$ is:
\begin{equation}
\Phi_g(\vec x_S,\vec x_A,t) = f_g(\vec x_A,t) \; \psi_g(\vec x_S)
\label{om.bmeasure2}
\end{equation}
We emphasize that $ f_g(\vec x_A,t)$ is a normalized function because $\Phi_g(\vec x_S,\vec x_A,t)$ and $\psi_g(\vec x_S)$ are also normalized functions in their respective configurations spaces. By its own construction, $f_{g_1}(\vec x_A,t) \cap f_{g_2}(\vec x_A,t) = 0$ during the measuring time. Thus, even if $\psi_{g_1}(\vec x_S)$ and $\psi_{g_2}(\vec x_S)$ overlap, the states $\Phi_{g_1}(\vec x_S,\vec x_A,t)$ and $\Phi_{g_1}(\vec x_S,\vec x_A,t)$ do not overlap in the larger configuration space. See \fref{om_measure}a. Thus, we can ensure that a general function in the quantum system, \eref{om.bmeasure1}, can be rewritten in the whole, that is, quantum plus measuring, configuration space associated to a good measuring apparatus as:
\begin{equation}
\Phi_T(\vec x_S,\vec x_A,t) = \sum_{g} c_g(t) \;\; f_g(\vec x_A,t) \; \psi_g(\vec x_S)
\label{om.bmeasure3}
\end{equation}
In summary, during the time of measurement, the only total wave
functions that can \textit{live} in the entire quantum system that
includes a good (in this case a projective or Von-Neumann type) measuring apparatus of the eigenvalues $g$ are the
ones written in \eref{om.bmeasure3}. An example of such
wave functions is depicted in \fref{om_measure}a. It is important to
notice that \eref{om.bmeasure3} implies no restriction on the wave
function $\psi_S(\vec x_S,t)$ but only on the total wave function
$\Phi_T(\vec x_S,\vec x_A,t)$. If these restrictions are not
respected, we can find other types of total wave functions in the
configuration space $\{\vec x_S,\vec x_A\}$, but they would be
incompatible with stating that we have an apparatus that is able to
measure the eigenavalue $g$ with certainty at time $t$.

After this discussion, we can show quite trivially how the von
Neumann measurement is exactly reproduced with Bohmian mechanics. As
we have mentioned, apart from the wave function in \eref{om.bmeasure3}, we
have to select an initial trajectory $\{\vec x_S[0],\vec x_A[0]\}$.
Such a trajectory will evolve with the total wave function, and
during the measurement, the particle trajectory $\{\vec x_S[t],\vec
x_A[t]\}$ will be situated in only one of the nonoverlapping wave
packets of \eref{om.bmeasure3}, for example $f_{g_a}(\vec x_A,t)
\psi_{g_a}(\vec x_S)$ as depicted in \fref{om_measure}a. Thus, the
pointer positions will be situated in $\vec x_A[t]\in S_{g_a}$ and
we will conclude with certainty that the eigenvalue of the quantum
system is $g_a$. In addition, the subsequent evolution of this
trajectory can be computed from $f_{g_a}(\vec x_A,t) \psi_{g_a}(\vec
x_S)$ alone. In other words, we do not need the entire wave function
\eref{om.bmeasure3} because the particle velocity can be computed
from $f_{g_a}(\vec x_A,t) \psi_{g_a}(\vec x_S)$. The rest of circles
of \fref{om_measure}a are empty waves that do not overlap with
$f_{g_a}(\vec x_A,t) \psi_{g_a}(\vec x_S)$ so that they have no
effect on the velocity of the Bohmian particle\footnote{Let us remind that, in a realistic measuring apparatus, the number of particles of the pointer is around the Avogadro's number. Therefore, the wave functions $f_{g}(\vec x_A,t) \psi_{g}(\vec x_S)$ \emph{live} in a $N=10^{23}$ configuration space where the possibility of a coincidence of two empty waves is almost zero.}. This is the simple
explanation of how the complicated orthodox collapse is interpreted
within Bohmian mechanics. This discussion is depicted in
\fref{om_measure}

We have mentioned before that the von Neumann measurement predicts that the probability of finding $g_a$ is $P_{g_a} = |c_{g_a}(t)|^2$. We can easily recover this result from our Bohmian picture. This probability is obtained after the repetition of the same experiment, with the same $\Phi_T(\vec x_S,\vec x_A,t)$, many times. The second (quantum equilibrium) postulate of Bohmian mechanics (see \sref{om.sec_many.5}) ensures that the modulus of the wave function will be reproduced by the trajectories when we repeat the same experiment many times. Then, the probability of finding the eigenvalue $g_a$, that is, the eigenstate $\psi_{g_a}(\vec x_S)$, is just:
\begin{eqnarray}
P_{g_a} &=& \frac {\int \int |c_{g_a}(t) f_{g_a}(\vec x_A,t) \psi_{g_a}(\vec x_S)|^2 d\vec x_S d\vec x_A} {\int \int \sum_{g} |c_g(t) f_g(\vec x_A,t) \psi_g(\vec x_S)|^2 d\vec x_S d\vec x_A}= \frac {|c_{g_a}(t)|^2} {\sum_{g} |c_g(t)|^2} = |c_{g_a}(t)|^2
\end{eqnarray}
\looseness-1where we have explicitly used the condition that the each
\textit{$\Phi_{g_1}(\vec x_S,\vec x_A,t)$} does not overlap with the others for the
computation of the modulus of the total wave function,
\textit{$|\Phi_T(\vec x_S,\vec x_A,t)|^2$}, in the denominator.
Additionally, we use $\int |\psi_g(\vec x_S)|^2 d\vec x_S = 1$,
$\int |f_g(\vec x_A,t)|^2 d\vec x_A = 1$, and $\sum_{g} |c_g(t)|^2 =
1$. Thus, the same probabilistic predictions are obtained from
Bohmian and orthodox quantum formulations when a von Neumann
measurement is performed.

We want to enlarge the explanation of the role played by the empty waves belonging to the wave packet of \eref{om.bmeasure3}, which do not contain the particle. In principle, one can argue that such empty waves can evolve and, in later times, overlap with the original wave that contains the particle. If we are interested in doing subsequent (i.e., two times) quantum measurements, a good measuring apparatus has to avoid these ``spurious'' overlaps. This can be understood as an additional condition for qualifying our measuring apparatus as a good apparatus. In the examples of the subsequent subsections \ref{om.measure_moment} and \ref{om.measure_transmission}, we will realize that the assumption of nonoverlapping between empty and full waves can be very reasonable in many practical implementations of the measuring apparatus. In fact, in a real system with a very large number of particles associated to the apparatus (or the environment), the probability that such overlap occurs in the configuration space of all the particles is almost null. See a discussion on page 79 in Ref. \cite{om.bomhhiley1993}. If the empty wave packets later coincide (by putting some reflected mirrors for example), then, the apparatus described above would not be a good apparatus to measure eigenvalues. On the contrary, this apparatus (with the mirrors) would be a very interesting/enlightening design to test quantum interference effects with waves and particles. The double-slit experiment is the most famous example of interference effects between empty and full wave packets.      

Finally, let us mention that we have only considered von Neumann
(i.e., projective) measurements. Other types of measurements are
also possible, which do not collapse into an eigenstate (for example, the generalized or  weak measurements). Obviously, such generalized measurements can also be explained within Bohmian mechanics with an extension of the simple ideas for projective measurements discussed here
\cite{om.bomhhiley1993,om.Durrnaive,om.Durrllibre}.

\subsubsection{Example: Bohmian measurement of the momentum}
\label{om.measure_moment}

Let us explain the momentum measurement of the stationary state of
an electron initially in an energy eigenstate of a square well of
size $L$. For example, $\psi_n(x) = C \; \sin(n \pi x/L)$ within the
well and zero elsewhere, $C$ being the normalization constant and
$n$ an integer (denoting the vibrational state). Since the wave
function is real, we obtain $p = \partial S(x)/\partial x = 0$.
Thus, the Bohmian particle is at rest, meaning that its velocity and
kinetic energy are zero. However, we know that for a high enough
value of $n$, the previous wave function $\psi(x)$ can be roughly
approximated by a sum of two momentum eigenstates with eigenvalues
$\pm n \hbar/L$. Therefore, when we do a (Copenhagen) momentum
measurement, we will obtain an outcome $\pm n \hbar/L$ and the
previous wave function will \textit{collapse} into one of the two
momentum eigenstates.

We have shown that the Bohmian measurement has to provide the same
eigenvalues and eigenstates. How? The answer is that according to
the Bohmian measurement, we have to specify in detail how the
measurement process develops; which is the Hamiltonian that provokes the measurement.

\looseness-1A measurement of the momentum, for example, can be undertaken by
removing the walls that form the well and detecting the electron
somewhere in a screen far from the initial walls. The time interval
between removing the walls and detecting the particles will allow us
a computation of the ``electron velocity.'' The time-dependent
process of removing the walls will imply that the stationary wave
function will evolve into two time-dependent wave packets moving on
opposite directions, which will become completely separated in
space. The particle will end in one wave packet or the other with a
momentum very close to $\pm n \hbar/L$, the sign depending on which
wave packet the initial position of the particle enters
\cite{om.bohm1952b}. Finally, the electron will be detected far from
the walls after the appropriate interval of time. For this
particular experiment, we are sure that the empty wave will not
affect the wave that carries the particle. As mentioned earlier, the
relevant point is that we have been able to measure the same
eigenstates and eigenvalues predicted by the orthodox interpretation
without invoking the \textit{collapse} of the wave function as a fundamental law.

In order to understand the Bohmian explanation of this experiment, it is important to discuss the roles played by the kinetic and quantum potentials. The initial Bohmian kinetic energy is zero, while the initial quantum potential is $Q(x,t) = (n\hbar/L)^{2}/(2m)$ for the initial state $\psi(x) = C \; sin(n \pi x/L)$. On the contrary, the final energy of the system appears in the form of kinetic energy of electrons $(n\hbar/L)^{2}/(2m)$, while the final quantum potential from \eref{om.hamitlon3} is negligible $Q(x,t)\sim 0$. See problem \ref{om.P4}.

The reader will realize that we have not follow all the indications of the previous section. In principle, for a formal Bohmian
discussion of measurement, we would have to discuss the wave function and particles of the whole quantum plus measuring apparatus system. However, in this section we have avoided the wave function and trajectory related to the measuring apparatus. In this example, the screen is the measuring apparatus of the position of the system's particle. The part of the measuring process involving the screen has not been explicitly described in our example (we could obviously explain why a proper coupling of the degrees of freedom of the system and the screen will imply that after impinging with the screen, the system wave function becomes a wave packet highly localized around the position of the measurement, but we have not done it because it is quite obvious). 

This example clearly shows that what we get from a measurement depends on the measurement itself. We have explained an experiment to measure the velocity of the particle. In fact, the Bohmian particle was at rest (with zero velocity) before we start the measurement. However, because of the type of measurement that we have designed, the Bohmian particle acquires a non-zero velocity. What we get from the measurement is exactly this non-zero value of the velocity, but this is not the pre-existing velocity of the particle before the measurement. Contrarily to a classical measurement, in the quantum world, in general, we cannot get information of a quantum system from a  measurement without providing some type of perturbation on it. This property of quantum phenomena is named contextuality. 

\enlargethispage{13pt}
\subsubsection{Example: Sequential Bohmian measurement of the transmitted and reflected particles}
\label{om.measure_transmission}

\begin{figure}
\centering
\includegraphics[width=0.47\columnwidth]{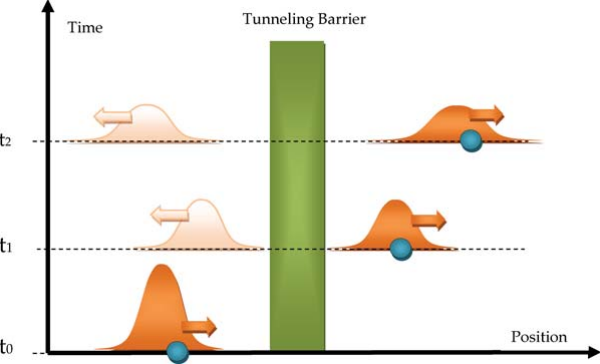}
\caption{Schematic explanation of the ability of Bohmian mechanics to discuss the
unitary and nonunitary evolution of a wave packet incident upon a
tunneling barrier for three different times $t_0 < t_1 < t_2$.}
\label{om_sequential}\vspace*{-6pt}
\end{figure}

We provide another didactic example of how Bohmian
mechanics explains sequential measurements. The wave packets in
\fref{om_sequential} represent the solution of the (unitary)
Schr\"odinger equation for a wave packet incident upon a tunneling
barrier, at three different times. The initial wave packet (with norm
equal to one) is divided into a transmitted plus a reflected wave
packet. According to the Copenhagen interpretation, when the system
is measured at time $t_1$, a nonunitary evolution appears in the
wave function. Thus, randomly, for example the reflected wave packet disappears.
Only the transmitted wave packet describes the electron at time
$t_1$. Then, when the system is measured again at $t_2$, the
electron is only represented by the transmitted wave packet.\enlargethispage{13pt}

Alternatively, the same evolution can be
explained with Bohmian mechanics. The initial position of the
Bohmian trajectory is selected according to the quantum equilibrium hypothesis at the initial time $t_0$.
Let us assume that we select an initial position that corresponds to a trajectory that is finally transmitted. Then, at times $t_1$ and $t_2$, the evolution of the trajectory is
only determined by the transmitted wave packet because the reflected wave
packet is an ``empty wave'' that has no effect on the evolution of
the trajectory.  Here, we have implicitly assumed that the
eigenstates of the measuring apparatus are transmitted wave packets
for positive currents, and reflected ones for negative currents.

As expected, the probability of measuring, first, the particle as
being transmitted at time $t_1$ and measuring, after, the same
particle as being reflected at time $t_2$ is zero either with
orthodox or Bohmian mechanics.

\begin{figure}
\centering
\includegraphics[width=0.57\columnwidth]{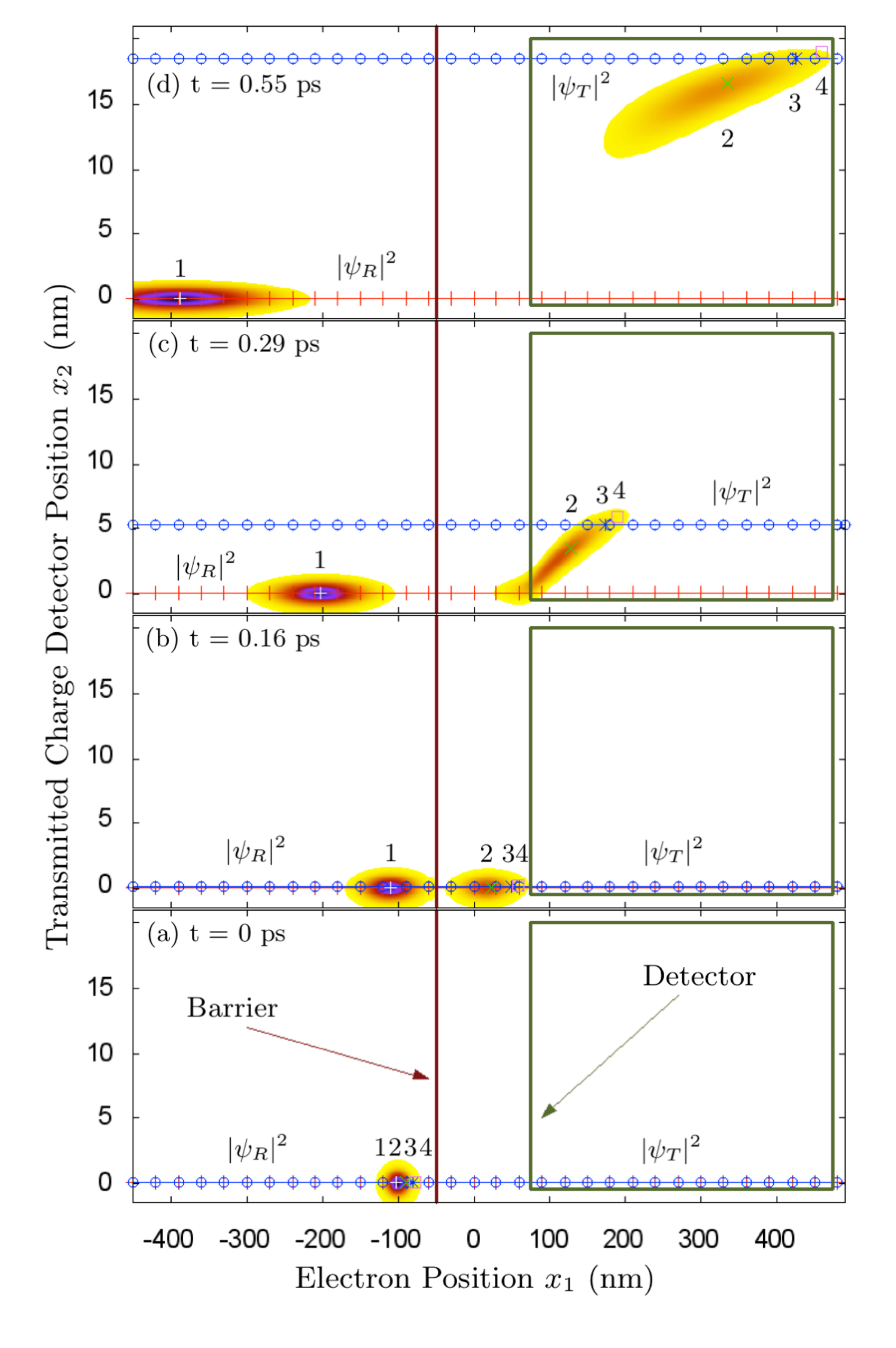}
\caption{Time evolution of the squared modulus of $\Psi(x_1,x_2,t)$ at four different times. The configuration space region where the \emph{transmitted charge detector} is present is indicated by a rectangle and the barrier by a solid line. The $+$ line indicates the squared modulus of the conditional wave function $|\psi_R|^2=|\Psi(x_1,x_2^{\alpha=1}[t],t)|^2$, while the  $\odot$ line corresponds to $|\psi_T|^2=|\Psi(x_1,x_2^{\alpha=3}[t],t)|^2$. Four trajectories $\{x_1^\alpha[t],x_2^\alpha[t]\}$ with different initial positions are presented with $[]$, $*$, $\times$ and $+$. The actual detector position associated with the reflected trajectory ($+$) with $\alpha=1$ does not move because there is no interaction between this trajectory and the detector. Reprinted with permission from \cite{om.albareda}. Copyright 2013 Springer Nature.}
\label{conditional wave function-figure1}
\end{figure}

Again, the reader can argue that we have not enlarged the system including the measuring apparatus. Next, we provide a more formal description of the Bohmian measurement of the  charge for a tunneling electron including the quantum system plus the measuring apparatus.  The quantum system is defined by an electron labeled as $x_1[t]$ impinging on a
 tunneling barrier. Behind the barrier there is a measuring device named the ``transmitted charge detector" and modeled as a single degree of freedom $x_2[t]$, which can detect the successful transmission of an electron. The dynamic of the system plus apparatus includes, first, an interaction of the electron with the potential barrier and, subsequently, an interaction with the transmitted charge detector. It is important to stress that both interactions are regarded at the very same level within Bohmian mechanics. The measurement interaction introduces a channelization of the wave function in the configuration space such that the desired property of the quantum system (here, whether the electron is reflected or transmitted) can be read off from the final position $x_2[t]$ of a particle, which plays the role of the pointer of the apparatus. 
The interaction between the electron and the pointer can be modeled as:
\begin{eqnarray}
H_{int}= \lambda Q(x_1) P_{x_2} = -i \hbar\lambda Q(x_1) \frac {\partial } {\partial x_2},
\label{eq-interaction2} 
\end{eqnarray}
where $P_{x_2} = -i\hbar \partial /\partial x_2 $ is the momentum operator of the detector and $\lambda =50\; nm/ps $ is the interaction constant. $Q(x_1)$ is a function that is equal to zero when the electron is outside the detector, ($x_1 < 75 \; nm$ in \fref{conditional wave function-figure1}), and is equal to one when the particle is inside the detector ($x_1 > 75 \; nm$).\footnote{The transition of $Q(x_1)$, from zero to one, is softly done in order to minimize the perturbation of the quantum system as explained in Ref. \cite{om.albareda,om.2marian}.}  In \fref{conditional wave function-figure1} the region in the configuration space in which $Q(x_1)$ is different from zero is represented by a rectangle in the right of the figure.
The many-particle Schr\"odinger equation reads

\begin{eqnarray}
 i \hbar \frac{\partial \Psi(x_1,x_2,t)}{\partial t}=
 \Big(- \frac{\hbar^2}{2m}\frac {\partial^2} {\partial x_1^2}  - \frac{\hbar^2}{2 M}\frac {\partial^2} {\partial x_2^2}   
 + U(x_1) - i \hbar \lambda Q(x_1) \frac {\partial} {\partial x_2}\Big) \Psi(x_1,x_2,t), \nonumber\\
\label{eq-schr2D}
\end{eqnarray}
where $m$ is the effective mass of the electron, $M$ is the mass of
the apparatus pointer and $U(x_1)$ is the external potential energy barrier seen as a vertical line in the middled of \fref{conditional wave function-figure1}.

The main feature of a transmitted charge detector is that the center of mass of the wave function in the $x_2$ direction has to move if the
electron is transmitted and it has to be at rest if the electron is reflected.  We solve \eref{eq-schr2D} numerically considering as initial wave function the products of two gaussian wave packets, i.e. $\Psi(x_1,x_2,0) = \psi(x_1,0)\phi(x_2,0)$. All details of this simulation can be found in Ref. \cite{om.albareda,om.2marian}. In particular we are considering $M=75000 \; m$. In \fref{conditional wave function-figure1} the numerical solution of the squared modulus of $\Psi(x_1,x_2,t)$ is plotted at four different times. At the initial time $t = 0$,  \fref{conditional wave function-figure1}~(a),  the entire wave function is at the left of the barrier.  At a later time $t_0$ the wave function has split up into reflected and transmitted parts due to the barrier, see \fref{conditional wave function-figure1}~(b).  Then, because the electron has not yet arrived at the transmitted charge detector, the wave function has the following form: 

\begin{equation}
\Psi(x_1,x_2,t_0) = \left[\psi_{T}(x_1,t_0)+\psi_{R}(x_1,t_0)\right]\phi(x_2,t_0).
\end{equation}

After that, Figs. \ref{conditional wave function-figure1} (c) and (d), the
interaction of the detector with the transmitted part of the wave function appears. For time $t>t_0$ the transmitted part of the wave
function is shifted up in the $x_2$ direction while the reflected part does not move. The interaction with the apparatus thus produces 
two channels in the configuration space, one corresponding to the electron being transmitted and the other corresponding to the electron being
reflected, getting an entangled superposition among the electron and the apparatus.

\begin{figure}
\centering
\includegraphics[width=0.57\columnwidth]{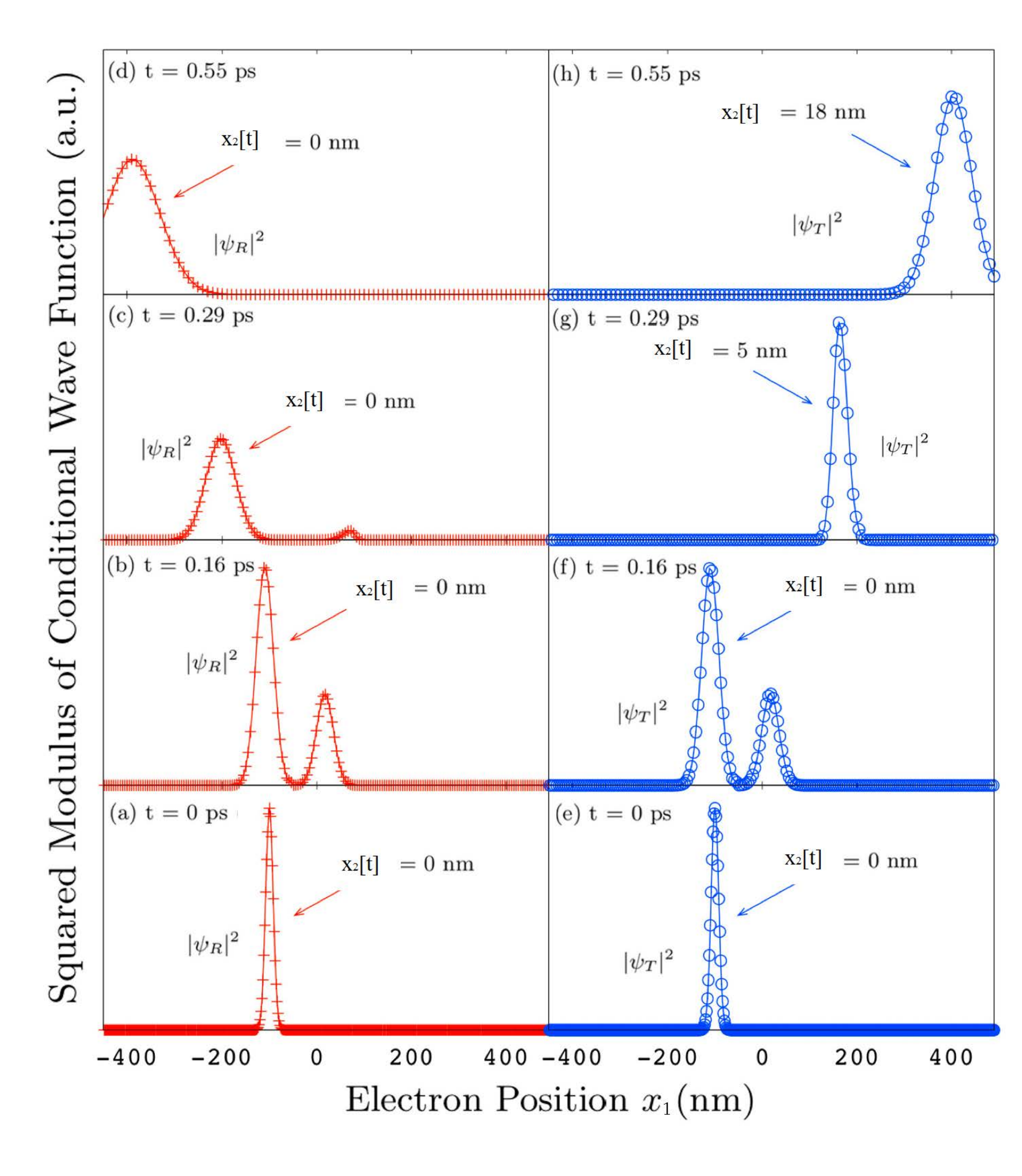}
\caption{ The $+$ line in (a), (b), (c) and (d) is the time evolution of the squared modulus of the conditional wave function associated to the trajectory $\alpha = 1$ in \fref{conditional wave function-figure1}, i.e. $\psi_{R} = |\Psi(x_1,x_2^{\alpha=1}[t],t)|$. The $\odot$ line in (e), (f), (g) and (h) is the squared modulus of the conditional wave function associated to the trajectory $\alpha = 3$ in \fref{conditional wave function-figure1}. i.e. $\psi_{T} = |\Psi(x_1,x_2^{\alpha=3}[t],t)|$. The actual detector position $x_2[t]$ is plotted at each time in order to compare these results with those in \fref{conditional wave function-figure1}. Reprinted with permission from \cite{om.albareda}. Copyright 2013 Springer Nature.}
\label{conditional wave function-figure2}
\end{figure}

In \fref{conditional wave function-figure1} we also plot the actual positions of the system and detector $\{x_1[t],x_2[t]\}$ for four different possible initial positions $\{x_1[0], x_2[0]\}$, corresponding to four distinct runs of the experiment (labelled by $\alpha=1,...,4$). Of the four possible evolutions shown, three show the electron being transmitted ($\alpha=2,3,4$) and one being reflected ($\alpha=1$). While the pointer position $x_2[t]$ does not move for the reflected particle, its evolution for the transmitted ones clearly shows a movement. In conclusion, looking at the \emph{detector} position we can perfectly certify if the particle has been reflected ($x_1[t]< -50\;nm$ and $x_2[t]=0\;nm $) or transmitted ($x_1[t]> -50\;nm$ and $x_2[t] \approx 15\;nm $). We hope the reader will realize how trivially we have been
able to explain the measurement, using only a \emph{channelized} (unitary) time-evolution of 2D wave function plus two Bohmian trajectories in the physical space, one trajectory for the system and another for the measuring apparatus. 

Once we have solved the complete problem of the measurement in the (2D) configuration space, we can describe the same measurement in (1D) physical space with the help of the \emph{conditional wave function}. The key point illustrated here is that the collapse of the one-particle wave function for the electron arises naturally and automatically in Bohmian mechanics.  It is simply a consequence of slicing the unitary-evolving (2D) wave function $\Psi$ along the (moving) line $x_2 = x_2[t]$, resulting $\psi_1(x_1,t)=\Psi(x_1,x_2[t],t)$. In \fref{conditional wave function-figure1} we have plotted two solid horizontal lines corresponding to a slice of the wave function at two different values of $x_2[t]$. In \fref{conditional wave function-figure2} we report the evolution of these (time-dependent) slices of the many-particle wave function, the \emph{conditional wave function} for the electron, for the trajectories $\alpha = 1$ and $\alpha = 3$ from \fref{conditional wave function-figure1}. We clearly see that if the particle is reflected, as it is the case for $\alpha = 1$, the position of the pointer does not change with time and, after the interaction with the detector has been performed, the electron's \emph{conditional wave function} includes only a reflected part. See Figs. \ref{conditional wave function-figure2} (c) and (d). On the other hand, when the particle is transmitted (e.g., $\alpha =3$), it is the reflected part of the \emph{conditional wave function} which collapses away, leaving only the transmitted packet.  See Figs. \ref{conditional wave  function-figure2} (g) and (h). Note that the evolution of $\psi_1(x_1,t)$ (the electron's \emph{conditional wave function}) is not unitary, even though the evolution of $\Psi$ is. 

While a wave function formulation of quantum mechanics provides only statistical information about the experimental results, with the help of the Bohmian trajectories, we have been able to recover the individual result of each experiment. In fact for each experiment the pointer of the detector is either moving (corresponding to a transmitted electron) or not (reflected electron), while an ensemble of repeated experiments (where the initial positions of the particles, both the electron $x_1[0]$ and the detector $x_2[0]$, are selected according to the squared modulus of the wave function at the initial time $|\Psi(x_1,x_2,0)|^2$) reproduce the same statistical results.

Finally, by looking at figures \ref{conditional wave function-figure1} and \ref{conditional wave  function-figure2}, we conclude that the Bohmian theory provides a quite simple explanation on when, why and how the (conditional) wave function collapses. The Bohmian explanation of the measurement, with waves and particles, demystifies the orthodox postulate of the collapse law by explaining it as a trivial result of making a transition from a large configuration space (quantum system plus apparatus) to a smaller one (quantum system alone). To be fair, in spite of its extraordinary simplicity to explain  the collapse law, the model explained here for measuring the transmitted charge has still an important approximation. We have assumed that the pointer is just one degree of freedom $x_2[t]$. As we have already mentioned, a realistic model for the pointer has to involve a very large number of particles (on the order of Avogadro's number $\approx 10^{23}$). Then, we will also be able to conclude that the macroscopic pointer has a classical behavior. However, such type of simulation involves facing again with the many body problem mentioned in \sref{om.sec_many.1}. Some preliminary discussions on the classical-quantum transition for macroscopic objects can be found in Refs. \cite{om.clasic1,om.clasic2,om.4marian,om.3xavier,om.1xavier}.

\subsection{The evaluation of a mean value in terms of Hermitian operators}
\label{om.sec_measurement.2}

\subsubsection{Why Hermitian operators in Bohmian mechanics?}

In the previous section, we have emphasized the ability of Bohmian
mechanics to address quantum phenomena just from the
Schr\"odinger equation and the equation of motion of the
trajectories. The operators, which are an indispensable tool in the
orthodox formulation of quantum mechanics to deal with  the measurement
process, become unnecessary in the Bohmian formulation. Once one
realizes that operators are unnecessary, the conceptual difficulties
associated with the wave function collapse of the orthodox
formulation disappear (as we have seen in passing from \fref{conditional wave function-figure1} to \fref{conditional wave function-figure2}). However, we have mentioned that the Bohmian measurement process explained above (with an idealized single-particle pointer) has limited practical utility. Including the measuring apparatus into the entire
Hamiltonian is not an easy task, because it implies increasing the
number of degrees of freedom. The exact Bohmian simulation of the measurement
process is, most of the time, a quite difficult computational task.

In these circumstances, the use of a Hermitian operator acting only
on the wave function of the quantum system with the ability of
providing the outcomes of the measurement process without the
explicit simulation of the measuring apparatus is very welcomed. Let
us emphasize that we are talking only at the computational level. In
simple words, operators are not needed in Bohmian mechanics (the operators are part of the Copenhagen ontology, but they are not part of the Bohmian ontology), but
they are a very helpful mathematical trick in practical
computational issues. These ideas are emphasized by Goldstein,
D\"{u}rr, Teufel, and coworkers when they refer to the ``naive
realism about operators'' \cite{om.Durrnaive,om.Durrllibre,om.goldstein}. Somehow, we are arguing here that the Copenhagen theory can be transformed into an effective model of the Bohmian theory where trajectories are deliberatively ignored  for computational/practical  proposals. Then, although we have a computational ignorance about the position of such electron, the idea of an (unmeasured) electron traversing a tunneling barrier would be fully supported by such effective quantum theory.

\subsubsection{Mean value from the list of outcomes and their probabilities}

We now want to reproduce the mean value, obtained from orthodox Hermitian
operators, within Bohmian mechanics. Someone who dislikes what we
have explained up to here about the Bohmian measurement can just
focus on how Bohmian trajectories exactly reproduce the mean value of orthodox
Hermitian operators. For simplicity, we will consider
single-particle systems. The generalization to many-particle systems
can be done straightforwardly.

First, according to the orthodox measurement, we assume that a particular experiment is described by the quantum system wave function $\psi(x,t)$ alone. After a large number, $M$, of repetitions of the same experiment, we can elaborate a large list $i = 1\ldots,I$ of the possible outcomes $\{a_i\}$ with their occurrence probabilities $\{P_i\}$. From this data we can compute the mean value as:
 \begin{equation}
 \label{om.meanvalue_probability}
 \avg{\hat{A}}_{\psi} = \sum_{i = 1}^{I}a_i P_i
 \end{equation}
This is the standard (classical or quantum) definition of the mean value.

\subsubsection{Mean value from the wave function and the operators}

According to the orthodox quantum theory, an observable is associated with a Hermitian operator $\hat{A}$ that describes the measurement process.
Such an operator fulfills $\hat{A} \ket{u_i} = a_i\ket{u_i}$, $\{\ket{u_i}\}$ being its eigenstates.
Therefore, $P_i = \braket{\psi}{u_i}\braket{u_i}{\psi}$ so that we can write:
 \begin{equation}
 \label{om.meanvalue_0bservable}
 \avg{\hat{A}}_{\psi} = \bra{\psi} \left( \sum_{i = 1}^{I}a_i \ket{u_i}\bra{u_i} \right) \ket{\psi} = \matrixel{\psi}{\hat{A}}{\psi}
\end{equation}
where we have used $a_i\ket{u_i} = \hat{A} \ket{u_i}$ and we have
identified the expression $\sum_{i = 1}^{I} \ket{u_i}\bra{u_i}$ as
the identity because of the orthonormality of the eigenstates.
Therefore, we can compute the mean value of an ensemble of
experiments from \eref{om.meanvalue_0bservable} by only knowing the
wave function and the operator (without knowing either the
eigenstates or the outcome values and their probabilities).  Thus, we have been able to compute
mean values within the orthodox quantum interpretation without
directly mentioning the wave function collapse.

\subsubsection{Mean value from  Bohmian mechanics in the position representation}

We can always write the Hermitian operator $\hat{A}$ and the mean value $\avg{\hat{A}}_{\psi}$ in the position representation. Then, the mean value of this operator over the wave function $\psi(x,t)$ is given by:
\begin{equation}
\label{om.orthodox_mean_value}
\avg{\hat{A}}_{\psi} = \int_{-\infty}^{\infty} \psi^{*}(x,t) \hat{A} \left( x,-i \hbar \frac {\partial} {\partial x} \right) \psi(x,t) dx
\end{equation}
Alternatively, the same mean value can be computed from Bohmian
mechanics by defining a spatial average of a ``local'' magnitude
$A_B(x)$ weighted by $R^2(x,t)$:
\begin{equation}
\label{om.Bohm_mean_value}
\avg{\hat{A}}_{\psi} = \int_{-\infty}^{\infty} R^{2}(x,t) A_B(x) dx
\end{equation}
In order to obtain the same value with Eqs. (\ref{om.orthodox_mean_value}) and (\ref{om.Bohm_mean_value}), we can easily identify the local mean value $A_B(x)$ as:
\begin{equation}
\label{om.local_Bohm_mean_value}
A_B(x) = Real \left( \left[\frac {\psi^{*}(x,t) \hat{A} \left( x,-i\hbar \frac {\partial} {\partial x} \right) \psi(x,t)} {\psi^{*}(x,t) \psi(x,t)} \right]_{\psi(x,t) = R(x,t) e^{i \frac{S(x,t)} {\hbar}}} \right)
\end{equation}
We take only the real part, $Real()$, because we know that the mean
value must be real, and \eref{om.local_Bohm_mean_value} without
$Real()$ could take complex values.

\subsubsection{Mean value from Bohmian trajectories}

For practical purposes, we will compute the mean value using
\eref{om.Bohm_mean_value} with a large $j = 1,\ldots,M$ number of
Bohmian trajectories with different initial positions. We will
select the initial position $x^j[t_0]$ of the Bohmian trajectories
according to the second postulate. Therefore, we can use
\eref{om.sum_0f_particles_to} to write $R^2(x,t)$ in
\eref{om.Bohm_mean_value}. Finally, we obtain:
\begin{equation}
\label{om.meanvalue_discrete}
\avg{\hat{A}}_{\psi}=\lim_{M\rightarrow\infty} \frac {1} {M} \sum_{j=1}^{M} A_B(x^j[t]).
\end{equation}
By construction, in the limit $M\rightarrow\infty$, the value of \eref{om.meanvalue_discrete} is identical to the value of \eref{om.Bohm_mean_value}.

Now, we provide several examples of how some common mean values are
computed from the orthodox quantum formalism and from Bohmian
trajectories. First, we compute the mean value of the position:
\begin{equation}
\label{om.position_mean_0}
\avg{x}_{\psi} = \int_{-\infty}^{\infty} \psi^{*}(x,t) x \psi(x,t) dx
\end{equation}
with $x_B(x) = x$ so that:
\begin{equation}
\label{om.position_mean_B}
\avg{x}_{\psi} = \int_{-\infty}^{\infty} R^2(x,t) x dx
\end{equation}
Identically, the mean value of the momentum:
\begin{equation}
\label{om.momentum_mean_0}
\avg{p}_{\psi} = \int_{-\infty}^{\infty} \psi^{*}(x,t) \left(-i\hbar \frac {\partial} {\partial x}\right) \psi(x,t) dx
\end{equation}
with $p_B(x) = {\partial S(x,t)}/{\partial x}$:
\begin{equation}
\label{om.momentum_mean_B}
\avg{p}_{\psi} = \int_{-\infty}^{\infty} R^2(x,t) \frac {\partial S(x,t)} {\partial x} dx
\end{equation}
For the classical potential, we have:
\begin{equation}
\label{om.Potential_energy_mean_0}
\avg{V}_{\psi} = \int_{-\infty}^{\infty} \psi^{*}(x,t) V(x,t) \psi(x,t) dx
\end{equation}
with $V_B(x,t) = V(x,t)$ so that:
\begin{equation}
\label{om.Potential_energy_mean_B}
\avg{V}_{\psi} = \int_{-\infty}^{\infty} R^2(x,t) V(x,t) dx
\end{equation}
Now, we compute the mean value of the kinetic energy:
\begin{equation}
\label{om.kinetic_energy_mean_0}
\avg{K}_{\psi} = \int_{-\infty}^{\infty} \psi^{*}(x,t) \left(-\frac {\hbar^2} {2m} \frac {\partial^2} {\partial x^2}\right) \psi(x,t) dx
\end{equation}
It is important to notice that the local mean value of the kinetic
energy takes into account the Bohmian kinetic energy plus the
quantum potential. In particular, $K_B(x)$ can be obtained from the
expression:
\begin{equation}
\label{om.local_kinetic_energy_mean_B}
K_B(x) = Real \left( -\frac {R(x,t) e^{-iS(x,t)/\hbar} \frac{\hbar^2} {2m} \left( \frac {\partial} {\partial x} \right)^2 R(x,t) e^{iS(x,t)/\hbar}} {R^2(x,t)} \right)
\end{equation}
The real part\footnote{It can be demonstrated quite easily that the imaginary part of \eref{om.local_kinetic_energy_mean_B} is equal to the spatial derivative of the current that becomes zero when integrated over all space. We use $J(x = \pm\infty,t) = 0$, which is always valid for wave functions that are normalized to unity, but it is not true for other types of wave functions such as plane waves.} of $K_B$ is:
\begin{equation}
\label{om.local_kinetic_energy_mean_B_bis}
K_B = \frac {1} {2m}\left(\frac {\partial S(x,t)} {\partial x} \right)^2 + Q(x,t)
\end{equation}
so that, finally, we obtain the Bohmian expression of the mean kinetic energy of the ensemble of trajectories:
\begin{equation}
\label{om.kinetic_energy_mean_B}
\avg{K}_{\psi} = \int_{-\infty}^{\infty} R^2(x,t) \left( \frac {1} {2m} \left(\frac {\partial S(x,t)} {\partial x} \right)^2 + Q(x,t)\right) dx
\end{equation}
See problem \ref{om.P4} for a detailed calculation and the explanation of the meaning of the two terms.

Finally, we compute the mean value of the current density operator. First, let us remark that the probability density operator can be written as $\ket{x}\bra{x}$ and its expected mean value is $\braket{\psi}{x} \braket{x}{\psi} = \abs{\psi(x,t)}^2$ or, in the Bohmian language, $\braket{\psi}{x} \braket{x}{\psi} = R^2(x,t)$. The (Hermitian) current operator can be written as $\hat{J} = 1/(2m) (\ket{x}\bra{x}\hat{p} + \hat{p}\ket{x}\bra{x})$. It can be easily demonstrated that:
\begin{equation}
\avg{J}_{\psi}=J(x,t)=v(x,t) R^2(x,t)=\lim_{M\rightarrow\infty} \frac {1} {M} \sum_{j=1}^{M} v(x^j[t]) \delta(x-x^j[t]).
\end{equation}
The average value of the current density depends on the position,
and it is equal to the average Bohmian velocity multiplied by the
squared modulus of $R(x,t)$. At a particular position ``$x$,'' this
current is just the sum of all particles that reside arround this
position $x = x^j[t]$ at time $t$, multiplied by the Bohmian velocity at such position and time. 

\subsubsection{On the meaning of local Bohmian operators \textit{A}$_\textit{B}$(\textit{x})}

It is important to emphasize that the local Bohmian operators
$A_B(x)$ are not the eigenvalues of the operator $\hat{A}$. In
general, the eigenvalues are not position dependent, while $A_B(x)$
are. The expression $A_B(x)$ is what is needed to compute the mean
values of $\hat{A}$ with \eref{om.Bohm_mean_value}. This is its
exact meaning. We have encapsulated all the effects of the measuring apparatus on the system into $A_B(x)$. Somehow, the use of $A_B(x)$ seems to emphasize that we are measuring a pre-existent property of the system, while we have learnt during this section devoted to the measurement that an interaction between system and apparatus during the time taken for doing the measurement (see \fref{conditional wave function-figure1}) is mandatory. 

There is a well-known example, mentioned in the previous
\sref{om.measure_moment}, that emphasizes the differences between
eigenvalues and local Bohmian operators. We consider the particle
between two walls, separated by a distance $L$, whose wave function
is $\psi(x) = C \; \sin(n\pi x/L)$ within the walls and zero
elsewhere. Since the wave function is real, we obtain $p = \partial
S(x)/\partial x = 0$. Thus, the local Bohmian momentum is $p_B(x) =
{\partial S(x,t)}/{\partial x} = 0$, and the mean value in
\eref{om.momentum_mean_B} is also zero. On the contrary, the wave
function can be written as:
\begin{equation}
\psi(x) \approx C' \left(e^{-i\pi x/L}-e^{i\pi x/L} \right)
\end{equation}
We know that $e^{i \; n \pi x/L}$ and $e^{-i \; n \pi x/L}$ are
eigenstates of the momentum operator with eigenvalues $a_1 = n \hbar
\pi/L$ and $a_2 = -n \hbar \pi/L$, respectively. Thus, the
probabilities for each of these eigenvalues are identical, and the
mean value of the momentum from \eref{om.meanvalue_probability} is
again zero.

In conclusion, in general, $A_B(x)$ cannot be identified with $a_i$.
However, by construction, the mean values computed from $a_i$ and
$A_B(x)$ are identical. A similar discussion about the evaluation of
the mean value of the  quantum power can be found in Ref.
\cite{om.quantumpower}.

In this \sref{om.sec_measurement.2}, we have only discussed how the
mean values (i.e., the first moment) obtained from orthodox quantum
mechanics can be exactly reproduced within Bohmian mechanics. More
complicated higher moments (e.g., the second moment called the
variance, whose positive square root is the standard deviation)
computed from orthodox quantum mechanics can be equivalently
reproduced within Bohmian mechanics, as we anticipated in
\sref{om.measure_transmission} \cite{om.Durrnaive,om.Durrllibre,
om.goldstein}. With the appropriate care on the contextuality of quantum phenomena,  two-time correlations functions can also be computed in the same way. See also Ref. \cite{om.Muga}.

\section{Concluding Remarks}

In this chapter we have provided a fully comprehensive and didactic overview of the Bohmian formulation of quantum mechanics that can be useful for understanding the following chapters and also for any newcomers interested in using Bohmian mechanics in their daily research activity.

To be fair, most of the research on quantum phenomena is developed
with the orthodox machinery, and few significative contributions
have been performed with Bohmian mechanics. Many researchers ignore
Bohmian mechanics, or they decide not to use it because they believe
that someone demonstrated that there is something strange about
Bohmian trajectories. In the introduction of this book, we illustrated
this situation using Hans Christian Andersen's tale ``The Emperor's
New Clothes.'' The tale explains how a naked emperor strolling in
procession on the streets believes he is well dressed, until a small
child cries out, ``The emperor is naked!'' Children tales end
happily with a moral that pretends to improve their behavior. Almost
a century has passed, and we are still somewhere in the middle of
the Bohmian tale. Bohm and Bell were the first to show to the
scientific community that ``Bohmian mechanics is a correct
formulation of quantum phenomena.'' As in the tale and despite the
whispering among the townsfolk, it seems that the idea that the
orthodox formulation is the only licit tool to analyze quantum
phenomena has gone on and the ``orthodox emperor'' continues to walk
more proudly than ever.

It is still not clear what ending awaits the Bohmian tale and what
moral it will teach us. In our opinion, Bohmian mechanics will only
be able to have a happy ending once it can be explicitly
demonstrated that a particular  meaningful quantum problem is better calculated
(in terms of computational abilities) or formulated within Bohmian
mechanics than within its orthodox counterpart. Perhaps, some new experiments will allow us to elucidate which one better captures the reality underneath. 
As we have mentioned in the section "What Is a Quantum Theory" in the introduction, even if we could demonstrate in the future that either the Copenhagen or the Bohmian theories is wrong (or both), the practical utility of these theories in their range of validity would not dismiss. 

We hope that a happy ending for the Bohmian theory is not very far and that this chapter (and the entire book) can contribute to shorten the time needed to arrive to it. If so, we might learn a moral similar to the one deduced from Andersen's tale: ``It is useful to look for answers outside of the main stream.'' Sometimes the scientific revolutions are hidden there.\vspace*{-9pt}\\

\section*{Acknowledgments}

This work has been partially supported by the FEDER and  ``Ministerio de Ciencia e Innovaci\'{o}n'' through the Spanish Project TEC2015-67462-C2-1-R,  the Generalitat de Catalunya (2014 SGR-384),  the European Union’s Horizon 2020 Grant agreement no: 604391 of the Flagship initiative  ``Graphene-Based Revolutions in ICT and Beyond'' and European Union’s Horizon 2020 research and innovation programme under the Marie Skłodowska-Curie grant agreement No 765426 (TeraApps).  We want to thank Fabio L. Traversa and Guillermo Albareda for their help with the discussion of Bohmian
computational algorithms. We also want to thank Alfonso Alarc\'{o}n and
Albert Benseny for their help in editing and reviewing much of this
chapter and Enrique Colom\'{e}s for the design and numerical simulations of the cover. Finally, we want to sincerely thank Ram\'on Corbal\'an,
Sheldon Goldstein, Basile Hiley, Hrvoje Nikoli\'c, Ward Struyve, Travis Norsen, Nino Zangh\`i, Roderick Tumulka, Damiano Marian, Adriano Orefice,
and David Peat for very fruitful discussions.\vspace*{-9pt}\\

\section{Problems and solutions}

\begin{problem} \label{om.p1}
Check that \eref{om.HJ1} is fulfilled for a free particle with
$V(x,t) = 0$. We define the initial time as $t_0 = 0$ and the initial
position as $x_0 = 0$.
\end{problem}

\begin{solution}
For a free particle, the Lagrangian is just the kinetic energy with
a constant velocity, $v[t] = v_0$. We have $S_{\rm p} = K(v_0) t_f$
so that $S_{\rm p} = m {x_f}^2/(2 t_f)$ and $S_{{\rm p}-\delta t} =
m {x_f}^2/( 2 (t_f + \delta t))$. Notice that modifying the final
time, without modifying the final position, means changing the
initial velocity. If we use the Taylor expansion  $1/(1 + x)\simeq
1-x$, then we find $S_{{\rm p} - \delta t} = m {x_f}^2 \; (1 -
\delta t/t_f)/(2 t_f)$. Finally, $S_{{\rm p} - \delta t} - S_{\rm p}
= -m {x_f}^2 \; \delta t/(2{t_f}^2)$. This is the kinetic energy
with a negative sign that, in our simple system, is also the
Hamiltonian with a negative sign. Therefore, \eref{om.HJ1} is
exactly recovered.
\end{solution}

\begin{problem} \label{om.p2}
Following the example of problem \ref{om.p1}, test \eref{om.HJ2}.
\end{problem}

\begin{solution}
The new value will be $S_{{\rm p} - \delta x} = m (x_f + \delta
x)^2/(2 t_f)$. If we neglect $(\delta x)^2$, we obtain $S_{{\rm
p} - \delta x} = m {x_f}^2/(2 t_f) + m {x_f} \delta x/t_f$. So that
$S_{{\rm p} - \delta x} - S_{\rm p} = m \; v \delta \; x$. Finally,
\eref{om.HJ2} is recovered.
\end{solution}

\begin{problem} \label{om.p3}
Solve \eref{om.hamilton_jacobi1D} for the free particle case developed in problems \ref{om.p1} and \ref{om.p2} to find $S(x_0,t_0;x,t)$.
\end{problem}

\begin{solution}
Equation \eref{om.hamilton_jacobi1D} can be rewritten as ${\partial S(x,t)}/{\partial t} = -H(x,t) = -E(v_0)$. The system is conservative, and the value of the Hamiltonian at any time is equal to the initial kinetic energy $E(v_0) = m v_0^2/2$. Therefore, we can transform \eref{om.hamilton_jacobi1D} into $(\partial S(x,t)/\partial x)^2/(2 m) = E(v_0)$ so that $\partial S/\partial x = \pm \sqrt{2 m E(v_0)} = m v_0$. If we assume that the initial velocity (energy) is independent of the initial (final) position, then we easily integrate the partial derivatives of $S$ to obtain finally, $S(x,t) = \pm m v_0 x - E(v_0) t$.

As expected, this action function corresponds to trajectories with
constant velocity $\pm v_0$, that is, $x[t] = x_0 \pm v_0 \; t$. The
important point of our calculation here is that we have found all
trajectories with an arbitrary initial position $x_0$ but a fixed
initial velocity $\partial S/\partial x = \pm m v_0$. Let us
emphasize that, although the classical function $S$ is multivalued,
if the initial velocity was $ + v_0$, then we can use this knowledge  and keep always the positive velocity for such
trajectories.
\end{solution}

\begin{problem} \label{om.p3bis}
Show that the two real Eqs. (\ref{om.hamilton_jacobi1D}) and (\ref{om.conservation_law1D}) are equivalent to the following complex classical wave equation, also written in \eref{om.classcho}:
\begin{eqnarray}
\label{om.p.classcho}
i \hbar \frac{ \partial \psi_{cl}(x,t)} {\partial t} &=& -\frac {\hbar^2}{2m} \frac{ {\partial}^2 \psi_{cl}(x,t)} {\partial x^2} + V(x,t) \psi_{cl}(x,t)+ \frac {\hbar^{2}} {2 m} \frac { {\partial}^2 |\psi_{cl}(x,t)|/ \partial x^2}{|\psi_{cl}(x,t)|} \psi_{cl}(x,t)\nonumber\\
\end{eqnarray}
\end{problem}

\begin{solution}
We start the formal demonstration by introducing the polar form of the wave function, $\psi_{cl}(x,t) = R(x,t) exp(i S(x,t)/\hbar)$, into \eref{om.classcho}. We have explicitly included the time in the scalar potential of \eref{om.classcho} in order to compare it with the Schr\"odinger equation. Then, we evaluate the temporal and spatial derivatives:
\begin{eqnarray}
\label{om.derivades_classcho1}
\frac {\partial} {\partial t} \left(Re^{i S/\hbar}\right)& = &\left( \frac {\partial R} {\partial t} + \frac {i} {\hbar} R\frac{\partial S} {\partial t} \right)e^{i S/\hbar}\\
\label{om.derivades_classcho2}
\frac {\partial^2} {\partial x^2} \left(Re^{i S/\hbar}\right) & = &\!\Bigg( \frac {\partial^2 R} {\partial x^2} \,{+}\, 2 \frac {i} {\hbar} \frac{\partial R} {\partial x} \frac{\partial S} {\partial x} \,{+}\, \frac {i} {\hbar} R \frac{\partial^2 S} {\partial x^2} \,{+}\, \frac {-1} {\hbar^2} R \left( \frac{\partial S} {\partial x}\right)^2\! \Bigg) e^{i S/\hbar}\nonumber\\
\end{eqnarray}
Now, if we introduce Eqs. (\ref{om.derivades_classcho1}) and
(\ref{om.derivades_classcho2}) into \eref{om.classcho} and
eliminate the factor $e^{i S/\hbar}$, which is present in all terms,
we obtain:
\begin{eqnarray}
\label{om.classcho_des}
i \hbar \frac {\partial R} {\partial t} - R\frac{\partial S}
{\partial t} &=& -\frac {\hbar^2}{2m} \frac {\partial^2 R} {\partial
x^2} - i\frac {\hbar}{m} \frac{\partial R} {\partial x}
\frac{\partial S} {\partial x} - i \frac {\hbar}{2m} R
\frac{\partial^2 S} {\partial x^2} +\, \frac {1} {2m} R \left(
\frac{\partial S} {\partial x}\right)^2+ V \; R + \frac {\hbar^{2}} {2 m} \frac { {\partial}^2 R} {\partial x^2}\nonumber\\
\end{eqnarray}
The real part of \eref{om.classcho_des} is:
\begin{equation}
\label{om.classcho_real}
 -R\frac{\partial S} {\partial t} = \frac {1} {2m} R \left( \frac{\partial S} {\partial x}\right)^2 + V \; R
\end{equation}
which is exactly the Hamilton--Jacobi \eref{om.hamilton_jacobi1D}. 

On the other hand, the imaginary part of equation \eref{om.classcho_des} gives:
\begin{equation}	
\label{om.classcho_imag}
 \frac {\partial R} {\partial t} = -\frac {1}{m} \frac{\partial R} {\partial x} \frac{\partial S} {\partial x} - \frac {1}{2m} R \frac{\partial^2 S} {\partial x^2}
\end{equation}
Now, if we multiply both terms of \eref{om.classcho_imag} by $2 R$,
we directly recover the conservation law of
\eref{om.conservation_law1D}. Notice how the constant $\hbar$ does
not appear either in \eref{om.classcho_real} or in
\eref{om.classcho_imag}.
\end{solution}

\begin{problem} \label{om.p4}
Show that $\psi_{cl}(x,t) = exp(i (\pm p x - E t)/\hbar)$ is a solution of the classical wave equation of \eref{om.classcho} for the free-particle system with $V(x,t) = 0$.
\end{problem}

\begin{solution}
We compute the first two terms of the classical wave \eref{om.classcho}.
\begin{eqnarray}
\label{om.derivades_classcho1_bis}
{i} \hbar \frac {\partial } {\partial t} \left(e^{i S/\hbar}\right) &=& {i} \hbar \frac {i} {\hbar} E e^{i S/\hbar}\\
\label{om.derivades_classcho2_bis}
-\frac {\hbar^2}{2m} \frac {\partial^2} {\partial x^2} e^{i S/\hbar} &=& -\frac {\hbar^2}{2m} \left(\frac {i} {\hbar}\right)^2 p^2 e^{i S/\hbar}
\end{eqnarray}
Since the other two terms are zero, we obtain $E = p^2/(2\; m)$. Thus, $\psi_{cl}(x,t) = exp(i (\pm p x - E t)/\hbar)$ is a solution of the classical wave \eref{om.classcho}.

Surprisingly, a plane wave, apart from a free-particle quantum solution of the Schr\"odinger equation, is also a licit solution of a wave equation for classical mechanics. The fact that both classical and quantum solutions are identical is due to the fact that for the particular solution discussed here, $R = 1$, the last term of \eref{om.classcho}, that is, the quantum potential, is zero. Whether we want to model a free particle as a plane wave or as an ensemble of trajectories is just a matter of convenience.
\end{solution}

\begin{problem} \label{om.p5}
Show the following identity for the computation of the Bohmian velocity.
\begin{equation}
v(x,t) = \frac {1} {m} \frac {\partial S(x,t)} {\partial x} = \frac{J(x,t)} {|\psi(x,t)|^2}\nonumber
\end{equation}
where $J(x,t)$ is defined by \eref{om.current}.
\end{problem}

\begin{solution}
We start by performing the time derivative of $S(x,t)$:
\begin{eqnarray*}
\label{om.polar1D}
\frac {\partial S(x,t)} {\partial x} &=& \hbar \frac {\partial } {\partial x} \tan^{-1} \left( \frac {\psi_{i}(x,t)} {\psi_{r}(x,t)} \right)\\
\frac {\partial S(x,t)} {\partial x} &= &\hbar \frac { \frac {\partial} {\partial x} \frac {\psi_{i}(x,t)} {\psi_{r}(x,t)} } {1 + \left( \frac {\psi_{i}(x,t)} {\psi_{r}(x,t)} \right)^2} \\
\frac {\partial S(x,t)} {\partial x} &=& \hbar \frac { \psi_r(x,t) \frac {\partial \psi_i(x,t)} {\partial x} - \psi_i(x,t) \frac {\partial \psi_{r}(x,t)} {\partial x} } {{\psi_{r}^2(x,t)} + {\psi_{i}^2(x,t)} }
\end{eqnarray*}

Finally, using the following identity:
\begin{eqnarray*}
&&\psi(x,t) \frac {\partial \psi^{*}(x,t)} {\partial x} - \psi^{*}(x,t) \frac {\partial \psi(x,t)} {\partial x}= \frac {2} {i} \left( \psi_r(x,t) \frac {\partial \psi_i(x,t)} {\partial x} - \psi_i(x,t) \frac {\partial \psi_{r}(x,t)} {\partial x} \right)\nonumber
\end{eqnarray*}
we obtain that both definitions of the velocity of quantum trajectories are identical:
\begin{equation}
v(x,t) = \frac {1} {m} \frac {\partial S(x,t)} {\partial x} = \frac{J(x,t)} {|\psi(x,t)|^2}\nonumber
\end{equation}
where $J(x,t)$ is defined by \eref{om.current}.
\end{solution}

\begin{problem} \label{om.ppre6}
Show that the Bohmian velocity definition
\begin{equation*}
v(x,t) = \frac{J(x,t)}{|\psi(x,t)|^2}\nonumber
\end{equation*}
given in \eref{om.velocity} fulfills the continuity equation, \eref{om.Schrodinger1D_4}:
\begin{equation*}
\frac{\partial |\psi(x,t)|^2} {\partial t} + \frac{\partial J(x,t)} {\partial x} = 0\nonumber
\end{equation*}
$J(x,t)$ being the current defined in \eref{om.current}:
\begin{equation*}
J(x,t) = i \frac {\hbar} {2 m} \left( \psi(x,t) \frac {\partial \psi^{*}(x,t)} {\partial x} - \psi^{*}(x,t) \frac {\partial \psi(x,t)} {\partial x} \right)\nonumber
\end{equation*}
\end{problem}
As a conclusion, we learn that a proper (i.e., quantum equilibrium) ensemble of Bohmian trajectories will reproduce $R(x,t)^2 = |\psi(x,t)|^2$ at any time.

\begin{solution}
We rewrite the continuity equation as:
\begin{equation}
\frac {\partial} {\partial t} \int_{x}^{x + \Delta x} R^2(x',t) dx' + J(x + \Delta x,t) - J(x,t) = 0\nonumber
\label{om.ppre6_1}
\end{equation}
By using the Bohmian definition of the velocity, we rewrite the above equation as:
\begin{eqnarray}
\frac {\partial} {\partial t} \int_{x}^{x + \Delta x} R^2(x',t) dx' = - { R^2(x + \Delta x,t) \; v(x + \Delta x,t) + R^2(x,t) \; v(x,t)}\nonumber
\label{om.ppre6_2}
\end{eqnarray}
Now, we make use of the ensemble of Bohmian trajectories by using
\eref{om.sum_0f_particles_to} to write $R^2(x + \Delta x,t)$ and
$R^2(x,t)$ as a sum of delta functions at $t = t_0$. Finally, we
obtain:
\begin{eqnarray}
\frac {\partial} {\partial t} \int_{x}^{x + \Delta x} R^2(x',t) dx' =
&&+ \lim_{M\rightarrow\infty} \frac {1} {M} \sum_{i = 1}^{M} \delta(x - x_i[t]) v_i(t) \nonumber\\&&-\lim_{M\rightarrow\infty} \frac {1} {M} \sum_{i = 1}^{M} \delta(x + \Delta x - x_i[t]) v_i(t)\nonumber\\
\label{om.ppre6_3}
\end{eqnarray}
We can rewrite $\delta$ as time derivatives of the Heaviside (or
unit step) functions $\Theta$:
\begin{eqnarray}
\frac {\partial} {\partial t} \int_{x}^{x + \Delta x} R^2(x',t) dx' &=&
+ \lim_{M\rightarrow\infty} \frac {1} {M} \sum_{i = 1}^{M} \frac {\partial} {\partial t} \left( \Theta(x_i[t] - x) \; \Theta (x + \Delta x - x_i[t]) \right)\quad\qquad\nonumber\\
\label{om.ppre6_4}
\end{eqnarray}
Thus, counting the number of Bohmian $i$ particles inside the region
$[x,x + \Delta x]$ does exactly reproduce the square modulus of the
wave function and its time-dependent variations. In conclusion, if
we know that the ensemble of Bohmian trajectories reproduces
$R(x,t_0)$ at the particular time $t_0$, then it will reproduce
$R(x,t)$ at any other time:
\begin{equation}
R^2(x,t) = \lim_{M\rightarrow\infty} \frac {1} {M}  \sum_{i = 1}^{M} \delta(x - x_i[t])\nonumber
\end{equation}
This is precisely the reason why it is claimed that Bohmian mechanics exactly reproduces the position measurement of orthodox quantum mechanics. Since an ensemble of Bohmian trajectories reproduces $\psi(x,t)$ at any time, the ensemble of trajectories will also reproduce the mean value of other observables when they are written in the position representation.\\
\end{solution}

\begin{problem} \label{om.p8}
Demonstrate using Bohmian trajectories that the (average) dwell time
$\left\langle \tau \right\rangle$ that a particle described by the
wave function $\psi(\vec r,t)$ spends in a particular region $\vec r
\in \Omega$ can be written in an orthodox language as:
\begin{equation}
\left\langle \tau \right\rangle = \int^{\infty}_{-\infty}dt \int_{\Omega} dv|\psi(\vec r,t)|^2
\label{om.dwell}
\end{equation}
\end{problem}

\begin{solution}
For a particular Bohmian trajectory $\vec r_k[t]$, the time spent in
a particular region $\tau_k$ for this $k$-particle can be
unambiguously written as:
\begin{equation}
\tau_k = \int^{t_k^o}_{t_k^i}dt = \int^{\infty}_{-\infty}dt \Theta(t - t_k^i)\Theta(t_k^o - t)\nonumber
\end{equation}
where $t_k^i$ and $t_k^o$ are the times when the electron enters and leaves the volume $\Omega$, respectively. The function $\Theta(t)$ is a Heaviside function (or unit step function). We can relate the Heaviside function to the delta function $\delta$ as:
\begin{equation}
\Theta(t - t_k^i)\Theta(t_k^o - t) = \int_{\Omega} dv \delta (\vec r - \vec r_k[t])\nonumber
\end{equation}
In order to obtain the average value of the dwell time, $\left\langle \tau \right\rangle$, we have to make a sum over the ensemble $k = 1,\ldots,M$ of Bohmian trajectories:
\begin{equation}
\left\langle \tau \right\rangle = \lim_{M \to \infty}\frac{1}{M}\sum^{M}_{k = 1} \tau_k = \lim_{M \to \infty}\frac{1}{M}\sum^{M}_{k = 1} \int^{\infty}_{ - \infty}dt \int_{\Omega} dv \delta (\vec r - \vec r_k[t])\nonumber
\end{equation}
Finally, by noting that the ensemble of Bohmian trajectories reproduces the modulus of the wave function:
\begin{eqnarray}
|\psi (\vec r,t)|^2 = \lim_{M \to \infty}\frac{1}{M}\sum^{M}_{k = 1}\delta (\vec r - \vec r_k[t])\nonumber
\end{eqnarray}
we obtain the expected results, \eref{om.dwell}. See more information on tunneling times in \cite{om.extra14,om.oriolstime}.
\end{solution}

\begin{problem} \label{om.P1}
Let us consider a single Bohmian trajectory $x_k[t]$ accounting for the quantum dynamics of a 1D particle in a potential $V(x)$. Show that the left probability density:
\begin{equation}
P_L \equiv \int_{-\infty}^{x_k[t]}\left| \psi (x,t) \right|^2 dx\nonumber
\end{equation}
is constant, which, in turn, means that 1D Bohmian trajectories do not cross each other.
\end{problem}

\begin{solution}
We are interested in showing that:
\begin{equation}
\frac {d P_L} {dt} = \frac {d } {dt} \int^{x_k[t]}_{-\infty} dx |\psi(x,t)|^2 = 0\nonumber
\end{equation}
We know that:
\begin{equation}
\frac {d } {dt} \int^{x_k[t]}_{-\infty} dx |\psi(x,t)|^2 = |\psi(x,t)|^2 \frac {d x_k[t]} {dt} + \int^{x_k[t]}_{-\infty} dx \frac {\partial |\psi(x,t)|^2} {\partial t} = 0\nonumber
\end{equation}
By the continuity equation, \eref{om.Schrodinger1D_4}, with the current density in \eref{om.current}, we obtain:
\begin{equation}
\frac {d } {dt} P_L = |\psi(x,t)|^2 \frac {d x_k[t]} {dt} - \int^{x_k[t]}_{-\infty} dx \frac {\partial J(x,t)} {\partial x} = 0\nonumber
\end{equation}
We finally obtain $\frac {d } {dt} P_L = 0$, that is, $P_L$  is a constant  when the Bohmian trajectory moves according to:
\begin{equation}
\frac {d x_k[t]} {dt} = \frac {J(x_k[t],t)} {|\psi(x_k[t],t)|^2 }\nonumber
\end{equation}
This is exactly the Bohmian velocity of \eref{om.velocity2}.
\end{solution}

\begin{problem} \label{om.P4}
Show that the expected value of the kinetic energy, where the kinetic operator is given by $\hat{T} = -(\hbar^2/2m) \nabla^2$, yields:
\begin{equation}
\left\langle \hat{T} \right\rangle = \left\langle Q \right\rangle + \frac{\left\langle p^2 \right\rangle}{2m} 
\label{kinetic}
\end{equation}
where $Q$ is the quantum potential. \Eref{kinetic} indicates that the quantum potential $Q$ could be also interpreted as an additional quantum kinetic energy.
\end{problem}

\begin{solution}

By writing the wave function in polar form, $\psi = R e^{iS/\hbar}$, we calculate the expression of the expected value of the kinetic energy by averaging over all the possible positions of the particle, that is, in a (infinite) volume $V$:
\begin{eqnarray}
\left\langle \hat{T} \right\rangle &=& \int \psi^* \hat{T} \psi dV = \int \psi^* \left[-\frac{\hbar^2}{2m} \nabla^2 \psi \right] dV \nonumber\\
&=& \int R e^{-iS/\hbar} \left[-\frac{\hbar^2}{2m} \nabla^2 (R e^{iS/\hbar}) \right] dV \nonumber\\
&=& \int R e^{-iS/\hbar} \left[-\frac{\hbar^2}{2m} \vec{\nabla} \cdot \left( \vec{\nabla} R e^{iS/\hbar} + \frac{i}{\hbar} R \vec{\nabla} S e^{iS/\hbar} \right) \right] dV \nonumber\\
&=& \int R \!\left[-\frac{\hbar^2}{2m}\! \left(\! \nabla^2 R +
\frac{2i}{\hbar} \vec{\nabla} R \cdot \vec{\nabla} S + \frac{i}{\hbar} R \nabla^2 S - \frac{1}{\hbar^2} R (\nabla S)^2 \!\right)\! \right] dV \nonumber\\
&=& \int R^2 \left(-\frac{\hbar^2}{2m} \frac{\nabla^2 R}{R} \right)
dV - \frac{i \hbar}{2m} \int \vec{\nabla} \cdot \left( R^2
\vec{\nabla} S\right) dV + \int R^2 \left( \frac{1}{2m} (\vec{\nabla} S)^2 \right) dV\nonumber\\
\label{om.kineticquantum}
\end{eqnarray}
The first and third terms in the r.h.s. of \eref{om.kineticquantum} are, respectively, $\left\langle Q \right\rangle$ and $\frac{\left\langle p^2 \right\rangle}{2m}$. The application of the divergence theorem on the integral in the second term, leads us to:
\begin{eqnarray}
\int \vec{\nabla} \cdot \left( R^2 \vec{\nabla} S\right) dV = \oint \left( R^2 \vec{\nabla} S \right) \cdot \vec{n} dS = 0, 
\label{om.TQKintegralzero}
\end{eqnarray}
$S$ being the surface boundary enclosing the volume of integration
and $\vec{n}$ the outward pointing unit vector perpendicular to it.
\Eref{om.TQKintegralzero} is zero since the population must be
conserved, and there can be no flux of probability outside the
(infinite) volume $V$.

Thus, we see that the expected value of the quantum kinetic energy operator corresponds to the sum of the expected values of the classical kinetic energy $\frac{\left\langle p^2 \right\rangle}{2m}$ and the quantum potential. This discussion is related to what we have explained in \sref{om.measure_moment}.

\end{solution}

\begin{problem} \label{om.P5}
Show that for a stationary state, $\psi(x,t) = \varphi(x) e^{-iE t/\hbar}$, that is, for an eigenstate of the Hamiltonian, it follows that:
\begin{equation}
P^2(x) = 2m \left[E - V(x) - Q(x) \right]
\label{momentumsquare}
\end{equation}
where $P(x) = {\partial S}/{\partial x}$ is the Bohmian momentum and $E = -\partial S / \partial t$ the energy eigenvalue of the corresponding energy eigenstate. Note that \eref{momentumsquare} suggests that the semiclassical (i.e., WKB) approximation for the particle momentum given by $p(x) = \pm \sqrt{2m[E-V(x)]}$ is accurate at those positions where the quantum potential is relatively small.
\end{problem}

\begin{solution}

For an eigenstate of the Hamiltonian, it follows that:
\begin{eqnarray}
\frac{\partial R}{\partial t} = 0 \nonumber\\
\frac{\partial S}{\partial t} = -E\nonumber
\end{eqnarray}
Then, by using the quantum Hamilton--Jacobi \eref{om.hamilton_jacobi1D_des}:
\begin{equation}
\frac{\partial S}{\partial t} = - V - \frac{1}{2m} \left( \frac{\partial S}{\partial x} \right)^2 - Q\nonumber
\end{equation}
with
\begin{equation}
Q = -\frac{\hbar^2}{2m} \frac{1}{R} \frac{\partial^2 R}{\partial x^2}\nonumber
\end{equation}
one obtains
\begin{equation}
P^2(x) = 2m \left[ E-V(x)-Q(x) \right]\nonumber
\end{equation}
where we have used the Bohmian definition of the momentum $P(x) = {\partial S}/{\partial x}$.
\end{solution}

\begin{problem} \label{om.complexaction}
Recently, there have been several investigations where a quantum trajectory extension of Bohmian mechanics has been performed taking the quantum action $S$ to be complex \cite{om.imaginaryaction,om.imaginaryactionprl}. Assuming the following expression for the wave function:
\begin{equation}
\psi (x,t) = e^{i\frac{S(x,t)}{\hbar} }\nonumber
\end{equation}
derive the complex quantum Hamilton--Jacobi equation associated with it.

Then, assuming that $v(x,t) \equiv (\vec{\nabla} S(x,t))/m$ and
taking the time derivate of this Hamilton--Jacobi equation, obtain the Newtonian-like
equation of motion for the velocity:
\begin{equation}
\frac{dv[x_0,t]}{dt} = - \frac{\vec{\nabla} V}{m} + \frac{i\hbar}{2m} \nabla^2 v 
\label{movelocity}
\end{equation}
where the quantum features, such as nonlocality, are manifested in the second term of the r.h.s. of \eref{movelocity}.
\end{problem}

\begin{solution}

Casting $\psi (x,t) = e^{i\frac{S(x,t)}{\hbar} }$ into the time-dependent Schr\"odinger equation yields:
\begin{equation}
\frac{\partial S}{\partial t} + \frac{1}{2 m} (\vec{\nabla} S)^2 + V = \frac{i\hbar}{2 m} \nabla^2 S \label{mo2}\nonumber
\end{equation}
where
\begin{equation}
Q_C \equiv - \frac{i \hbar}{2m} \nabla^2 S\nonumber
\end{equation}
is the ``new'' quantum potential.
The first step to obtain the ``new'' Newtonian-like equation of motion is to take the spatial derivative of the complex quantum Hamilton--Jacobi equation:
\begin{equation}
m \left( \frac{\partial v}{\partial t} + v \vec{\nabla} v \right) - \frac{i \hbar}{2} \vec{\nabla}^2 v = - \vec{\nabla} V\nonumber
\end{equation}
Then, by identifying the total derivative of the velocity:
\begin{equation}
\frac{d v}{d t} = \frac{\partial v}{\partial t} + v \vec{\nabla} v\nonumber
\end{equation}
one finally arrives at
\begin{equation}
\frac{dv[x_0,t]}{dt} = - \frac{\vec{\nabla} V}{m} + \frac{i\hbar}{2m} \nabla^2 v\nonumber
\end{equation}
The terms at the r.h.s. can be identified as the classical and the quantum force acting on the particle, respectively.

\end{solution}

\begin{problem} \label{om.P7}
The Gross--Pitaevskii equation governs the dynamics of a Bose--Einstein condensate (BEC) in the so-called mean field approximation. In the zero temperature limit, it reads:
\begin{equation} 
\label{moGrossPitaevskii}
i \hbar \frac{\partial \psi (\vec{r},t)}{\partial t} = \left[-{\frac{\hbar^2}{2m}}\nabla^2 + V(\vec{r},t) + g \left| \psi (\vec{r},t) \right|^2 \right] \psi (\vec{r},t)
\end{equation}
where $g$ is the nonlinear interaction between the atoms of the BEC, $V(\vec{r},t)$ is the trapping potential, and the BEC wave function is normalized to the number of atoms as $\int d^3\vec{r} \left| \psi (\vec{r},t) \right|^2 = N$. Note that for $g = 0$, one formally recovers the Schr\"odinger equation. Apply Madelung's formulation to the BEC by writing down its wave function as $\psi({\vec{r}},t) = \sqrt{n({\vec{r}},t)}e^{i\theta(\vec{r},t)}$ and obtain the continuity and Hamilton--Jacobi equations associated with \eref{moGrossPitaevskii}.
\end{problem}

\begin{solution}
First of all, notice that  following the usual  nomenclature of the literature on this \eref{moGrossPitaevskii}, $\psi({\vec{r}},t) = \sqrt{n({\vec{r}},t)}e^{i\theta(\vec{r},t)}$, here we use different functions for the role of $R(\vec{r},t)$ and $S(\vec{r},t)$ that are related, but not identical, to $ = n({\vec{r}},t)$ and $\theta(\vec{r},t)$, respectively. Following the same steps as in \sref{sec.quantum_HJ.om}, it is straightforward to obtain:
\begin{eqnarray}
m\frac{\partial \vec{v}}{\partial t} & = & - \vec{\nabla}\left(V + ng + \frac{1}{2}mv^2-\frac{\hbar}{2m} \frac{1}{\sqrt{n}}\nabla^2\sqrt{n}\right) \label{mommadelung} \\
\frac{\partial n}{\partial t} & = & - \vec{\nabla}({n{\vec{v}}}) 
\label{momcont}
\end{eqnarray}
where we have defined ${\vec{v}({\vec{r}},t)}\equiv \frac{\hbar}{m}\vec{\nabla} \theta(\vec{r},t)$.
\Eref{mommadelung} is the classical Euler equation for an inviscid compressible fluid with an additional pressure term whose origin is purely quantum (the last term in this expression) that is very often called \textit{quantum pressure}. These equations will be used later in chapter 3.
\end{solution}

\begin{problem} \label{om.P8}
Use the many-particle Schr\"odinger \eref{om.schordingerND} to deduce a many-particle local continuity equation.
\end{problem}

\begin{solution}

In order to find a local continuity equation, let us work with $\psi(\vec{x},t)$ and its complex conjugate $\psi^*(\vec x,t)$. In particular, we can rewrite \eref{om.Schrodinger1D} as:
\begin{eqnarray}
\psi^*(\vec x,t) i \hbar \frac{\partial \psi(\vec x,t)} {\partial t}
&=& \psi^*(\vec x,t)\left( \sum_{k = 1}^N -\frac{\hbar^2}{2m}\frac
{\partial^2} {\partial x_k^2} \psi(\vec x,t)\right) +\, \psi^*(\vec x,t) V(x,t) \psi(\vec x,t)\nonumber
\end{eqnarray}
and the complex conjugate as:
\begin{eqnarray}
-\psi(\vec x,t)i \hbar \frac{\partial \psi^*(\vec x,t)} {\partial t} &=& \psi(\vec x,t)\left( \sum_{k = 1}^N - \frac{\hbar^2}{2m}\frac {\partial^2} {\partial x_k^2} \psi^*(\vec x,t) \right)+\, \psi(\vec x,t)V(x,t) \psi^*(\vec x,t)\nonumber
\end{eqnarray}
When the first equation is subtracted from the second, we obtain:
\begin{equation}
\frac{\partial |\psi(\vec{x},t)|^2} {\partial t} + \sum_{k = 1}^{N} J_k(\vec{x},t) = 0\nonumber
\end{equation}
where we have used:
\begin{eqnarray}
- \psi^*(\vec x,t)\frac{\hbar^2}{2m}\frac {\partial^2} {\partial x_k^2} \psi(\vec x,t) + \psi(\vec x,t)\frac{\hbar^2}{2m}\frac {\partial^2} {\partial x_k^2} \psi^*(\vec x,t) \nonumber\\= \frac{\hbar^2}{2m} \frac {\partial} {\partial x_k} \left( \psi(\vec x,t)\frac {\partial} {\partial x_k} \psi^*(\vec x,t)-\psi^*(\vec x,t)\frac {\partial} {\partial x_k} \psi(\vec x,t) \right)\qquad\nonumber
\end{eqnarray}
and defined:
\begin{equation}
J_k(\vec{x},t) = i \frac {\hbar} {2 m} \left( \psi(\vec{x},t) \frac {\partial \psi^{*}(\vec{x},t)} {\partial x_k}- \psi^{*}(\vec{x},t) \frac {\partial \psi(\vec{x},t)} {\partial x_k} \right)\nonumber
\end{equation}
as the $k$-th component of the current density.
\end{solution}

\begin{problem} \label{om.P9}
Use the continuity equation to show the following relation for the modulus of a wave function $\psi(x,t)$ evaluated at the position $x^{\alpha}[t_1]$ at time $t_1$ and at the position $x^{\alpha}[t_2]$ at time $t_2$:
\begin{eqnarray}
|\psi(x^{\alpha}[t_2],t_2)|=exp\left(-\int_{t_1}^{t_2}\left[\frac{1}{2}\frac{\partial v(x,t)}{\partial x}\right]_{x^{\alpha}[t]}dt \right) |\psi(x^{\alpha}[t_1],t_1)|\nonumber
\end{eqnarray}
Notice that we are discussing the evolution of $|\psi(x^{\alpha}[t],t)|$ along a single trajectory  $x^{\alpha}[t]$.
\end{problem}

\begin{solution}
From the continuity equation, we get:
\begin{eqnarray}
\frac{\partial |\psi(x,t)|^2}{\partial t}+\frac{\partial }{\partial x} \left(|\psi(x,t)|^2 v(x,t)\right)=0\nonumber
\end{eqnarray}
which can be rewritten as: 
\begin{eqnarray}
2|\psi(x,t)|\frac{\partial |\psi(x,t)|}{\partial t}+|\psi(x,t)|^2\frac{\partial v(x,t)}{\partial x}+2v(x,t)|\psi(x,t)|\frac{\partial |\psi(x,t)|}{\partial x}=0\nonumber
\end{eqnarray}
For one particular trajectory $x^{\alpha}[t]$ then $v(x^{\alpha}[t],t)=\frac {dx^{\alpha}[t]}{dt}$. Then, we get: 
\begin{eqnarray}
\frac{d |\psi(x^{\alpha}[t],t)|}{dt}=&&\frac{\partial |\psi(x^{\alpha}[t],t)|}{\partial t}+v(x^{\alpha}[t],t)\left[\frac{\partial |\psi(x,t)|}{\partial x}\right]_{x^{\alpha}[t]}=\nonumber\\=&&-|\psi(x^{\alpha}(t),t)|\frac{1}{2}\left[\frac{\partial v_B(x,t)}{\partial x}\right]_{x^{\alpha}[t]}\nonumber
\end{eqnarray}
Then, finally:
\begin{eqnarray}
\frac{d}{dt}  ln \left(|\psi(x^{\alpha}[t],t)|\right)=-\frac{1}{2}\left[ \frac{\partial v(x,t)}{\partial x}\right]_{x^{\alpha}[t]}\nonumber
\end{eqnarray}
which means that:
\begin{eqnarray}
|\psi(x^{\alpha}[t_2],t_2)|=exp\left(-\int_{t_1}^{t_2}\left[\frac{1}{2}\frac{\partial v(x,t)}{\partial x}\right]_{x^{\alpha}[t]}dt \right) |\psi(x^{\alpha}[t_1],t_1)|\nonumber
\end{eqnarray}
\end{solution}

\begin{problem} \label{om.P10}
Use the Hamilton--Jacobi equation to show the following relation for the quantum action $S(x,t)$ evaluated at the position $x^{\alpha}[t_1]$ at time $t_1$ and at the position $x^{\alpha}[t_2]$ at time $t_2$:
\begin{eqnarray}
S(x^{\alpha}[t_2],t_2)=S(x^{\alpha}[t_1],t_1) +\int_{t_1}^{t_2}\left( \frac{1}{2}m\cdot v(x^{\alpha}[t],t)^2 -V(x^{\alpha}[t],t)-Q(x^{\alpha}[t],t)\right)dt \nonumber
\end{eqnarray}
\end{problem}

\begin{solution}
We get: 
\begin{eqnarray}
\frac {dS(x^{\alpha}[t],t)}{dt}=\left[\frac {\partial S(x,t)}{\partial x}\right]_{x^{\alpha}[t]}v(x^{\alpha}[t],t)+\left[\frac {\partial S(x,t)}{\partial t}\right]_{x^{\alpha}[t]}\nonumber
\end{eqnarray}
where $\frac {\partial S(x,t)}{\partial x}=m\cdot v(x,t)$ and $\frac {\partial S(x,t)}{\partial t}=-m\cdot v(x,t)^2/2-V(x,t)-Q(x,t)$. Then, we get:
\begin{eqnarray}
\frac {dS(x^{\alpha}[t],t)}{dt}=m \cdot v(x^{\alpha}[t],t)^2-m\cdot v(x^{\alpha}[t],t)^2/2-V(x^{\alpha}[t],t)-Q(x^{\alpha}[t],t)\nonumber
\end{eqnarray}
and finally:
\begin{eqnarray}
S(x^{\alpha}[t_2],t_2)=S(x^{\alpha}[t_1],t_1) +\int_{t_1}^{t_2}\left( \frac{1}{2}m\cdot v(x^{\alpha}[t],t)^2 -V(x^{\alpha}[t],t)-Q(x^{\alpha}[t],t)\right)dt \nonumber
\end{eqnarray}
\end{solution}

\begin{problem} \label{om.P11}
Use the continuity equation and the Hamilton--Jacobi equation to show the following relation for the wave function $\psi(x,t)$ evaluated at the position $x^{\alpha}[t_1]$ at time $t_1$ and at the position $x^{\alpha}[t_2]$ at time $t_2$:
\begin{eqnarray}
\frac{\psi(x^{\alpha}[t_2],t_2)}{\psi(x^{\alpha}[t_1],t_1)}=exp\left( \int_{t_1}^{t_2}\left[ \frac{i}{\hbar} \left(\frac{1}{2}m\cdot v(x,t)^2 -V(x,t)-Q(x,t)\right)-\frac{1}{2}\frac{\partial v(x,t)}{\partial x}\right]_{x^{\alpha}[t]} dt \right) \nonumber
\end{eqnarray}
Notice that we are discussing the evolution of $\psi(x^{\alpha}[t],t)$ along a single trajectory  $x^{\alpha}[t]$.
\end{problem}

\begin{solution}

We write $\psi(x^{\alpha}[t_2],t_2)=|\psi(x^{\alpha}[t_2],t_2)|exp(\frac{i}{\hbar}S(x^{\alpha}[t_2],t_2))$ and use the results  of problem \ref{om.P9} for the modulus and the results of problem \ref{om.P10} for the quantum action to write: 
\begin{eqnarray}
\frac{\psi(x^{\alpha}[t_2],t_2)}{\psi(x^{\alpha}[t_1],t_1)}=exp\left( \int_{t_1}^{t_2}\left[ \frac{i}{\hbar} \left(\frac{1}{2}m\cdot v(x,t)^2 -V(x,t)-Q(x,t)\right)-\frac{1}{2}\frac{\partial v(x,t)}{\partial x}\right]_{x^{\alpha}[t]} dt \right) \nonumber
\end{eqnarray}
Once we have finished the demonstration, it is interesting to realize the differences between the expression above and the typical Feynman path integral developed in \sref{om.sec_single.5} evaluated at $x^{\alpha}[t_2]$:
\begin{equation}
\psi(x^{\alpha}[t_2],t_2) = \int_{-\infty}^{\infty} G(x_0,t_0;x^{\alpha}[t_2],t_2) \psi(x_0,t_0) dx_0\nonumber
\end{equation}
\end{solution}
The evaluation of $\psi(x^{\alpha}[t_2],t_2)$ does only require the value of $\psi(x^{\alpha}[t_1],t_1)$ and the knowledge of $v(x,t)$, $V(x,t)$, $Q(x,t)$ and $\frac{1}{2}\frac{\partial v(x,t)}{\partial x}$ along the path of the single trajectory $x^{\alpha}[t]$. No information about other trajectories departing from different positions $x\neq x^{\alpha}[t_1]$ at time $t_1$ are required.


\section[Appendix]{Appendix: Numerical Algorithms for the Computation of Bohmian Mechanics}\label{om.sec_comput}

In this section, we will describe the basic numerical algorithms
available in the literature for the computation of Bohmian
mechanics. An excellent source of information on this issue can be
found in Ref. \cite{om.wyatt2005}. There are many possible
algorithms that can be classified following different schemes.
One can distinguish between analytical or synthetic algorithms.\footnote{The names ``analytical'' and ``synthetic'' are related to philosophical concepts. The original name of analytic propositions referred to those propositions that were true simply by virtue of their meaning, while synthetic propositions were not.}
An analytical algorithm, first, solves the Schr\"odinger equation and, then, evaluates the Bohmian velocity directly from the numerical wave function. In such algorithms, the computational difficulties are directly related to the numerical solution of the Schr\"odinger equation. Alternatively, one can choose computational algorithms where the Schr\"odinger equation is substituted by the quantum Hamilton--Jacobi equation (synthetic approach). Then, the difficulties of the algorithms are intrinsic to the numerical resolution of the (nonlinear) Hamilton--Jacobi scheme. In addition,
the quantum Hamilton--Jacobi equation can be solved with or without trajectories. Another classification divides
algorithms between time-dependent and time-independent ones. From a computational point of view, the
former is basically related to an eigenvalue problem, while the latter is related to an initial
value problem. Finally, a classification between 1D spatial grids and 2D (or even 3D)
is also possible.\footnote{Certainly, the direct solution of more than three degrees of freedom is very computationally demanding. To the best of our knowledge, the only Bohmian algorithm that takes into account correlations for more than three degrees of freedom is the one mentioned in Sec. 1.3.6.     
This many-particle Bohmian algorithm does not fit with the previous
classification because it is a mixed analytical and synthetic
algorithm.} In \tref{tblA.1}, we have summarized the characteristics
of the numerical algorithms that we discuss.

\begin{table}
\centering
\caption{1D and 2D Bohmian computational algorithms (TD $=$ time dependent; TI $=$ time independent; SE $=$ Schr\"odinger equation; QHJE $=$ quantum Hamilton--Jacobi equation; BT $=$ Bohmian trajectries).}{%
\tabcolsep3.5pt\begin{tabular}{@{}cccccc@{}} \toprule
 & & \multicolumn{2}{c}{Analytical} & \multicolumn{2}{c}{Synthetic} \\
 \cline{3-6}\\[-10pt]
 & & no-trajectory & trajectory & no-trajectory & trajectory \\ \colrule
 Spatial 1D & Time-dependent & --- & TDSE$_{1D}$ $+$ BT & TDQHJE$_{1D}$ & TDQHJE$_{1D}$ $+$ BT\\
 & Time-independent & --- & TISE$_{1D}$ $+$ BT & TIQHJE$_{1D}$ & TIQHJE$_{1D}$ $+$ BT\\ \botrule
 Spatial 2D & Time-dependent & --- & TDSE$_{2D}$ $+$ BT & TDQHJE$_{2D}$ & TDQHJE$_{2D}$ $+$ BT\\
 & Time-independent & --- & TISE$_{2D}$ $+$ BT & TIQHJE$_{2D}$ & TIQHJE$_{2D}$ $+$ BT\\ \botrule
\end{tabular}}\label{tblA.1}
\end{table}

Much of our attention in this section will be devoted to
time-dependent scenarios. In such scenarios, either the
Schr\"odinger equation or the quantum Hamilton--Jacobi equation can
be written as a first-order differential equation in time of the
form:\enlargethispage{-1pc}
\begin{equation}
\label{om.equationdt}
\frac {dW(t)} {dt} = f(W(t),t)
\end{equation}
where $W(t)$ can take the role of $\psi(x,t)$, $R(x,t)$ and
$S(x,t)$). $f(W(t),t)$ is a generic function that depends on $W(t)$
and time $t$.  We define a mesh for the time variable $t_j = j
\Delta t$ for $j = 1,\ldots,M$ and assume that the initial value of
$W(0)$ is known. Then, one can use an explicit finite difference of
the time derivative:
\begin{equation}
\left.\frac {dW(t)} {dt}\right|_{t = t_j} \approx \frac {W(t_{j + 1})-W(t_j)} {\Delta t}
\end{equation}
giving a simple and explicit numerical solution of the time-dependent \eref{om.equationdt} as:
\begin{equation}
\label{om.explicit}
W(t_{j + 1}) = W(t_j) + f(W(t_j),t_j) \Delta t
\end{equation}
where the solution $W(t_{j + 1})$ at $t_{j + 1}$ can be found
directly from the knowledge of $W(t_j)$ and $f(W(t_j),t_j)$ at the
previous time $t_j$. These kinds of solutions are named explicit
solutions  (also known as forward Euler method) and are, in general,
unstable because the accumulated error during successive time integrations grows
exponentially. Only for a very dense time mesh, that is, a very
small $\Delta t$, the solution becomes conditionally stable.

On the contrary, we can look for an \textit{implicit} approximation of the derivative:
\begin{equation}
\left.\frac {dW(t)} {dt}\right|_{t = t_j} \approx \frac {W(t_{j})-W(t_{j-1})} {\Delta t}
\end{equation}
that provides an implicit solution (also known as backward Euler method) of the time-dependent \eref{om.equationdt} as:
\begin{equation}
\label{om.implicit}
W(t_{j}) = W(t_{j-1}) + f(W(t_j),t_j) \Delta t
\end{equation}
The practical numerical solution of \eref{om.implicit} depends on
the type of equation. For instance, for the Schr\"odinger equation,
the function $f(W(t_j),t_j)$ is linear with respect to $W(t_j)$, and
then, the relationships between $W(t_{j})$ and $W(t_{j-1})$ can be
written as a set of coupled linear equations, one for each grid
point. Then, the entire system of linear equations can be written as
the product of a (known) matrix and an (unknown) vector. The
(unknown) vector is formed by the unknown value of $W(t_j)$ at each
grid point. The solution follows from inverting the matrix. On the
contrary, for the quantum Hamilton--Jacobi equation, where
$f(W(t_j),t_j)$ is nonlinear, one has to look for the numerical
solution of $W(t_{j})-W(t_{j-1})- f(W(t_j),t_j) \Delta t = 0$
through a Newton-like procedure. Apart from the important increase\enlargethispage{-1pc}
of the computational burden, both types of implicit solutions 
provide a nongrowing error independently of the value of $\Delta t$.
For this reason, they are called unconditionally stable algorithms.
See a summary of the classification in \tref{tblA.2}.

\begin{table}
\centering
\caption{Bohmian time-dependent computational algorithms.}{%
\tabcolsep5pt\begin{tabular}{@{}cccc@{}} \toprule
 & Type of solution & SE$\rightarrow$ linear $f(W(t),t)$ & QHJE$\rightarrow$ nonlinear $f(W(t),t)$ \\ \colrule
Implicit $\frac {dW(t)}{dt}$ & Unconditional stable & Matrix inversion & Newton-like algorithm \\
Explicit $\frac {dW(t)}{dt}$ & Conditional stable & Explicit solution & Explicit solution \\
 & Unstable & No solution & No solution \\
\botrule
\end{tabular}}\label{tblA.2}
\end{table}

Hereafter, we will explain in detail some of these numerical
algorithms, discussing their main advantages and drawbacks. We will
mention some very simple algorithms that can be useful for those
newcomers who are just interested in computing Bohmian trajectories.\enlargethispage{-1pc}

\subsection{Analytical computation of Bohmian trajectories}

According to the first postulate of Bohmian mechanics, the
computation of trajectories is quite simple once the wave function
is known.  As we mentioned in the introduction, here, the utility of
using Bohmian trajectories is that allows for an intuitive and different
explanation of many quantum phenomena. In particular, in this
section, we will discuss three simple numerical methods for
solving the 1D Schr\"odinger equation:
\begin{equation}
\label{om.Schrodinger1Dappendix}
i \hbar \frac{\partial \psi(x,t)} {\partial t} = -\frac {\hbar^2} {2m} \frac{ {\partial}^2 \psi(x,t)} {\partial x^2} + V(x,t) \psi(x,t)
\end{equation}

\subsubsection{Time-dependent Schr\"odinger equation for a 1D space (TDSE$_{1D}$-BT) with an explicit method}
\label{timedependentexplicit}

In this subsection, we will explain a very simple explicit algorithm to compute the time-dependent Schr\"odinger equation. Apart from a mesh in the time variable, an additional mesh is imposed on the 1D spatial degree of freedom with a constant step, $\Delta x$, that is, $x_k = k \; \Delta x $ for $k = 1,\ldots,N$. Then, we define $\psi(x,t)|_{x = x_k;t = t_j} = \psi_{j}(x_k)$, and the temporal and spatial derivatives present in the time-dependent Schr\"odinger equation can be approximated by:
\begin{eqnarray}
\left.\frac{\partial \psi \left(x,t \right)}{\partial t}\right|_{x = x_k;t = t_j} &\approx& \frac{\psi_{j + 1}(x_k) - \psi_{j - 1}(x_k)} {2\Delta t} \label{om.finite-difference_t} \\
\left.\frac{{{\partial }^{2}}\psi \left( x,t \right)}{\partial {{x}^{2}}}\right|_{x = x_k;t = t_j} &\approx& \frac{\psi_{j}(x_{k + 1}) - 2\psi_{j}(x_{k}) + \psi_{j}(x_{k - 1})}{{\Delta x}^{2}} \label{om.finite-difference_x}
\end{eqnarray}
Inserting Eqs. (\ref{om.finite-difference_t}) and (\ref{om.finite-difference_x}) into \eref{om.Schrodinger1Dappendix}, we obtain the following simple recursive expression:
\begin{eqnarray}
\psi_{j + 1}(x_{k}) &=& \psi_{j - 1}(x_{k}) + i\frac{\hbar \Delta t}{{{\Delta x}^{2}}m}\left(\psi_{j}(x_{k + 1}) - 2\psi_{j}(x_{k}) + \psi_{j}(x_{k - 1})\right)-\,i\frac{2\Delta t}{\hbar }V_{j}(x_k) \psi_{j}(x_{k})\nonumber\\
\label{om.finite-difference_recurs}
\end{eqnarray}
Once we know the wave function at the particular times $t_j$ and
$t_{j - 1}$ for all spatial positions in the mesh, we can compute
the wave function at next time $t_{j + 1}$, using
\eref{om.finite-difference_recurs}. This algorithm solves an
initial-value problem. The recursive application of
\eref{om.finite-difference_recurs} provides the entire time
evolution of the wave packet. \Eref{om.finite-difference_recurs} is
not valid for the first, $x_1$, and last, $x_N$, points. To avoid
discussions on the boundary conditions, we can use a very large
spatial simulating box so that the entire wave packet is contained
in it at any time. Then the wave function at the borders is
negligible.\enlargethispage{-1pc}

\begin{figure}
\centering
\includegraphics{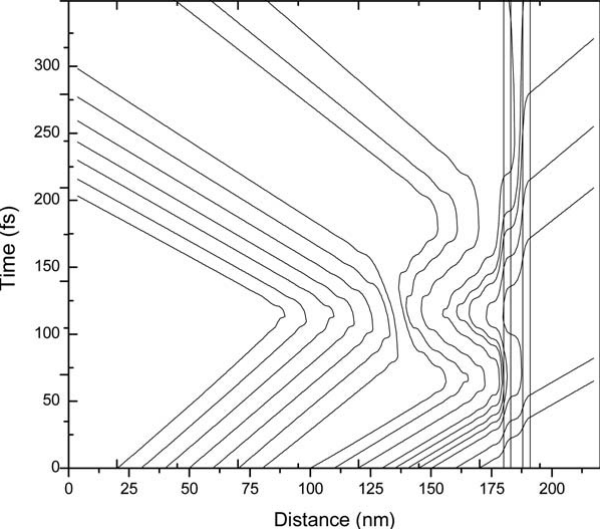}
\caption{Representative Bohmian trajectories associated with double-packet scattering in a double-barrier potential. The position of the barriers is indicated by vertical lines. Reprinted with permission from \cite{om.oriolstime}. Copyright 1996 American Physical Society.}
\label{om_fig_twopaq1}
\end{figure}

\begin{figure}
\centering
\includegraphics{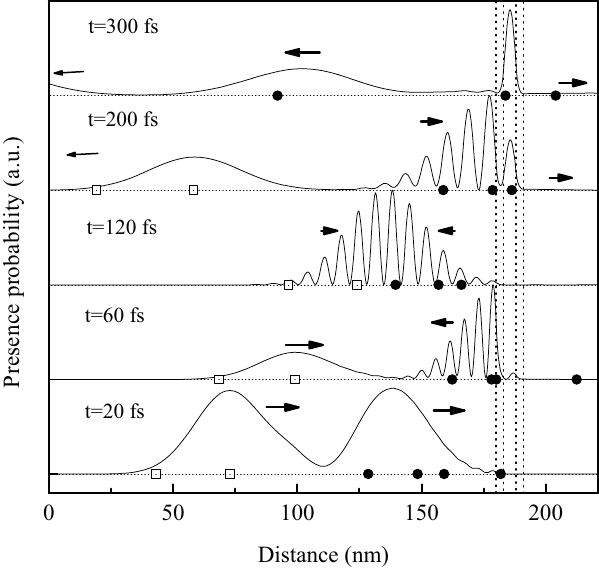}
\caption{Picture motion of the same double-packet wave function considered in
\fref{om_fig_twopaq1}, calculated by numerical integration of the
time-dependent Schr\"odinger equation. Five representative
``snapshots'' obtained at different times are shown with the
vertical scale arbitrarily changed in each case for clarity. Although
the norm of the wave function is always unity, it does not seem so
because of the changes in the scale. The marks are visual aids that
indicate the position of some related Bohmian trajectories shown in
\fref{om_fig_twopaq1}. The double-barrier position is indicated by
the vertical dashed lines, and the arrows indicate the sense of
motion of the two packets. Reprinted with permission from
\cite{om.oriolstime}. Copyright 1996 American Physical Society.}
\label{om_fig_twopaq2}
\end{figure}

As indicated in Tables \ref{tblA.1} and \ref{tblA.2}, this explicit
solution can be unstable, and its error grows in each recursive
application of \eref{om.finite-difference_recurs}. To provide a
(conditional) stable solution we have to deal with small
values of $\Delta t$ and $\Delta x$. For example, to study electron
transport in nanoscale structures as depicted in
Figs. \ref{om_fig_twopaq1} and \ref{om_fig_twopaq2}, this recursive
procedure provides accurate results (the norm of the wave packets is
conserved with high precision) when $\Delta x$ is on the order of
$1$ or $2$ \AA \  \ and the temporal step, $\Delta t$, is around
$10^{-16}$s.

In order to define the initial value of the wave function, we can consider that the wave packet at times $t = \{t_1,t_1 + \Delta t\}$ evolves in a flat potential region contained in a much larger simulation box. Then, for example, we can define the  initial wave function as a time-dependent Gaussian wave packet \cite{om.cohen}:
\begin{eqnarray}
\psi (x,t) &=& {{\left( \frac{2{{a}^{2}}}{\pi } \right)}^{1/4}}\frac{{{e}^{i\phi }}{e}^{i(k_c(x-x_0))}}{{{\left( {{a}^{4}} + \frac{4{{\hbar }^{2}}{{(t-{{t}_{1}})}^{2}}}{{{m}^{2}}} \right)}^{1/4}}}\exp \left(-\frac{{{\left[x - {{x}_0} - \frac{\hbar {{k}_{c}}}{m}(t - {{t}_{1}}) \right]}^{2}}}{{{a}^{2}} + \frac{2i\hbar (t - {{t}_{1}})}{m}} \right)\nonumber\\
\label{om.finite-difference_innitial}
\end{eqnarray}
where $a$ is the spatial dispersion of the wave packet, $m$ the particle effective mass, $x_0$ the central position of the wave packet at the initial time $t_1$, ${k}_{c} = \sqrt{\frac{2mE}{{\hbar }^{2}}}$ the central wave vector in the $x$ direction related with the central energy $E$, and $\phi = -\theta - {\hbar {k_C}^{2}t}/{(2m)}$ with $\tan (2 \theta ) = {2 \hbar t}/{(m{{a}^{2}})}$ (see \cite{om.cohen}). In particular, at the initial time $t = t_1$, we obtain the simplified expression:
\begin{equation}
\psi (x,t = t_1) = {{\left( \frac{2}{\pi {{a}^{2}}} \right)}^{1/4}}{{e}^{i\left( {{k}_{c}}(x-{{x}_0}) \right)}} \exp \left( -\frac{{{(x-{{x}_0})}^{2}}}{{{a}^{2}}} \right)
\end{equation}
We define the wave packet spatial dispersion ${{\sigma }_{x}} =
a/\sqrt{2}$ and the wave packet width in the reciprocal  space as
${{\sigma }_{k}} = 1/{{\sigma }_{x}}$.

To finish this practical computation of Bohmian trajectories from
the direct solution of the time-dependent wave function, we can
discuss some features of Bohmian trajectories related to the fact
that they do not cross in the configuration space. We consider a
quite ``exotic'' initial wave packet. We use a sum of two Gaussian
wave packets defined by \eref{om.finite-difference_innitial} with different central
positions and central (momenta) wave vectors \cite{om.oriolstime}.
The wave packet is certainly quite exotic because it describes just
one particle. In Figs. \ref{om_fig_twopaq1} and \ref{om_fig_twopaq2} we
see that Bohmian trajectories can be reflected for two different
reasons: first, because of their interaction with the classical
potential (the particles collide with the barrier) and, second,
because of the collision with other trajectories traveling in the
opposite direction (Bohmian trajectories do not cross). The second
process is responsible for the reflection of those particles of the
first packet, which never reach the barrier, and for the reflection
of the entire second packet. These collisions between Bohmian
particles are related to the quantum  potential
in regions where the classical potential is zero, but for them to
occur, there should be particles coming from right to left. In this
regard, if the initial wave packet is prepared as a superposition of
eigenstates incident from left to right (as is always assumed in
scattering thought experiments), and the classical potential is zero
on the left-hand side (l.h.s.) of the barrier, then finding
particles coming from the r.h.s. in this region will be at least
very uncommon.

\subsubsection{Time-independent Schr\"odinger equation for a 1D space (TISE$_{1D}$) with an implicit (matrix inversion) method}
\label{timeindependentexplicit}

Next, we consider another very simple computational algorithm similar to the one discussed in the previous section (1D, analytic and explicit) but for time-independent systems. First, let us discuss under which circumstances we can expect that a time-independent Schr\"odinger equation can provide a reasonable physical description. Let us assume that the wave function $\psi(x,t)$ can be written as a product of a spatial function $\varphi(x)$ and a temporal function $\zeta(t)$, that is, $\psi(x,t) = \varphi(x) \zeta(t)$. Then, the Schr\"odinger equation, \eref{om.Schrodinger1Dappendix}, with a time-independent potential $V(x)$ can be written as:
\begin{equation}
\label{om.schrodinger1D_tind}
\frac {i \hbar \frac{\partial \zeta(t)} {\partial t}} {\zeta(t)} = \frac {-\frac {\hbar^2} {2m} \frac{ {\partial}^2 \varphi(x)} {\partial x^2} + V(x) \varphi(x)} {\varphi(x)}
\end{equation}
Since the l.h.s. of \eref{om.schrodinger1D_tind} only depends on $t$ and the r.h.s. of \eref{om.schrodinger1D_tind} only depends on $x$, both sides are equal to a constant that we refer as $E$. Therefore, we obtain two separate equations:
\begin{eqnarray}
\label{om.schrodinger1D_tind_t}
i \hbar \frac{\partial \zeta(t)} {\partial t} = E\zeta(t) \\
\label{om.schrodinger1D_tind_x}
-\frac {\hbar^2} {2m} \frac{ {\partial}^2 \varphi(x)} {\partial x^2} + V(x) \varphi(x) = E \varphi(x)
\end{eqnarray}
The first of these equations has a trivial analytical solution $\zeta(t) = e^{i\frac {E t} {\hbar}}$. However, the time-independent Schr\"odinger equation needs a numerical method to be solved for arbitrary shapes of the potential $V(x)$.

The second-order derivative can be approximated as:
\begin{equation}
\left.\frac{{{\partial }^{2}}\varphi\left( x \right)}{\partial {{x}^{2}}}\right|_{x = x_k} \approx \frac{\varphi(x_{k + 1})-2\varphi(x_{k}) + \varphi(x_{k-1})}{{\Delta x}^{2}}
\label{om.finite-difference_x_senset}
\end{equation}
Then, \eref{om.schrodinger1D_tind_x} can be written in a matrix representation as:
\begin{equation}
\left(-\frac {\hbar^2} {2m} \hat{\nabla}^2 + \left(V(x_k) - E\right) \hat{I} \right) \vec \varphi = 0
\label{om.matrix_inversion}
\end{equation}
where $\vec \varphi = (\varphi(x_2),\ldots,\varphi(x_{N - 1}))^T$ is a $N - 2$ vector, $\hat{{I}}$ is the $(N - 2)\times(N - 2)$ identity matrix and $\hat{\nabla}^2$ is the $(N - 2)\times(N - 2)$ matrix constructed from \eref{om.finite-difference_x_senset} as:
\begin{equation}
\hat{\nabla}^2 = \frac{1}{{\Delta x}^2}\left(
\begin{array}
[c]{ccccc}%
-2 & 1 & 0 & 0 &\ddots \\
 1 & -2 & 1 & 0 &\ddots\\
\ddots & \ddots & \ddots & \ddots & \ddots\\
\ddots & 0 & 1 & -2 & 1\\
\ddots & 0 & 0 & 1 & -2
\end{array}
\right)
\end{equation}
The solution of $\vec \varphi$ can be found by inverting the matrix equation, \eref{om.matrix_inversion}. The (nonzero) solution of \eref{om.matrix_inversion} requires that the determinant of the matrix be equal to zero. The determinant requirement is called the characteristic equation (less often, the secular equation). In our case, the determinant can only be zero for some particular values of the energy $E$ (i.e., quantized energies). Thus, this algorithm finds the eigenvalues $E$ and eigenvectors $\vec \varphi$, typical of a boundary value problem.\footnote{The practical implementation of this algorithm can be done, for example, quite easily with the software MATLAB. The instruction \textbf{[V,D] = eig(A)} produces matrices of eigenvalues (D) and eigenvectors (V) of matrix A, so that $A*V = V*D$. \Eref{om.matrix_inversion} can be rewritten as $A*V = V*D$.}

In the previous vector definition, $\vec \varphi$, only $N - 2$ unknowns are considered. We have assumed that $\varphi(x_1) = 0$ and $\varphi(x_N) = 0$. This implies that $\vec \varphi$ is a real vector. Then, if we select an eigenstate,  we realize that the current density is zero, $J(x,t) = 0$, because $\varphi(x_k) = \varphi(x_k)^*$. All Bohmian velocities are zero, $\partial S/\partial x = 0$, meaning that Bohmian trajectories do not ``move.'' This result is obviously consistent with the fact that the amplitude $R(x,t)$ of the global wave function $\psi(x,t) = \varphi(x) e^{-iE t/\hbar}$ is time independent. One is tempted to affirm that there is no kinetic energy for these Bohmian trajectories. However, as we see in \eref{om.kinetic_energy_mean_B}, the evaluation of the mean value of the kinetic energy takes into account not only the Bohmian velocity but also the quantum potential. We can explain this point, differently, from the quantum Hamilton--Jacobi equation, \eref{om.hamilton_jacobi1D_des}. In our particular time-independent case, we have an energy conservation law:
\begin{equation*}
V(x) - \frac{\hbar^2} {2 m} \frac{{\partial}^2 R(x)/ \partial x^2} {R(x)} = E
\end{equation*}
where we have used $\partial S/\partial t = -E$ in \eref{om.hamilton_jacobi1D_des}. Then, we can obtain the following (time-independent Schr\"odinger-like) equation for the real wave function $R(x)$:
\begin{equation}
\label{om.schrodinger1D_tind_x_for R}
-\frac{\hbar^2} {2 m} \frac{{\partial}^2 R(x)} {\partial x^2} + V(x) R(x) = E \; R(x)
\end{equation}
Comparing this expression with the previous
\eref{om.schrodinger1D_tind_x}, we realize that in this particular
problem (for which all Bohmian trajectories have zero velocity), the quantum
potential takes the role of a ``kinetic'' energy.

\subsubsection{Time-independent Schr\"odinger equation for a 1D space (TISE$_{1D}$) with an explicit method}

We can also look for a very simple explicit algorithm to find the solution $\vec \varphi$ of the time-independent Schr\"odinger equation. \Eref{om.schrodinger1D_tind_x} is a homogeneous second-order differential equation that can be written as:
\begin{equation}
\frac{{d}^{2}}{d{x}^{2}}\varphi(x) = f(x)\psi (x)
\label{om.numerov1}
\end{equation}
where $f(x)$ is the constant of proportionality between the solution and the second derivative at each point:
\begin{equation}
f(x) = \frac{ 2 m} {{\hbar }^{2}}(V(x) - {E})
\end{equation}

From the Taylor expansion for $\varphi(x_{k + 1})$ around $x_k$, we get:
\begin{eqnarray}
\varphi(x_{k + 1}) =&& \varphi(x_{k}) + \Delta x \left.\frac {d\varphi(x)}{dx}\right|_{x = x_k} + \frac{\Delta x^2}{2!} \left.\frac {d^2\varphi(x)}{dx^2}\right|_{x = x_k} +\, \left.\frac{\Delta x^3}{3!} \frac{d^3\varphi(x)}{dx^3}\right|_{x = x_k} \nonumber\\&&+ \frac{\Delta x^4}{4!} \left.\frac{d^4\varphi(x)}{dx^4}\right|_{x = x_k}+\, \frac{\Delta x^5}{5!} \left.\frac{d^5\varphi(x)}{dx^5}\right|_{x = x_k} + {O} (h^6)
\end{eqnarray}
Identically, for $\varphi(x_{k-1})$ around $x_k$, we get:
\begin{eqnarray}
\varphi(x_{k - 1}) =&& \varphi(x_{k}) - \Delta x \left.\frac {d\varphi(x)}{dx}\right|_{x = x_k} + \frac{\Delta x^2}{2!} \left. \frac {d^2\varphi(x)}{dx^2}\right|_{x = x_k}-\, \frac{\Delta x^3}{3!} \left.\frac{d^3\varphi(x)}{dx^3}\right|_{x = x_k}
 \nonumber\\ &&+ \frac{\Delta x^4}{4!} \left.\frac{d^4\varphi(x)}{dx^4}\right|_{x = x_k}-\, \frac{\Delta x^5}{5!} \left.\frac{d^5\varphi(x)}{dx^5}\right|_{x = x_k} + {O} (\Delta x^6)
\end{eqnarray}
The sum of these two equations gives:
\begin{eqnarray}
\varphi(x_{k + 1}) + \varphi(x_{k-1}) =&& 2\varphi(x_{k}) + {\Delta x^2} \left.\frac {d^2\varphi(x)}{dx^2}\right|_{x = x_k}+\, \frac{\Delta x^4}{12} \left.\frac {d^4\varphi(x)}{dx^4}\right|_{x = x_k} + {O} (\Delta x^6)\nonumber\\
\end{eqnarray}
We use \eref{om.numerov1} to define, directly, the second-order derivative. We compute the fourth-order derivative from a second-order derivative written as a difference  quotient. Finally, we find a recursive expression for the wave function $\varphi(x_{k-1})$:
\begin{equation}
\varphi(x_{k - 1}) = \frac{\left[2 + \frac{10}{12}{\Delta x}^{2}f(x_k) \right]\varphi(x_{k}) - \left[1 - \frac{{\Delta x}^{2}}{12}f(x_{k + 1}) \right]\varphi(x_{k + 1})}{1 - \frac{{\Delta x}^{2}}{12}f(x_{k - 1})}
\label{om.numerov}
\end{equation}

We have explicitly written $\varphi(x_{k - 1})$ in \eref{om.numerov}
because later, in a numerical example, we will be interested in
specifying the boundary conditions of $\varphi(x)$ at $x\rightarrow
\pm \infty$. The Numerov method \cite{om.numrov1,om.numrov2}
provides a numerical algorithm to solve ordinary differential
equations of the second order in which the first-order term does not
appear. It is a fourth-order linear multistep method. The above
expression, \eref{om.numerov}, allows a simple solution of
$\varphi(x_{k-1})$ once we know $\varphi(x_{k})$, $\varphi(x_{k +
1})$, $f(x_k)$, $f(x_{k + 1})$, and $f(x_{k-1})$. Finally, repeating
the process, we can determine the entire solution.

We will apply this last explicit computational algorithm for
tunneling scenarios. For the numerical simulations, we will use
$\Delta x = 1$ {\AA}. Again, the amplitude of the global wave
function $\psi(x,t) = \varphi(x) e^{-iE t/\hbar}$ is time
independent. Thus, the current density $J(x,t)$, from
\eref{om.current}, is also time independent. Therefore, from the
continuity equation, \eref{om.conservation_law1D}, we deduce that
$J(x,t)$  is also uniform (position-independent). Then, the velocity
$v(x) = $ constant$/R^2(x)$ does not depend on time but depends on position.
We have $S(x,t)\ne 0$. The velocity can be computed from:
\begin{equation}
\label{om.tunneling}
\frac {\partial S(x,t)} {\partial x} = \sqrt{2 m (E-V(x)-Q(x))}
\end{equation}
See problem \ref{om.P5} for a detailed calculation. This quantum expression, \eref{om.tunneling}, is different from the similar expression extracted from the classical Hamilton--Jacobi equation ${\partial S(x,t)}/ {\partial x} = \sqrt{2 m (E - V(x))}$. When $E < V(x)$, we will not be able to find a (real) velocity in a classical system (classical particles are unable to pass through spatial regions where the potential barrier is higher than its energy). Interestingly, when $E < V(x)$ in \eref{om.tunneling}, we will still find (real) velocities because of the term $Q(x)$ in \eref{om.tunneling} (Bohmian particles are able to pass through spatial regions where the potential barrier is higher than its energy). This is the well-known tunneling effect.

As a numerical example, let us concentrate on (time-  independent) scattering states traveling  from left to right in a double-barrier structure. For a simulation box $0<x<40\; \rm nm$, these states can be written as:
\begin{eqnarray}
\label{om.scattering_sates}
\varphi(x) = \frac {1} {\sqrt{2 \pi}} \left( e^{ik x} + r(k) e^{-ik x} \right) \;\; ; \; \; \; x<0\; \rm nm\\
\varphi(x) = \frac {1} {\sqrt{2 \pi}} \left( t(k) e^{ik x} \right) \;\; ; \; \; \; x>40\; \rm nm
\end{eqnarray}
where $r(k)$ and $t(k)$ are (unknown) complex coefficients accounting for reflection and transmission, respectively. However, up to an irrelevant (complex) constant that can be fixed by ``normalization,'' we know the results $\varphi(x)$ at $x\rightarrow \pm \infty$. In the central region, we can use the Numerov numerical method using the fact that the wave function and its derivative have to be continuous in the matching points between the regions where we have analytical or numerical knowledge of the wave functions.

\begin{figure}
\centering
\includegraphics[width=23pc]{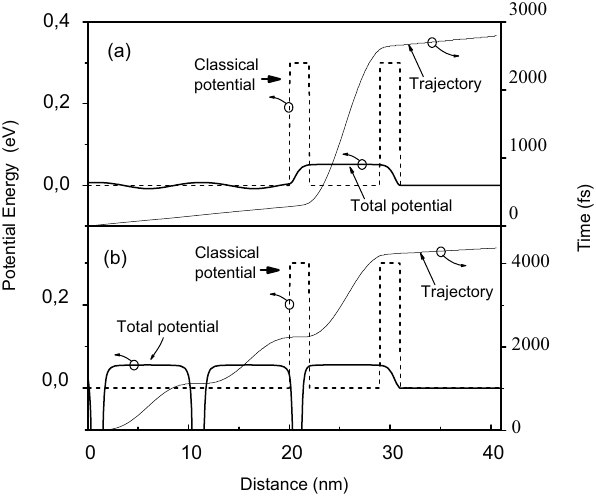}
\caption{Bohmian trajectories associated with stationary scattering states impinging on a typical
GaAs/AlGaAs RTD with 2 nm barriers of 0.3 eV separated by a 7 nm
well. The classical potential and the total potential (the sum of
classical and quantum terms) are also depicted in dashed and solid
lines, respectively. (a) Resonant eigenstate, $E_k = 0.05$ eV; (b)
nonresonant eigenstate, $E_k = 0.0$6 eV. Reprinted with permission
from \cite{om.oriols_sst}. Copyright 1996 Elsevier.
\textit{Abbreviation}: RTD, resonant tunneling diode.}\vspace*{-9pt}
\label{om_fig_single_tind}
\end{figure}

In \fref{om_fig_single_tind}, we have represented one Bohmian trajectory for a state incident from left to right upon a double-barrier potential profile. Its energy $E = 0.05$ eV corresponds to the first resonant level of the double barrier\cite{om.oriols_sst}. In this particular case, where the transmission coefficient is very close to unity, the results obtained within the Bohmian approach are quite compatible with our intuitive understanding of the tunneling phenomenon: the particle is transmitted, and its velocity decreases in the well. We do also plot the total potential-that is, the sum $V(x) + Q(x)$-which is lower than the electron energy $E = 0.05$ eV.

However, the situation is far from intuitive for nonresonant states. In \fref{om_fig_single_tind}b, we have represented one of these nonresonant trajectories for the same potential profile. Now, although the transmission coefficient is much smaller than unity for the nonresonant states, they present the same features as the resonant ones: all Bohmian trajectories are transmitted through the barrier. This fact can be easily understood from a mathematical point of view. As we mentioned, the current and particle densities are time independent and positive everywhere. So, the Bohmian velocity is always positive. In this regard, although Bohmian trajectories associated with scattering states perfectly reproduce the presence probability and the current density, they do not reproduce our particle-intuitive understanding of the tunneling phenomenon, since in principle, we would expect reflected as well as transmitted trajectories. The origin of this lack of intuitive behavior of Bohmian trajectories is only due to the initial wave function that we select. The velocity of Bohmian particles is neither related to the incident electron nor related to the reflected one but to the average of both because the wave function itself is defined as a weighted sum of waves with positive and negative momentum. In particular, if the barrier is infinitely high, then the reflection coefficient is equal to one and Bohmian's velocity is zero (one would expect positive and negative electron velocities). Transport is better formulated as a time-dependent process. A time dependent picture of transport involves a wave packet with a well defined initial velocity and, after the interaction, a well-defined reflected (or transmitted) velocity, but not a unique velocity with an average of incident and reflected velocities. 

All the 1D numerical algorithms previously discussed can be easily
generalized to 2D and 3D systems (see, for example, the solution of
the Schr\"odinger equation for computing Bohmian trajectories in
atomic scenarios in Refs. \cite{om.mompart1,om.mompart2}). However,
the computational burden of such techniques with finite-difference
meshes is enormous, and the solution using adaptative meshes seems interesting. For example, the triangular mesh generated using
an algorithm inspired by Ref. \cite{om.Persson}, with Delaunay refinement
algorithms, yields a final mesh that is surprisingly well shaped,
resulting in an excellent support to solve partial differential
equations, as it was remarked in Ref. \cite{om.Pinto}.\vspace*{-9pt}

\subsection{Synthetic computation of Bohmian trajectories}

\looseness-1Up to here, we have only discussed analytical algorithms to find
Bohmian trajectories. However, it is possible to compute Bohmian
trajectories without knowing the wave function. The key point is
using the polar form  $\psi(\vec r,t) = R(\vec r,t)\exp(iS(\vec
r,t)/\hbar)$ to obtain the quantum Hamilton--Jacobi equations. As
we anticipated, these new equations are nonlinear in $R(\vec r,t)$
and $S(\vec r,t)$. We will now discuss the advantages and
disadvantages of numerically solving the quantum Hamilton--Jacobi
equation, \eref{om.hamilton_jacobi1D_des}. Hereafter, we will
analyze two different algorithms to solve numerically
\eref{om.hamilton_jacobi1D_des}. First, directly solving these
equations (the Eulerian formalism) or some equivalent variable
change . Second rewriting the Hamilton--Jacobi equations within
the Langrangian formalism and solving them.

The first point that we want to clarify is that, although the
function $S(\vec r,t)$ is in principle multivalued, the numerical
solution of $S(\vec r,t)$ has a unique value at each point $\vec r$
and $t$. The reason is because of an important theorem on the
existence and uniqueness of solutions to first-order equations with
given initial-value problems. See page 70 in
\cite{om.perko}.

\subsubsection{Time-dependent quantum Hamilton--Jacobi equations (TDQHJE$_{1D}$) with an implicit (Newton-like fixed Eulerian mesh) method}

In the Eulerian formalism, we use a fixed grid. We are interested in providing a discussion valid for 1D, 2D, and 3D spatial scenarios. Therefore, we define a grid in real space through a set of discrete vectors $\vec p_i$ for $i = 1,\ldots,N$-in particular, $\vec p_i = (x_i,y_i,z_i)$ for a 3D space, $\vec p_i = (x_i,y_i)$ for a 2D space, and $\vec p_i = (x_i)$ for 1D. In order to provide a compact notation, we define:
\begin{eqnarray}
\vec {R}_{j} = \left(R_j(\vec{p}_{1}),\ldots,R_j(\vec{p}_{N})\right)^{T}\label{Rdef}\\
\vec {S}_{j} = \left(S_j(\vec{p}_{1}),\ldots,S_j(\vec{p}_{N})\right)^{T}\label{Sdef}\\
\vec {V}_{j} = \left(V_j(\vec{p}_{1}),\ldots,V_j(\vec{p}_{N})\right)^{T}\label{Vdef}
\end{eqnarray}
where $R_j(\vec p_i) = R(\vec r,t)$ at $t = t_j$ and $\vec r = \vec p_i$ (identically for $S_j(\vec p_i) = S(\vec r,t)$ and $V_j(\vec p_i) = V(\vec r,t)$). We have also used a mesh in time $t_j = j \Delta t$ for $j = 1,\ldots,M$

In the definition of the temporal derivatives of the quantum Hamilton--Jacobi, we can use an \textit{implicit} expression, \eref{om.implicit}, or an \textit{explicit} one, \eref{om.explicit}. The explicit expression will lead to an algorithm similar to the one described in \sref{timedependentexplicit} with the same advantages (simplicity) and disadvantages (i.e., conditional stable). In this section, we will explore the \textit{implicit} route. Then, the discrete version of the quantum Hamilton--Jacobi equations, \eref{om.hamilton_jacobi1D_des}, reads:
\begin{gather}
\frac{\vec {R}_{j}^{2}-\vec {R}_{j-1}^{2}}{\Delta t} + \hat{\nabla}\left(
\frac{\vec {R}_{j}^{2}\hat{\nabla}\vec {S}_{j}}{m}\right) = 0\label{QHJED}%
\nonumber\\[-10pt]\\
\frac{\vec {S}_{j}-\vec {S}_{j-1}}{\Delta t} + \frac{(\hat{\nabla}%
\vec {S}_{j})^{2}}{2m}-\frac{\hbar^{2}}{2m}\frac{\hat{\nabla}%
^{2}\vec {R}_{j}}{\vec {R_{j}}} + \vec {V_{j}} = 0\nonumber
\end{gather}
The operations such as power, multiplication, and division between the vectors $\vec X_j$ and $\vec Y_j$ are understood over each point in the spatial mesh, that is:
\begin{equation}
\vec {X}_{j}\vec {Y}_{j} = [X_j(\vec{p}_{1})Y_j(\vec{p}_{1}),\ldots,X_j(\vec{p}_{N})Y_j(\vec{p}_{N})]^{T}\label{prod def}%
\end{equation}
Here the symbol $\hat{\nabla}$ is the matrix representation of the differential operator $\vec{\nabla}$ in the 1D, 2D, or 3D spatial mesh. The representation of $\hat{\nabla}$ depends in general on the choice of the spatial mesh and on how the spatial derivatives are approximated. For example, in a 1D system with a uniform mesh step size $\Delta x$, the discrete version of the spatial derivative in one dimension is:
\begin{equation}
\frac {\partial R(x,t)} {\partial x} = \frac {R_j(\vec p_{i + 1}) - R_j(\vec p_{i - 1})} {2 \Delta x}
\label{om.derivada2},
\end{equation}
with $\vec p_i = x_0 + i \Delta x $. Then, the nabla operator in one dimension $\hat{\nabla}_{1D}$ is represented by the following matrix:
\begin{equation}
\hat{\nabla}_{1D} = \frac{1}{2\Delta x}\left(
\begin{array}
[c]{ccccc}%
2 & -2 & 0 & 0 &\ddots\\
1 & 0 & -1 & 0 & \ddots\\
\ddots & \ddots & \ddots & \ddots & \ddots\\
\ddots & 0 & 1 & 0 & -1\\
\ddots & 0 & 0 & -2 & 2
\end{array}
\right)
\end{equation}
Notice that we have changed the definition of the spatial derivative from \eref{om.derivada2} in $x_1$ and $x_N$. In two or three dimensions the matrices still have many zeros with the entries $1$ and $ - 1$ properly located.

Hereafter, as we mentioned, we only discuss the computational
problems of the \textit{implicit} solution of the Hamilton--Jacobi
equations. Such an equation is nonlinear with respect to $\vec
{R}_{j}$ and $\vec {S}_{j}$, and the matrix inversion explained in
\sref{timeindependentexplicit} cannot be done. Then, in order to
find the numerical solution of the discrete version of \eref{QHJED},
we have to employ some Newton-like method. If we rewrite in compact
form the \eref{QHJED} as:
\begin{equation}
F(\vec {R}_{j},\vec {S}_{j}) = 0 \label{QHJEDC}%
\end{equation}
then we have to find the roots (zeros) of the nonlinear expression in \eref{QHJEDC}. It is proved that the sequence:
\begin{equation}
\left(
\begin{array}
[c]{c}%
\vec {R}_{j}^{(h + 1)}\\
\vec {S}_{j}^{(h + 1)}%
\end{array}
\right) = -J_{F}^{-1}(\vec {R}_{j}^{(h)},\vec {S}_{j}^{(h)})\left(
\begin{array}
[c]{c}%
\vec {R}_{j}^{(h)}\\
\vec {S}_{j}^{(h)}%
\end{array}
\right) \label{Newton}%
\end{equation}
where $J_{F}(\vec {R}_{j}^{(h)},\vec {S}_{j}^{(h)})$ is the Jacobian
of $F(\vec {R}_{j}^{(h)},\vec {S}_{j}^{(h)})$, for suitable initial
vectors $\vec {R}_{j}^{(0)}$ and $\vec {S}_{j}^{(0)}$, converges quadratically  to
the solution of \eref{QHJEDC} for $h\rightarrow\infty$. Thus the key
of this approach is to solve the linear system in \eref{Newton}.
However evaluating explicitly the Jacobian:
{\arraycolsep0pt\begin{eqnarray}
&&J_{F}(\vec {R}_{j},\vec {S}_{j}) =\left(\!
\begin{array}
[c]{@{}c@{\ }c@{}}%
\frac{2}{\Delta t}\hat{diag}(\vec {R}_{j}) + \hat{\nabla}\left( \hat{diag}\left(\frac
{2\vec {R}_{j}\hat{\nabla}\vec {S}_{j}}{m}\right)\right) & \hat{\nabla
}\hat{diag}\left( \frac{\vec {R}_{j}^{2}}{m}\right) \hat{\nabla}\\
-\frac{\hbar^{2}}{2m}\hat{diag}\left(\! \frac{1}{\vec {R}_{j}}\!\right)
\hat{\nabla}^{2} - \hat{diag}\left(\! \frac{\hat{\nabla}^{2}\vec {R}_{j}}%
{\vec {R}_{j}^{2}}\!\right) & \frac{1}{\Delta t}\hat{diag}(\vec {S}%
_{j}) + \hat{diag}\left( \frac{\hat{\nabla}\vec {S}_{j}}{m}\right) \hat{\nabla}
\end{array}\!
\right)\nonumber\\ 
\end{eqnarray}}
where:
\begin{equation}
\hat{diag}(\vec {X}_{j}) = \left(
\begin{array}
[c]{ccccc}%
{X}_{j}(\vec p_1) & 0 & 0 & 0 &\ddots\\
0 & {X}_{j}(\vec p_2) & 0 & 0 & \ddots\\
\ddots & \ddots & \ddots & \ddots & \ddots\\
\ddots & 0 & 0 & {X}_{j}(\vec p_{N - 1}) & 0\\
\ddots & 0 & 0 & 0 & {X}_{j}(\vec p_N)
\end{array}
\right)
\end{equation}
we can readily recognize that for small components of $\vec {R}_{j}$, the Jacobian is ill conditioned and the system of \eref{Newton} is impossible to be solved numerically. Unfortunately this is the case of localized particles where $\vec {R}_{j}$ become exponentially small far from the central position of the particles. This problem does not appear with the \textit{explicit} algorithm for solving the quantum Hamilton--Jacobi equations. The problem can also be avoided by choosing the variable change $\psi(\vec r,t) = \exp(C(\vec r,t) + iS(\vec r,t)/\hbar)$ allowing to the equations:
\begin{gather}
\frac{\partial}{\partial t}C(\vec r,t) = -\frac{1}{2m}\left( \nabla^{2}S(\vec r,t) + 2\vec{\nabla}
S(\vec r,t)\cdot\vec{\nabla} C(\vec r,t)\right) \label{QHJE2}
\nonumber\\[-10pt]\\
\frac{\partial}{\partial t}S(\vec r,t) = -\frac{(\vec{\nabla} S(\vec r,t))^{2}}{2m} + \frac{\hbar^{2}}%
{2m}\left( \nabla^{2}C(\vec r,t) + (\vec{\nabla} C(\vec r,t))^{2}\right) -V(\vec r,t)\nonumber
\end{gather}
However, the advantage of this method carries out also its
disadvantage. The limitation of this approach is that the functions
$S(\vec r,t)$ and $C(\vec r,t)$ strongly vary in the entire
simulation space. Thus when we account for a realistic situation for
which a large simulation space is used, the computational problem
can become very hard because the number of space mesh points
$\vec{p}_{k}$ will be quite large to obtain an accurate solution.

\subsubsection{Time-dependent quantum Hamilton--Jacobi equations (TDQHJE$_{1D}$) with an explicit (Lagrangian mesh) method}

A different approach to solve the quantum Hamilton--Jacobi equation
is based on the Lagrangian picture. The basic idea is to use as a
grid the Bohmian trajectories themselves. In this picture, the point
$\vec{p}_k$ is no longer a fixed spatial point but a time-dependent
point $\vec{p}_k = \vec{p}_k(t)$ with $k = 1,\ldots,N$ moving with
instantaneous velocity $\vec{v}_k(t)$. For a detailed analysis see
Refs. \cite{om.Wyatt1, om.Wyatt2, om.Wyatt3, om.Frederick}.

To pass from the Eulerian to the Lagrangian hydrodynamic picture, we
must  be careful with the time derivative of the functions
$R(\vec{p}_{k}(t),t)$, $S(\vec{p}_{k}(t),t)$ and
$V(\vec{p}_{k}(t),t)$. We have to use the chain rule:
\begin{equation}
\frac{d}{dt} = \frac{\partial}{\partial t} + \vec v(\vec r,t)\cdot\vec{\nabla}
\end{equation}
which includes the \textit{convective} term \textit{$\vec{v}(\vec r,t)\cdot\vec{\nabla}$} with the velocity defined:
\begin{equation}
\vec{v}(\vec r,t) = \frac{1}{m}\vec{\nabla} S(\vec r,t)
\end{equation}
The new equations derived from the quantum Hamilton--Jacobi equations (or equivalently from \eref{QHJE2}) can be written as:
\begin{gather}
\frac{d}{dt}\vec{p}_k(t) = \vec{v}(\vec{p}_k(t),t)\label{traj1}\\[0.1in]
\frac{d}{dt}S(\vec{p}_k(t),t) = \frac{(\vec{\nabla} S(\vec{p}_k(t),t))^{2}}{2m} + \frac{\hbar^{2}%
}{2m}\frac{\nabla^{2}R(\vec{p}_k(t),t)}{R(\vec{p}_k(t),t)} - V(\vec{p}_k(t),t)\label{traj2}\\[0.1in]
\vec{v}(\vec{p}_k(t),t) = \frac{1}{m}\vec{\nabla} S(\vec{p}_k(t),t)\label{traj3}\\[0.1in]
\frac{d}{dt}R(\vec{p}_k(t),t) = -\frac{1}{2}R(\vec{p}_k(t),t)\vec{\nabla}\cdot\vec{v}(\vec
{p}_k(t))\label{traj4}%
\end{gather}
The differential system in Eqs. (\ref{traj1}--\ref{traj4}) is written for the trajectory $\vec{p}_k(t)$ with a particular initial position $\vec{p}_k(0)$. However to (numerically) reproduce the wave function $\psi(\vec r,t) = R(\vec r,t)\exp(iS(\vec r,t)/\hbar)$ defined in the entire space, we have to deal with $k = 1,\ldots,N$ trajectories with $N\rightarrow\infty$. The distribution of their initial positions $\vec{p}_k(0)$ has to be obtained according to the squared modulus of the initial wave function $|\psi(\vec p_k(0),0)|^2$, that is, the quantum equilibrium hypothesis described in \sref{om.sec_single.6}. See also Ref. \cite{om.Holand1993}. Thus, by construction we are using a mesh that is continuously adapting to the wave function time evolution because the Bohmian trajectories themselves follow the wave function.

In this picture the strategy to solve Eqs. (\ref{traj1}--\ref{traj4})
can be very complicated \cite{om.Wyatt1, om.Wyatt2, om.Wyatt3,
om.Frederick}. However, some simple consideration can be done in
order to understand the limits of this approach. The method widely
used \cite{om.Wyatt1, om.Wyatt2, om.Wyatt3, om.Frederick} to
integrate Eqs. (\ref{traj1}--\ref{traj4}) is to solve them in
cascade with the following approximations: at time $t_{j}$ the new
trajectories positions are computed through:
\begin{equation}
\vec {p}_{j,k} = \vec {p}_{j-1,k} + \Delta t\vec {v}_{j - 1,k}
\end{equation}
with $\vec {p}_{j,k} = \vec {p}_{k}(t_j)$. In the updated grid defined by $\vec {p}_{j,k}$, the variables ${S}_{j}(\vec {p}_{j,k})$, $\vec {v}_{j}(\vec {p}_{j,k})$, and $\vec {R}_{j}(\vec {p}_{j,k})$ are evaluated by:
\begin{eqnarray}
\hspace*{-24pt}{S}_{j}(\vec {p}_{j,k}) & =& {S}_{j - 1}(\vec {p}_{j,k}) + \Delta t\Bigg[ \frac{(\hat{\nabla}%
_{j - 1} {S}_{j - 1}(\vec {p}_{j,k}))^{2}}{2m}+\, \frac{\hbar^{2}}{2m}\frac{\hat{\nabla
}_{j - 1}^{2}{R}_{j - 1}(\vec {p}_{j,k})}{{R}_{j - 1}(\vec {p}_{j,k})} - \vec {V}_{j - 1}(\vec {p}_{j,k})\Bigg]\nonumber\\
\hspace*{-24pt}{R}_{j}(\vec {p}_{j,k}) & = &{R}_{j - 1}(\vec {p}_{j,k}) - \Delta t\left[ \frac{1}{2m}%
{R}_{j - 1}(\vec {p}_{j,k})\hat{\nabla}_{j - 1}^{2}\vec {S}_{j - 1}(\vec {p}_{j,k})\right] \nonumber\\
\hspace*{-24pt}{v}_{j}(\vec {p}_{j,k}) & = &\frac{1}{m}\hat{\nabla}_{j}\vec {S}_{j}(\vec {p}_{j,k})\nonumber
\end{eqnarray}
where $\hat{\nabla}_{j}$ is the gradient operator evaluated in $\vec {p}_{j,k}$ through efficient interpolation methods \cite{om.Wyatt1, om.Wyatt2, om.Wyatt3, om.Frederick}. It can be realized that we have used an (unstable) explicit (forward Euler) time integration method
 \cite{om.Striwerda} to avoid the computational burden associated with implicit methods with nonlinear equations discussed before. This limitation (small temporal step to ensure stable solutions) is made worse since the time step must be further reduced when some trajectory approaches points for which the potential $V$ and/or the quantum potential $Q$ strongly varies provoking large variations of some $\vec{v}_{j}$.

The  variables $S$ and
$\vec{v}$ vary in the entire simulation space; thus when we account
for a realistic situation such that a large simulation space is
used, the computational problem can become very hard due to the fact
that the number of space mesh points $\vec{p}_{k}$ can diverge to
obtain an accurate solution. On the contrary, this method has some
fundamental advantage over methods using fix grid points. In the
latter, the extension of 1D algorithms to 2D and 3D algorithms means
increasing exponentially the number of grip points. However, in the
former methods (i.e., when the grid points are Bohmian
trajectories), the number of  points can be
chosen by the user. Thus, for exemple, one can simulate a 1D system
with 1,000 Bohmian trajectories (grid points) and a 2D system with
~3,000 Bohmian trajectories (grid points).

\subsection{More elaborated algorithms}

We have presented very simple and easily implementable codes for
computing Bohmian trajectories in simple scenarios. These algorithms
can be used for newcomers to be introduced to the computation of
quantum dynamics with Bohmian mechanics. Certainly, there are many
more elaborated and ``professional'' computational algorithms  in the
literature. An excellent source of information on this issue can be
found in the book written by R. E. Wyatt entitled \textit{Quantum
Dynamics with Trajectories: Introduction to Quantum Hydrodynamics}
\cite{om.wyatt2005}.

To end this section, we want to briefly mention some original and
powerful techniques, inspired by Bohmian mechanics, available in the
literature, whose explicit description is far from the scope of this
introductory appendix. Recently, B. Poirier showed how Bohmian
trajectories can be computed for 1D eigenstates without the wave
function \cite{om.poirier}. The work of N. Makri shows the possibility of using
imaginary time algorithms to estimate eigenstates and eigenvalues
\cite{om.imaginarytime}. With a similar goal, I. P. Christov uses
quantum trajectories to develop time-dependent quantum Monte Carlo
algorithms \cite{om.ivan,om.ivan2}. We also mention the work of E.
L. Bittner \cite{om.extra16} developed in chapter 5 in this book.
The work presented in \sref{om.sec_many.6} can also be interpreted as an
approximating technique for computing many-particle systems
\cite{om.oriolsprl}. This last work has been used by the group of
Oriols to develop a powerful quantum electron transport simulator. (
http://europe.uab.es/bitlles). See Refs.
\cite{om.oriolsexample1,om.oriolsexample2} or chapter 6 for
practical results on electron correlation in nanoelectronics.
Another line of work is also devoted to the approximation of the
quantum potential. See, for example, the work of S. Garashchuk and
V. A. Rassolov \cite{om.gara}. The correlation between classical and
quantum systems has also been studied extensively. See, for example,
Ref. \cite{om.extra13,om.5marian,om.6colomes}. We do also highlight the work by D. J.
Tannor and coworkers on the use of imaginary quantum trajectories
(complex action) for quantum dynamics
\cite{om.imaginaryaction,om.imaginaryactionprl}. We also want to notice the important amount of work done by many people in the field of ab-initio molecular dynamics, specially for tackling the coupled electron-nuclear motion in nonadiabatic electronic transitions \cite{om.extra18,om.extra19,om.extra20,om.extra21,om.extra22,om.extra23,om.extra24}. Finally, we mention the more recent work by L. B. Madsen and co-workers in the field of bosonic systems \cite{om.extra25}.

All the works mentioned have been elaborated during the past 18
years after the pioneering work of Wyatt
\cite{om.Wyatt1,om.extra17}. Much work is still needed to evaluate
the real capability of synthetic applications of Bohmian mechanics
as a computational tool.

\markright{References}
\setcounter{enumiv}{0}

\end{document}